\documentclass[prb,twocolumn,superscriptaddress,10pt,longbibliography]{revtex4-1}
\usepackage{morewrites}
\usepackage{stackrel}
\usepackage[binary-units]{siunitx}
\usepackage{color}
\definecolor{mygreen}{rgb}{0,0.5,0} 
\definecolor{mygrey}{rgb}{0.5,0.5,0.5} 
\definecolor{myred}{rgb}{0.75,0,0} 
\definecolor{myblue}{rgb}{0,0,0.75} 
\definecolor{mymagenta}{cmyk}{0,1,0,0.12} 
\definecolor{mycyan}{cmyk}{1,0,0,0.12} 
\definecolor{myorange}{rgb}{1,0.5,0} 
\definecolor{myviolet}{rgb}{0.5,0.0,0.75} 

\newcommand{\btext}[1]{{\color{myblue}#1}}
\newcommand{\newbtext}[1]{{\color{myblue}#1}}
\renewcommand{\newbtext}[1]{#1}

\newcommand{\stext}[1]{{\color{myred}\sout{#1}}}
\renewcommand{\btext}[1]{#1}
\renewcommand{\stext}[1]{}

\usepackage[pdftex]{graphicx}
\usepackage{ulem}
\usepackage{hyperref}
\usepackage{bm}
\usepackage{pifont}
\graphicspath{{./}{./}}
\usepackage{anyfontsize}
\newcommand{\circled}[1]{\textcircled{\raisebox{0.25pt}{\fontsize{6}{7}\selectfont\hspace{-2.5pt} #1}}}
           
\usepackage{sfmath}

\usepackage[]{multibib}
\newcites{SI,EQUS,GU,USTC,IQOQI,ROMA,LMU,ETHZ,NICE,ICFOONE,ICFOTWO,BA,UdeC,NIST}{null,null,null,null,null,null,null,null,null,null,null,null,null,null}

\newcommand{\ket}[1]{|#1\rangle}

\newcommand{\GRIFFITH}{\circled{1}}
\newcommand{\EQUS}{\circled{2}}
\newcommand{\USTC}{\circled{3}}
\newcommand{\IQOQI}{\circled{4}}
\newcommand{\SAPIENZA}{\circled{5}}
\newcommand{\LMU}{\circled{6}}
\newcommand{\ETHZ}{\circled{7}}
\newcommand{\NICE}{\circled{8}}
\newcommand{\ICFOONE}{\circled{9}}
\newcommand{\ICFOTWO}{\circled{10}}
\newcommand{\UBA}{\circled{11}}
\newcommand{\UC}{\circled{12}}
\newcommand{\NIST}{\circled{13}}

\usepackage{geometry}
\newcommand{\mcite}[1]{\cite{#1}}
\newcommand{\pcite}[1]{\cite{#1}.}
\newcommand{\ccite}[1]{\cite{#1},}

\newcommand{\mciteSI}[1]{~\citeSI{#1}}
\newcommand{\pciteSI}[1]{~\citeSI{#1}.}
\newcommand{\cciteSI}[1]{~\citeSI{#1},}

\usepackage{array}
\newcolumntype{L}[1]{>{\raggedright\let\newline\\\arraybackslash\hspace{0pt}}m{#1}}
\newcolumntype{C}[1]{>{\centering\let\newline\\\arraybackslash\hspace{0pt}}m{#1}}
\newcolumntype{R}[1]{>{\raggedleft\let\newline\\\arraybackslash\hspace{0pt}}m{#1}}

\usepackage[font=small,labelfont=bf,
   justification=raggedright,
   format=plain]{caption}
\usepackage{subcaption}

\usepackage{mathtools}

\usepackage{array}
\newcolumntype{L}[1]{>{\raggedright\let\newline\\\arraybackslash\hspace{0pt}}m{#1}}
\newcolumntype{C}[1]{>{\centering\let\newline\\\arraybackslash\hspace{0pt}}m{#1}}
\newcolumntype{R}[1]{>{\raggedleft\let\newline\\\arraybackslash\hspace{0pt}}m{#1}}

\newcommand{\Bellsters}{\textit{Bellsters}}
\newcommand{\Bellster}{\textit{Bellster}}

\newcommand{\BBTGlobalTitle}{{Challenging local realism with  human choices}}

\begin{document}



\newcounter{NAuthors}
\newcommand{\BBTAuthor}[1]{\stepcounter{NAuthors}\author{#1}}
\setcounter{page}{0}
\title{ \Large \BBTGlobalTitle}

\newcommand{\ICFOAddress}{ICFO-Institut de Ciencies Fotoniques, The Barcelona Institute of Science and Technology, 08860 Castelldefels (Barcelona), Spain} 
\newcommand{\ICREAAddress}{ICREA -- Instituci\'{o} Catalana de Re{c}erca i Estudis Avan\c{c}ats, 08010 Barcelona, Spain} 
\newcommand{\IQOQIAddress}{Institute for Quantum Optics and Quantum Information (IQOQI), Austrian Academy of Sciences, Boltzmanngasse 3, 1090 Vienna, Austria}
\newcommand{\UdeCAddress}{Departamento de F\'{i}sica, Universidad de Concepci\'{o}n, 160-C Concepci\'{o}n, Chile}
\newcommand{\CEFOPAddress}{Center for Optics and Photonics, Universidad de Concepci\'{o}n, Casilla 4016, Concepci\'{o}n, Chile}
\newcommand{\MSIAddress}{MSI-Nucleus for Advanced Optics, Universidad de Concepci\'{o}n, 160-C Concepci\'{o}n, Chile}
\newcommand{\CEFOPMSIAddress}{Center for Optics and Photonics and MSI-Nucleus for Advanced Optics, Universidad de Concepci\'{o}n, Casilla 4016, Concepci\'{o}n, Chile}
\newcommand{\DIEAddress}{Departamento de Ingenier\'{i}a El\'{e}ctrica, Universidad de Concepci\'{o}n,160-C Concepci\'{o}n, Chile}
\newcommand{\SevillaAddress}{Departamento de F\'{i}sica Aplicada II, Universidad de Sevilla, 41012 Sevilla, Spain}
\newcommand{\LinkopingAddress}{Department of Electrical Engineering, Link\"{o}ping University,  581 83 Link\"{o}ping, Sweden}
\newcommand{\SapienzaAddress}{Dipartimento di Fisica, Sapienza Universit\`{a} di Roma, I-00185 Roma, Italy}
\newcommand{\EQUSAddress}{Centre for Engineered Quantum Systems, Centre for Quantum Computation and Communication Technology, School of Mathematics and Physics, University of Queensland, Brisbane, Queensland 4072, Australia}
\newcommand{\GriffithAddress}{Griffith University, Brisbane, Queensland 4111, Australia}
\newcommand{\CQDAddress}{Centre for Quantum Dynamics, Griffith University, Brisbane, Queensland 4111, Australia}
\newcommand{\CQCCTAddress}{Centre for Quantum Computation and Communication Technology, Griffith University, Brisbane, Queensland 4111, Australia}
\newcommand{\INFINIAddress}{Laboratoire de Physique de la Matire Condens\'{e}e, Universit\'{e} Nice Sophia Antipolis, CNRS UMR 7336, Parc Valrose, 06108 Nice Cedex 2, France}
\newcommand{\INPHYNIAddress}{Universit\'{e} C\^{o}te d'Azur, CNRS UMR 7010, \newbtext{Institut} de Physique de Nice (INPHYNI), Parc Valrose, 06108 Nice Cedex 2, France}
\newcommand{\ETHQDLAddress}{Quantum Device Lab, Department of Physics, ETH Zurich, CH-8093 Zurich, Switzerland}
\newcommand{\ETHAddress}{Department of Physics, ETH Zurich, CH-8093 Zurich, Switzerland}
\newcommand{\LMUAddress}{Ludwig-Maximilians-Universit\"{a}t, 80799 M\"{u}nchen, Germany}
\newcommand{\MPQAddress}{Max-Planck-Institut f\"{u}r Quantenoptik, Hans-Kopfermann-Strasse 1, Garching 85748, Germany}
\newcommand{\UBAAddress}{Departamento de F\'{i}sica, FCEyN, UBA and IFIBA, Conicet, Pabell\'{o}n 1, Ciudad Universitaria, 1428 Buenos Aires, Argentina}

\newcommand{\DdCUBAAddress}{Departamento de Computaci\'{o}n, FCEyN, UBA and ICC, CONICET, Pabell\'{o}n 1, Ciudad Universitaria, 1428 Buenos Aires, Argentina}

\newcommand{\CITEDEFAddress}{DEILAP, CITEDEF \& CONICET, J.B. de La Salle 4397, 1603 Villa Martelli, Buenos Aires, Argentina}
\newcommand{\NISTAddress}{National Institute of Standards and Technology, 325 Broadway, Boulder, CO 80305, USA}
\newcommand{\IIPURFNAddress}{International Institute of Physics, Federal University of Rio Grande do Norte, 59070-405 Natal, Brazil}
\newcommand{\CENSEAddress}{CAS Center for Excellence in Superconducting Electronics, Shanghai 200050, China}
\newcommand{\SIMITAddress}{State Key Laboratory of Functional Materials for Informatics, Shanghai Institute of Microsystem and Information Technology, Chinese Academy of Sciences, Shanghai 200050, China}
\newcommand{\PMOAddress}{Purple Mountain Observatory and Key Laboratory of Radio Astronomy, Chinese Academy of Sciences, 2 West Beijing Road, Nanjing 210008, China}
\newcommand{\TsinghuaAddress}{Center for Quantum Information, Institute for Interdisciplinary Information Sciences, Tsinghua University, Beijing 100084, China}
\newcommand{\USTCAddress}{Shanghai Branch, National Laboratory for Physical Sciences at Microscale and Dept. of Modern Physics, University of Science and Technology of China, Shanghai 230026, China}
\newcommand{\CASAddress}{Shanghai Branch, CAS Center for Excellence and Synergetic Innovation Center in Quantum Information and Quantum Physics, University of Science and Technology of China, Shanghai 230026, China}
\newcommand{\NUDTAddress}{School of Computer, NUDT, 410073 Changsha, China}

\newcommand{\BBTAuthorCoord}[1]{\BBTAuthor{#1} \affiliation{\ICFOAddress}}

\BBTAuthor{$^\dagger$The BIG Bell Test Collaboration: C. Abell\'{a}n}
\affiliation{\ICFOAddress}

\BBTAuthor{A. Ac\'{i}n}
\affiliation{\ICFOAddress}
\affiliation{\ICREAAddress}

\BBTAuthor{A. Alarc\'{o}n}
\affiliation{\CEFOPMSIAddress}
\affiliation{\DIEAddress}

\BBTAuthor{O. Alibart}
\affiliation{\INPHYNIAddress}


\BBTAuthor{C. K. Andersen}
\affiliation{\ETHQDLAddress}

\BBTAuthor{F. Andreoli}
\affiliation{\SapienzaAddress}

\BBTAuthor{A. Beckert} 
\affiliation{\ETHQDLAddress}

\BBTAuthorCoord{F. A. Beduini}

\BBTAuthor{A. Bendersky}
\affiliation{\DdCUBAAddress}

\BBTAuthor{M. Bentivegna}
\affiliation{\SapienzaAddress}

\BBTAuthor{P. Bierhorst}
\affiliation{\NISTAddress}

\BBTAuthor{D. Burchardt}
\affiliation{\LMUAddress}

\BBTAuthor{A. Cabello}
\affiliation{\SevillaAddress}

\BBTAuthor{J. Cari\~{n}e}
\affiliation{\CEFOPMSIAddress}
\affiliation{\DIEAddress}
\affiliation{\UdeCAddress}

\BBTAuthorCoord{S. Carrasco}

\BBTAuthor{G. Carvacho}
\affiliation{\SapienzaAddress}

\BBTAuthor{D. Cavalcanti}
\affiliation{\ICFOAddress}

\BBTAuthor{R. Chaves}
\affiliation{\IIPURFNAddress}

\BBTAuthor{J. Cort\'{e}s-Vega}
\affiliation{\CEFOPMSIAddress}
\affiliation{\UdeCAddress}

\BBTAuthor{A. Cuevas}
\affiliation{\SapienzaAddress}

\BBTAuthor{A. Delgado}
\affiliation{\CEFOPMSIAddress}
\affiliation{\UdeCAddress}

\BBTAuthor{H. de Riedmatten}
\affiliation{\ICFOAddress}
\affiliation{\ICREAAddress}

\BBTAuthor{C. Eichler}
\affiliation{\ETHQDLAddress}

\BBTAuthor{P. Farrera}
\affiliation{\ICFOAddress}


\BBTAuthor{J. Fuenzalida}
\affiliation{\CEFOPMSIAddress}
\affiliation{\UdeCAddress}
\affiliation{\IQOQIAddress}

\BBTAuthorCoord{M. Garc\'{i}a-Matos}

\BBTAuthor{R. Garthoff}
\affiliation{\LMUAddress}

\BBTAuthor{S. Gasparinetti}
\affiliation{\ETHQDLAddress}

\BBTAuthor{T. Gerrits}
\affiliation{\NISTAddress}

\BBTAuthor{F. Ghafari Jouneghani}
\affiliation{\CQCCTAddress}
\affiliation{\CQDAddress }


\BBTAuthor{S. Glancy}
\affiliation{\NISTAddress}

\BBTAuthor{E. S. G\'{o}mez}
\affiliation{\CEFOPMSIAddress}
\affiliation{\UdeCAddress}

\BBTAuthor{P. Gonz\'{a}lez}
\affiliation{\CEFOPMSIAddress}
\affiliation{\UdeCAddress}

\BBTAuthor{J.-Y. Guan}
\affiliation{\USTCAddress}
\affiliation{\CASAddress}

\BBTAuthor{J. Handsteiner}
\affiliation{\IQOQIAddress}

\BBTAuthor{J. Heinsoo}
\affiliation{\ETHQDLAddress}

\BBTAuthor{G. Heinze}
\affiliation{\ICFOAddress}

\BBTAuthorCoord{A. Hirschmann}

\BBTAuthor{O. Jim\'{e}nez}
\affiliation{\ICFOAddress}

\BBTAuthor{F. Kaiser}
\affiliation{\INPHYNIAddress}

\BBTAuthor{E. Knill}
\affiliation{\NISTAddress}
%
%
%
%

\BBTAuthor{L. T. Knoll}
\affiliation{\CITEDEFAddress}
\affiliation{\UBAAddress}

\BBTAuthor{S. Krinner}
\affiliation{\ETHQDLAddress}

\BBTAuthor{P. Kurpiers}
\affiliation{\ETHQDLAddress}

\BBTAuthor{M. A. Larotonda}
\affiliation{\CITEDEFAddress}
\affiliation{\UBAAddress}

\BBTAuthor{J.-\AA. Larsson}
\affiliation{\LinkopingAddress}

\BBTAuthor{A. Lenhard}
\affiliation{\ICFOAddress}

\BBTAuthor{H. Li}
\affiliation{\SIMITAddress}
\affiliation{\CENSEAddress}

\BBTAuthor{M.-H. Li}
\affiliation{\USTCAddress}
\affiliation{\CASAddress}

\BBTAuthor{G. Lima}
\affiliation{\CEFOPMSIAddress}
\affiliation{\UdeCAddress}

\BBTAuthor{B. Liu}
\affiliation{\NUDTAddress}
\affiliation{\IQOQIAddress}

\BBTAuthor{Y. Liu}
\affiliation{\USTCAddress}
\affiliation{\CASAddress}

\BBTAuthor{I. H. L\'{o}pez Grande}
\affiliation{\CITEDEFAddress}
\affiliation{\UBAAddress}

\BBTAuthor{T. Lunghi}
\affiliation{\INPHYNIAddress}

\BBTAuthor{X. Ma}
\affiliation{\TsinghuaAddress}

\BBTAuthor{O. S. Maga\~{n}a-Loaiza}
\affiliation{\NISTAddress}

\BBTAuthor{P. Magnard}
\affiliation{\ETHQDLAddress}

\BBTAuthor{A. Magnoni}
\affiliation{\UBAAddress}

\BBTAuthorCoord{M. Mart\'{i}-Prieto}

\BBTAuthor{D. Mart\'{i}nez}
\affiliation{\CEFOPMSIAddress}
\affiliation{\UdeCAddress}

\BBTAuthor{ P. Mataloni}
\affiliation{\SapienzaAddress}

\BBTAuthor{A. Mattar}
\affiliation{\ICFOAddress}

\BBTAuthor{M. Mazzera}
\affiliation{\ICFOAddress}

\BBTAuthor{R. P. Mirin}
\affiliation{\NISTAddress}

\BBTAuthor{M. W. Mitchell}
\email{morgan.mitchell@icfo.eu}
\affiliation{\ICFOAddress}
\affiliation{\ICREAAddress}

\BBTAuthor{S. Nam}
\affiliation{\NISTAddress}

\BBTAuthor{M. Oppliger}
\affiliation{\ETHQDLAddress}


\BBTAuthor{J.-W. Pan}
\affiliation{\USTCAddress}
\affiliation{\CASAddress}

\BBTAuthor{R. B. Patel}
\affiliation{\CQCCTAddress}
\affiliation{\CQDAddress }

\BBTAuthor{G. J. Pryde}
\affiliation{\CQCCTAddress}
\affiliation{\CQDAddress }

\BBTAuthor{D. Rauch}
\affiliation{\IQOQIAddress}

\BBTAuthor{K. Redeker}
\affiliation{\LMUAddress}

\BBTAuthor{D. Riel\"{a}nder}
\affiliation{\ICFOAddress}

\BBTAuthor{M. Ringbauer}
\affiliation{\EQUSAddress}

\BBTAuthor{T. Roberson}
\affiliation{\EQUSAddress}

\BBTAuthor{W. Rosenfeld}
\affiliation{\LMUAddress}

\BBTAuthor{Y. Salath\'{e}}
\affiliation{\ETHQDLAddress}

\BBTAuthor{L. Santodonato}
\affiliation{\SapienzaAddress}

\BBTAuthor{G. Sauder}
\affiliation{\INPHYNIAddress}

\BBTAuthor{T. Scheidl}
\affiliation{\IQOQIAddress}

\BBTAuthor{C. T. Schmiegelow}
\affiliation{\UBAAddress}

\BBTAuthor{F. Sciarrino}
\affiliation{\SapienzaAddress}

\BBTAuthor{A. Seri}
\affiliation{\ICFOAddress}

\BBTAuthor{L. K. Shalm}
\affiliation{\NISTAddress}

\BBTAuthor{S.-C. Shi}
\affiliation{\PMOAddress}

\BBTAuthor{S. Slussarenko}
\affiliation{\CQCCTAddress}
\affiliation{\CQDAddress }

\BBTAuthor{M. J. Stevens}
\affiliation{\NISTAddress}

\BBTAuthor{S. Tanzilli}
\affiliation{\INPHYNIAddress}


\BBTAuthor{F. Toledo}
\affiliation{\CEFOPMSIAddress}
\affiliation{\UdeCAddress}

\BBTAuthor{J. Tura }
\affiliation{\ICFOAddress}
\affiliation{\MPQAddress}

\BBTAuthor{R. Ursin}
\affiliation{\IQOQIAddress}

\BBTAuthor{P. Vergyris}
\affiliation{\INPHYNIAddress}

\BBTAuthor{V. B. Verma}
\affiliation{\NISTAddress}

\BBTAuthor{T. Walter}
\affiliation{\ETHQDLAddress}

\BBTAuthor{A. Wallraff}
\affiliation{\ETHQDLAddress}

\BBTAuthor{Z. Wang}
\affiliation{\SIMITAddress}
\affiliation{\CENSEAddress}

\BBTAuthor{H. Weinfurter}
\affiliation{\LMUAddress}
\affiliation{\MPQAddress }

\BBTAuthor{M. M. Weston}
\affiliation{\CQCCTAddress}
\affiliation{\CQDAddress }


\BBTAuthor{A. G. White}
\affiliation{\EQUSAddress}

\BBTAuthor{C. Wu}
\affiliation{\USTCAddress}
\affiliation{\CASAddress}

\BBTAuthor{G. B. Xavier}
\affiliation{\CEFOPMSIAddress}
\affiliation{\DIEAddress}
\affiliation{\LinkopingAddress}

\BBTAuthor{L. You}
\affiliation{\SIMITAddress}
\affiliation{\CENSEAddress}

\BBTAuthor{X. Yuan}
\affiliation{\TsinghuaAddress}

\BBTAuthor{A. Zeilinger}
\affiliation{\IQOQIAddress}

\BBTAuthor{Q. Zhang}
\affiliation{\USTCAddress}
\affiliation{\CASAddress}

\BBTAuthor{W. Zhang}
\affiliation{\SIMITAddress}
\affiliation{\CENSEAddress}

\BBTAuthor{J. Zhong}
\affiliation{\PMOAddress}

\date{\today}
\maketitle
\vspace{1cm}
\newpage ~
\clearpage ~ 

\onecolumngrid
\noindent 
{\huge Challenging local realism with human \btext{choices} \stext{randomness}}\\ ~ \\
{\Large  The BIG Bell Test Collaboration$^{\dagger}$}\\ ~ \\ 
\twocolumngrid

\noindent 

\textbf{
A Bell test is a randomized trial that compares experimental
observations against the philosophical worldview of local realism\ccite{BellP1964}
in which the properties of the physical world are independent of our
observation of them and no signal travels faster than light. A Bell test
requires spatially distributed entanglement, fast and high-efficiency
detection and unpredictable measurement settings\pcite{LarssonJPA2014Official,KoflerPRA2016}  Although
technology can satisfy the first two of these requirements\ccite{HensenN2015, GiustinaPRL2015, ShalmPRL2015, RosenfeldPRL2017} the use
of physical devices to choose settings in a Bell test involves making
assumptions about the physics that one aims to test. Bell himself
noted this weakness in using physical setting choices and argued that
human `free will' could be used rigorously to ensure unpredictability
in Bell tests\pcite{BellBook2004Ch7} Here we report a set of local-realism tests using human
choices, which avoids assumptions about predictability in physics.
We recruited about 100,000 human participants to play an online
video game that incentivizes fast, sustained input of unpredictable
selections and illustrates Bell-test methodology\pcite{BBTWebsite} The participants
generated 97,347,490 binary choices, which were directed via a
scalable web platform to 12 laboratories on five continents, where
13 experiments tested local realism using photons\ccite{GiustinaPRL2015,ShalmPRL2015} single atoms\ccite{RosenfeldPRL2017}
atomic ensembles\ccite{FarreraNComms2016} and superconducting devices\pcite{WallraffN2004} Over a 12-hour
period on 30 November 2016, participants worldwide provided a
sustained data flow of over 1,000 bits per second to the experiments,
which used different human-generated data to choose each
measurement setting. The observed correlations strongly contradict
local realism and other realistic positions in bipartite and tripartite\mcite{CarvachoNC2017}
scenarios. Project outcomes include closing the `freedom-of-choice
loophole' (the possibility that the setting choices are influenced by
`hidden variables' to correlate with the particle properties\mcite{ScheidlPNAS2010}), the
utilization of video-game methods\mcite{SorensenN2016} for rapid collection of human generated
randomness, and the use of networking techniques for
global participation in experimental science.
}

Bell tests, like Darwin's studies of finches and Galileo's observations of the moons of Jupiter, bring empirical methods to questions previously accessible only by other means, e.g. by philosophy or theology\pcite{ShimonySEP2005} Local realism, i.e., realism plus relativistic limits on causation, was debated by Einstein and Bohr using metaphysical arguments, and recently has been rejected by Bell tests\mcite{HensenN2015, GiustinaPRL2015, ShalmPRL2015, RosenfeldPRL2017} that closed all technical `loopholes.'  For example, the `detection-efficiency loophole' describes the possibility that the observed statistics are inaccurate due to selection bias, and is closed by high efficiency detection and statistical methods that analyse all trials. Recent work on device-independent quantum information\mcite{ColbeckThesis2007} shows how Bell inequality violation (BIV) can also challenge causal determinism\ccite{HoeferSEP2005} a second topic formerly accessible only by metaphysics\pcite{AcinN2016}  Central to both applications is the use of free variables to choose measurements: in the words of Aaronson\mcite{AaronsonAS2014}  ``Assuming no preferred reference frames or closed timelike curves, if Alice and Bob have genuine `freedom' in deciding how to measure entangled particles, then the particles must also have `freedom' in deciding how to respond to the measurements.''   \stext{Provable indeterminism is useful in communications security\pcite{PironioN2010}}

Prior Bell tests used physical devices\mcite{AbellanPRL2015,FurstMOE2010} to `decide' for Alice and Bob, and thus demonstrated only a relation among physical processes: if some processes are `free' in the required sense (see Methods \ref{Sec:Freedom}), then other processes are similarly `free.' In the language of strong Bell tests, this conditional relation leaves open the freedom-of-choice
loophole (FOCL), which describes the possibility that `hidden
variables' influence the setting choices.
Because we cannot guarantee such freedom within local realism, the tests must assume  physical indeterminacy in the hidden-variable theory\pcite{LarssonJPA2014Official}  Laboratory methods can tighten but never close this loophole\ccite{LarssonJPA2014Official,HensenN2015, GiustinaPRL2015, ShalmPRL2015,KoflerPRA2016}.

Gallicchio, Friedman, and Kaiser \mcite{GallicchioPRL2014} have proposed choosing settings by observation of cosmic sources at the edge of the visible universe. A Bell inequality violation under such conditions could only be explained within local realism if events across all of history conspire to produce the measured outcomes\pcite{HandsteinerPRL2017,WuPRL2017}  Bell himself argued that human choices could be considered `free variables' in a Bell test 
\mcite{BellBook2004Ch7} (see Methods \ref{Sec:JSB}), and noted the impracticality of using humans with 1970's technologies. Here we implement Bell's idea, using modern crowd-sourcing, networking, and gamification\mcite{SorensenN2016} techniques.  In this BIG Bell Test (BBT) the Alice and Bob of Aaronson's formulation are real people.  Assuming no faster-than-light communication, such experiments can prove the conditional relation: if human will is free,\stext{there are physical events with no causes} \btext{there are physical events (the measurement outcomes in the Bell tests) that are intrinsically random, i.e., impossible to predict\pcite{BeraRPP2017}  We note that this argument in no way uses the theory of quantum mechanics, and yet arrives to one of that theory's most profound claims. Intrinsic randomness supported by a BIV is central to so-called device-independent quantum technologies\pcite{ColbeckThesis2007, PironioN2010} }  

It is perhaps surprising that human choices, which are known to contain statistical regularities\ccite{BarHillelAAM1991} are \stext{sufficiently random} \btext{suitably unpredictable} for a Bell test. Recent works on the statistical {analysis} of Bell tests\mcite{BierhorstJPA2015, ElkoussNPJQI2016, KoflerPRA2016} clarify this:  \btext{sequence randomness, i.e., the absence of patterns and correlations in the sequence of choices, is not, \textit{per se}, a requirement for a rejection of local realism. Rather, statistical independence of choices from the hidden variables that describe possible measurement outcomes is required (see Methods \ref{Sec:Freedom}).  
This independence can fail in different ways, categorized by named loopholes: The FoCL describes the possibility that hidden variables  influence the setting choices.  The `locality loophole' describes the possibility that a choice at one station could influence a measurement result at the other station.  The term `locality' reflects one way of blocking this possibility, by space-like separation of the choice and measurement events (see Methods \ref{Sec:FoCLandLL}).
} 

Patterns strongly affect statistical strength in experiments that \btext{aim} to close LL by space-like separation \btext{--} they allow current choices to be predicted from earlier choices\btext{, which have had more time to reach the distant measurement}. As described below, the BBT tightens LL using many independent experiments rather than space-like separation.  \btext{Furthermore, the human capacity for free choice removes the need for assumptions about physical indeterminism, allowing the FoCL to be closed. Thus, although human choices show imperfect sequence randomness, they nonetheless enable a strong rejection of local realism with the BBT strategy.}

\stext{It is perhaps surprising that human choices, which are known to contain statistical regularities\ccite{BarHillelAAM1991} are {sufficiently random} for a Bell test. Recent works on statistical analysis of Bell tests\mcite{BierhorstJPA2015, ElkoussNPJQI2016, KoflerPRA2016} clarify this: provided statistical independence of settings and hidden variables (see Methods), patterns do not strongly influence a BIV's  \mbox{p-value}. Statistical interdependence of settings and hidden variables can arise due to hidden variables influencing the choices (FoCL), or vice versa (locality loophole = LL). Patterns do strongly affect \mbox{p-values} in experiments that try to close LL by space-like separation, as they allow current choices to be predicted from earlier choices. As described later, the BBT tightens LL using many independent experiments rather than timing. }

\begin{figure*}[t]
\parbox[b]{0.3\linewidth}{
}
\includegraphics[width=0.85\linewidth]{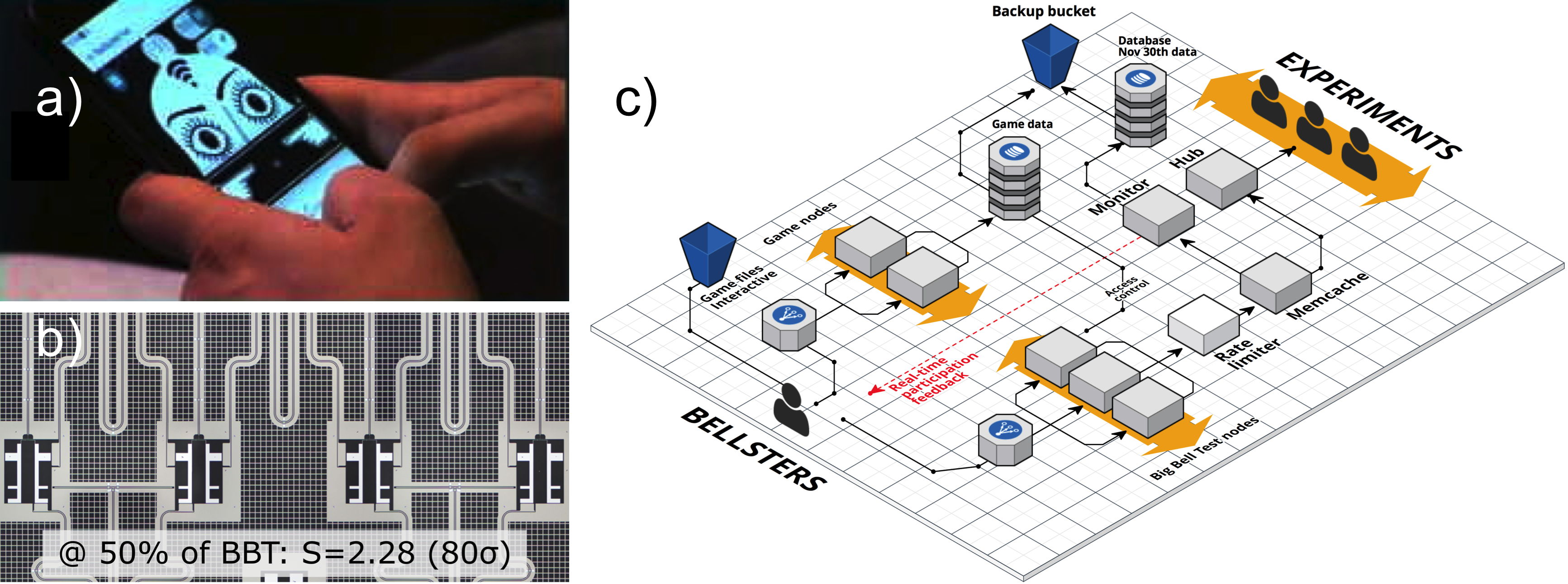}
\caption{\label{Fig:Architecture}Structure of The BIG Bell Test.  a) human participants, or \Bellsters, enter `0's and `1's in an online video game that incentivizes sustained generation of unpredictable bits.  \btext{Image: ICFO/Maria Pascual (Kaitos Games).} b) experiments use \Bellster-generated bits to control measurement-defining elements, such as wave-plates for photons or microwave pulses for matter qubits. Shown is a micrograph of superconducting qubits used in \ETHZ, 
with the measured ClauserÐHorneÐShimonyÐHolt
Bell parameter $S$ mid-way through the BBT.
 c) A cloud-based networking system integrates the activities a) and b), serving game elements to \Bellsters, distributing input bits to connected laboratories, and providing in-game feedback about experimental use of the player's input. Through this system, \Bellsters~are given direct, if brief, control of the experimental apparatus, so that each measurement setting is determined by a single human choice,  traceable to a given user ID and time of entry.  See Methods \ref{Sec:Networking}.
}
\end{figure*}

A major obstacle to a Bell test with humans has been the difficulty of  generating enough choices for a statistically significant test. A person can generate roughly three random bits per second, while a strong test may require millions of setting choices in a time span of minutes to hours, depending on the speed and stability of the experiment.  To achieve such rates, we crowd-sourced the basis choices, recruiting in total about 100,000 participants, the \textit{\Bellsters}, over the course of the project.  Each choice by a participant, encoded as a bit, `0' or `1',  was entered in an internet-connected device such as the participant's mobile phone. Servers relayed the entered bits to the 13 experiments, see Fig. \ref{Fig:Architecture}. The same bits were sent to many experiments, which used them for individual settings without re-use (except experiment \EQUS).
To encourage participants to contribute a larger number of more unpredictable bits, the input was collected in the context of a video game, The BIG Bell Quest (available at \href{https://museum.thebigbelltest.org/quest/}{https://museum.thebigbelltest.org/quest/}), implemented in javascript  to run directly in a device's web browser.  

The BIG Bell Quest is designed to reward sustained, high-rate input of unpredictable bits, while also being engaging and informative (see Methods \ref{Sec:Gamification}).  An interactive explanation first describes quantum nonlocality and the role played by participants and experimenters in the BBT. The player is then tasked with entering a given number of unpredictable bits within a limited time. {A machine learning algorithm (MLA) attempts to predict each input bit, modelling the user's input as a Markov process and updating the model parameters using reinforcement learning (see Methods \ref{Sec:PredictionEngine}).} 
Scoring and level completion reflect the degree to which the MLA predicts the player's input, motivating players to consider their own predictability and take conscious steps to reduce it, but the MLA does not act as a filter: all input is passed to the experiments.  Bellster input showed unsurprising deviations from ideal randomness\ccite{BarHillelAAM1991} e.g., $P(0) \approx 0.5237$ (bias toward `0' ) while adjacent bits show $P(01) + P(10) \approx 0.6406$ (excess of alternation).

Modern video-game elements were incorporated to boost engagement (animation, sound), to encourage persistent play (progressive levels, power-ups, boss battles, leaderboards) and to recruit new players (group formation, posting to social networks). Different level scenarios illustrate key elements of the BBT: human input, global networking, and measurements on quantum systems, while boss battles against the Oracle (see Methods) convey the conceptual challenge of unpredictability.  Level completion is rewarded with 1) a report on how many bits from that level were used in each experiment running at that time, 2) a `curious fact' about statistics, Bell tests, or the various experiments, and, if the participant is lucky,  3) one of several videos recorded in the participating laboratories, explaining the experiments. The game and BBT website  (preserved at \href{http://museum.thebigbelltest.org}{http://museum.thebigbelltest.org}) are available in Chinese, English, Spanish, French, German, Italian and Catalan, making them accessible to roughly three billion first- and second-language speakers.

To synchronize participant activity with experimental operation, the Bell tests were scheduled for a single day, Wednesday 30 November 2016. The date was chosen so that most schools worldwide would be in session, and to avoid competing media events such as the US presidential election.  Participants were recruited by a variety of channels, including traditional and social media and school and science museum outreach, with each  partner institution handling recruitment in 
their familiar geographical regions and
languages.
The media campaign focused on the nature of the experiment and the need for human participants. The press often communicated this with headlines such as ``Quantum theory needs your help'' (China Daily). 
A first, small campaign in early October was made to seed \stext{``viral''} \btext{``word-of-mouth''} diffusion of the story and a second, large campaign 29-30 November was made to attract a wide participant base. The media campaign generated at least 230 headlines in printed and online press, radio and television.

%
%
%
%

\begin{figure*}[t]
\includegraphics[width=0.75\linewidth]{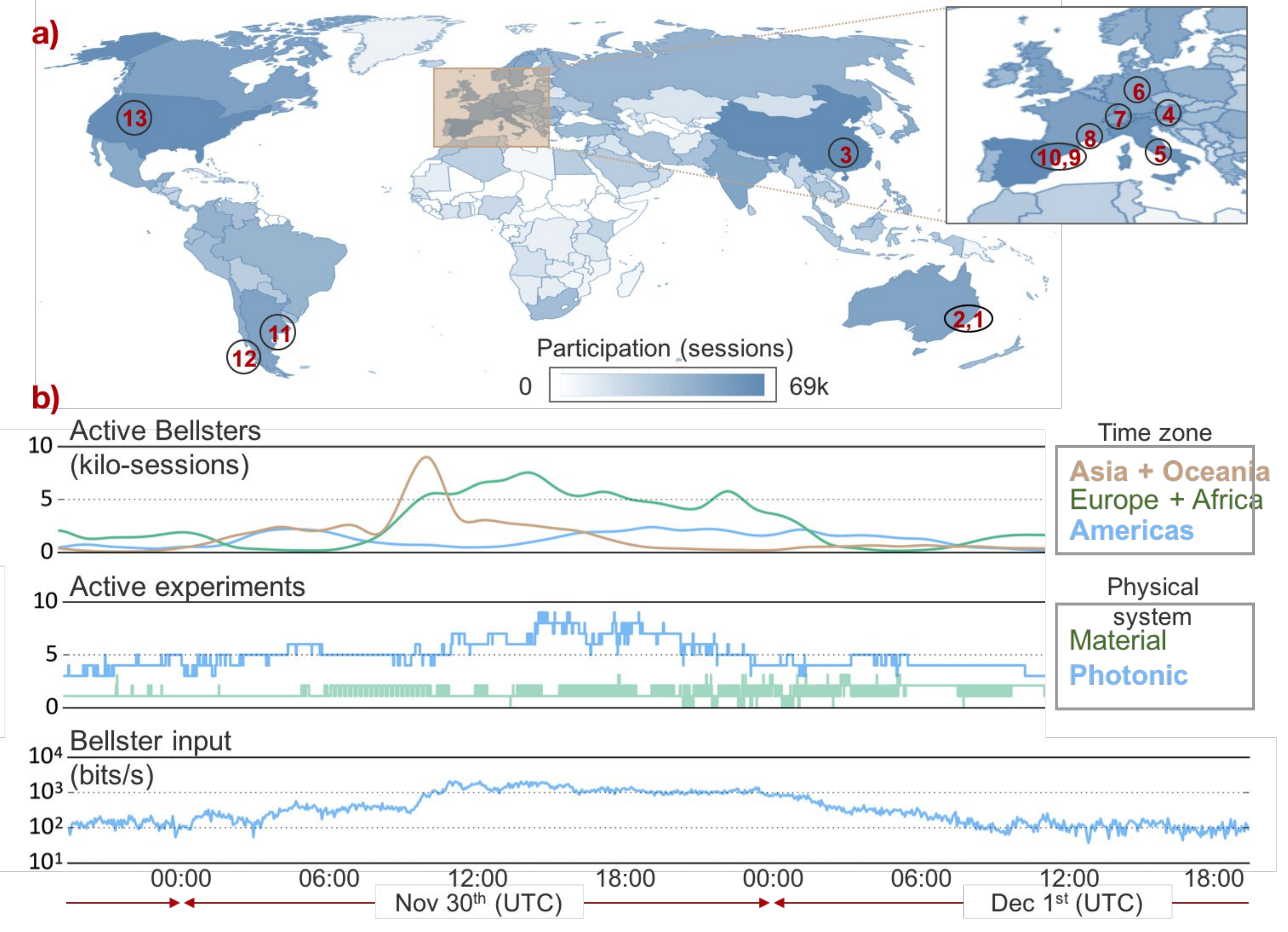}
\caption{\label{Fig:MapAndTiming} Geography and timing of the BBT.  a) Locations of the 13 BBT experiments, ordered from East to West. \stext{See Table \ref{Tab:Experiments}}  \btext{Numbers index the experiments, summarized in Table \ref{Tab:Experiments}}. Shading shows total sessions by country. Eight sessions from Antarctica are not shown. \btext{Map by Giorgio Colangelo using data from OpenStreetMaps, rendered in Wolfram Mathematica.} b) Temporal evolution of the project.  (top) live sessions versus time for different continent groups, showing a strong drop-off in the local early morning in each region. The spike in Asian participation around 11:00 UTC coincides with a live-streamed event in Barcelona, \newbtext{hosted by 
D. Jim\{e}nez and the CosmoCaixa science museum, re-broadcast live in
Chinese by L.-F. Yuan and the University of Science and Technology
of China (USTC).
 (middle) number of connected labs versus time, divided into experiments using only photons and experiments with at least one material component, e.g. atoms or superconductors.  (bottom) input bitrate versus time. Data flow remains nearly constant despite regional variations, with Asian \Bellsters~handing off to \Bellsters~from the Americas in the critical period 12:00-00:00 UTC. Session data from Google analytics.}
\end{figure*}

{The data networking architecture of the BBT, shown in  Fig.~\ref{Fig:Architecture},  includes elements of instant messaging and online gaming, and is designed to efficiently serve a 
fluctuating number of simultaneous users
that is not known in advance and could range from 10 to 100,000.
A gaming component handles the BBT website, participant accounts management, delivery of game code (javascript and video), score records and leaderboards. In parallel, a messaging component handles data conditioning, streaming to experiments, and reporting of participant choices generated via the game.  Horizontal scaling is used in both components: participants connect not directly to servers but rather to dynamic load balancers that spread the input among a pool of servers dynamically scaled in response to load. 
 The timing of input bits (but not their values) was used to identify robot participants and remove their input from the data stream. Game operation was unchanged, to avoid alerting the robots' masters.
A single, laboratory-side server received data from the participant-side servers, concatenated the user input and streamed it to the labs at laboratory-defined rates.  See Methods \ref{Sec:Networking} for details. }

By global time zoning, November 30th defines a 51-hour window, from 0:00 UTC+14h (e.g. Samoa) to 23:59 UTC-12h (e.g. Midway island). Nevertheless, most participants contributed during a 24-hour window centred on 18:00 UTC.   Recruitment of participants was geographically uneven, with a notable failure to recruit large numbers of participant from Africa.  Despite this, the latitude zones of Asia/Oceania, Europe/Africa, and the Americas had comparable participation, which proved important for the experiment. As shown in Fig.~\ref{Fig:MapAndTiming}, input from any single region dropped to low values during the local early morning, but was compensated by high input from other regions, resulting in a high sustained global bit rate. Over the 12-hour period from 09:00 UTC to 21:00 UTC, 30 November 2016, the input exceeded $10^3$ bits per second, allowing a majority of the experiments to run at their full speed.   Several experiments posted their results live on social networks.  Due to their high speeds, \UC~and \NIST~accumulated human bits to use in short bursts. As determined by independent measurements, \ICFOONE~and \UC~were not in condition to observe a BIV on the day of the event; they report later results using stored human bits.

\newcommand{\TBA}{--}

\begin{table*}[t]
\caption{\label{Tab:Experiments}%
Experiments carried out as part of the BBT, ordered by longitude, from East to West. Descriptions of the experiments are given in \stext{Methods} \btext{Supplementary Information}.   Stat. Sig. (statistical significance); indicates number of standard deviations assuming 
independent and identically distributed
trials, unless otherwise indicated. 
$\gamma$ signifies photon. Rate indicates the peak rate at which bits were used by the experiments. Due to the limited rate of Bellster input, some experiments had dead times.  \btext{$B$, $K$, $S$, $S_{AB}$, $S_{BC}$, $S_{\rm HRN}$, $S_{\rm QRN}$, indicate Bell parameters for the respective experiments, $S_{16}$ is the steering parameter (see Supplementary Information). $l_0$ indicates the minimum 
P\"{u}tz-Rosset-Barnea-Liang-Gisin measure of settingÐchoice independence
consistent with the observed BIV. 
USTC, University of Science and Technology of China; EQUS, Centre for Engineered Quantum Systems; IQOQI, Institute for Quantum Optics and Quantum Information; INFYNI, Institut de Physique de Nice; ICFO, Institut de Ciencies Fotoniques; LMU, Ludwig-Maximilians-Universit\"{a}t; ETHZ, ETH Zurich; CITEDEF, Institute of Scientific and Technical Research for Defence; UdeC, University of
Concepci\'{o}n; NIST, National Institute of Standards and Technology.} 
}
\begin{ruledtabular}
\begin{tabular}{cccccccc}
ID & Lead Institution  & Location  & Entangled system  & Rate & Inequality & Result & Stat. Sig.      \\
\hline 
\GRIFFITH &GRIFFITH & Brisbane, AU  & $\gamma$ polarisation \cite{} & \SI{4}{bps} & $S_{16} \le 0.511 $ & $S_{16} = 0.965 \pm 0.008$ & $ \SI{57}{\sigma}$   \\
\EQUS &EQUS & Brisbane, AU  & $\gamma$ polarisation \cite{} & \SI{3}{bps} & $|S| \le 2$ & $ \begin{array}{c} S_{AB} = 2.75 \pm 0.05 \\  S_{BC} = 2.79 \pm 0.05 \end{array}$&$\begin{array}{c} \SI{15}{\sigma} \\ \SI{16}{\sigma} \end{array}$   \\
\USTC & USTC & Shanghai, CN  & $\gamma$ polarisation  \cite{} & \SI{1}{kbps} & 
PRBLG\mcite{PutzPRL2014} 
& ${l}_0 = 0.10 \pm 0.05 $
& ${\mathrm {N/A}}$ \\
\IQOQI &IQOQI & Vienna, AT  & $\gamma$ polarisation & \SI{1.61}{kbps} & $|S|\le 2$ & $ \begin{array}{c} S_{HRN} = 2.639 \pm 0.008\\  S_{QRN} =2.643  \pm 0.006  \end{array}$ & $\begin{array}{c} \SI{81}{\sigma} \\  \SI{116}{\sigma} \end{array}$  \\
\SAPIENZA &SAPIENZA & Rome, IT  & $\gamma$ polarisation  \cite{} & \SI{0.62}{bps} & $B\le 1$ & $B = 1.225 \pm 0.007 $ & $\SI{32}{\sigma} $   \\
\LMU &LMU & Munich, DE  & $\gamma$-atom \cite{} & \SI{1.7}{bps} & $|S| \le 2$ \cite{} & $ \begin{array}{c} S_{HRN} = 
2.427 \pm 0.0223 \\ S_{QRN} =2.413 \pm 0.0223 \end{array}$ & $\begin{array}{c} \SI{19}{\sigma} \\ \SI{18.5}{\sigma}  \end{array}$   \\
\ETHZ &ETHZ & Zurich, CH  & transmon qubit \cite{} & \SI{3}{kbps} &  $|S|\le 2$ & $S =  2.3066 \pm 0.0012 $ & $ p < 10^{-99}  $  \\
\NICE &INPHYNI & Nice, FR  & $\gamma$ time-bin \cite{} & \SI{2}{kbps} & $|S| \le 2$ & $S = 2.431  \pm 0.003 $ & $\SI{140}{\sigma} $  \\
\ICFOONE &ICFO & Barcelona, ES  & $\gamma$-atom ensemble \cite{} & \SI{125}{bps} & $|S| \le 2$ & $S = 2.29  \pm 0.10 $ & $\SI{2.9}{\sigma}$ \\
\ICFOTWO &ICFO & Barcelona, ES  & $\gamma$ multi-frequency-bin \cite{} & \SI{20}{bps} & $|S|\le 2$ & $S = 2.25 \pm 0.08 $ & $ \SI{3.1}{\sigma} $  \\
\UBA &CITEDEF & Buenos Aires, AR  & $\gamma$ polarisation \cite{} & \SI{1.02}{bps} & $|S| \le 2$ & $S = 2.55  \pm 0.07$ & $\SI{7.8}{\sigma} $  \\
\UC &CONCEPCION & Concepcion, CL  & $\gamma$ time-bin \cite{} & \SI{52}{kbps} & $|S| \le 2$ & $S = 2.43 \pm 0.02$ & $  \SI{20}{\sigma} $    \\
\NIST &NIST & Boulder, US  & $\gamma$ polarisation \cite{} & \SI{100}{kbps} &  $K\le 0$ & $K =( 1.65 \pm 0.20 ) \times 10^{-4}$ & $\SI{8.7}{\sigma}$    \\
\end{tabular}
\end{ruledtabular}
\end{table*}

The Earth is only 43 light-ms in diameter, so human choices are too slow to be space-like separated from the measurements. 
This leaves open the locality loophole regarding the influence
of choices on remote detection.
 Influence   of Alice's measurement setting on Bob's detection (and vice-versa) is nonetheless excluded by space-like separation in experiments \USTC, 
and \NIST~(see Supplementary Information).  To tighten LL, we take a strategy we call the BIG test:  many simultaneous Bell tests in widely-separated locations using different physical systems, with each experiment's apparatus constructed and operated by different experimental teams. \stext{In this BIG test, a hidden-variable theory can only exploit LL if it has mechanisms by which the choices simultaneously influence hidden variables in all of these experiments, bringing them each to a result mimicking quantum predictions.} \btext{The only hidden-variable theories that escape this tightening are ones in which choices can simultaneously influence hidden variables in many differently-constructed experiments, to produce in each one a BIV.}   This strategy is strengthened by using the same bits in many experiments, as described above.

The suite of 13 BBT experiments, including true Bell tests and other realism tests requiring free choice of measurement, are summarized in Table \ref{Tab:Experiments} and described in the Supplementary Information. Experiments \GRIFFITH, \EQUS, \USTC, \IQOQI, \SAPIENZA, \NICE, \UBA, \UC, and \NIST~used entangled photon pairs,  \LMU~used single-photon/single-atom entanglement, \ICFOONE~used single-photon/atomic ensemble entanglement,  and \ETHZ~used entangled superconducting qubits.    Experiments \ETHZ~ and \NIST~used high-efficiency detection to avoid the fair sampling assumption, thus closing simultaneously the detection efficiency and freedom-of-choice loopholes. 
\SAPIENZA~demonstrated a violation of bi-local realism, while \ICFOTWO~violated a  Bell inequality for multi-mode entanglement. \GRIFFITH~demonstrated quantum steering and \EQUS~demonstrated \stext{entanglement in time} \btext{temporal quantum correlations} with a three-station measurement. 
\UC~closed the post-selection loophole typically present in Bell tests based on energy-time entanglement.  Analysis of \USTC~puts bounds on how well a measurement-dependent local model would have to predict Bellster behaviour to produce the observed results\pcite{PutzPRL2014}  \USTC, \IQOQI, \LMU~ and \NIST~tested
whether human-generated measurement choices gave different results
from machine-generated ones.
Most experiments observed statistically strong violations of their respective inequalities, justifying rejection of local realism in a multitude of systems and scenarios.  

In summary, on 30 November 2016, a suite of 13 Bell tests and similar experiments, using photons, single atoms, atomic ensembles and superconducting devices, demonstrated strong disagreement with local realism, using measurement settings chosen by tens of thousands of globally-distributed human participants. 
The results also show empirically that measurement settings independence, here provided by human agency, is in strong disagreement with causal determinism\ccite{HoeferSEP2005, AaronsonAS2014, AcinN2016} a topic formerly accessible only by metaphysics.
The experiments reject local realism in a wide variety of  physical systems and scenarios, set the groundwork for Bell-test based applications in quantum information,  introduce gamification of \btext{randomness generation} \stext{Bell tests and unpredictability concepts}, and demonstrate global networking techniques by which hundreds of thousands of individuals can directly participate in experimental science.  
 
Acknowledgements:  We are grateful to the many people and organizations
who contributed to this project, starting with the Bellsters (details at http://
thebigbelltest.org). We thank the Departament d'Ensenyament de la Generalitat
de Catalunya, Ministerio de Educaci—n, Cultura y Deporte of Spain, CosmoCaixa,
Fundaci—n Bancaria ``la Caixa'', INTEF, Optical Society of America, European
Centers for Outreach in Photonics (ECOP), Muncyt-Museo de Ciencia y
Tecnolog'a de Madrid, Investigaci—n y CiŽncia, Big Van, Crea Cincia, Tencent
news, Micius Salon, Politecnico di Milano, University of WaterlooÐInstitute
for Quantum Computing, Universidad Aut—noma de Barcelona, Universidad
Complutense de Madrid, Universitˆ degli Studi di Milano, Toptica, EPS Young
Minds, Real Sociedad Espa–ola de F'sica, Ajuntament de Barcelona, Ajuntament
de Castelldefels, Universitˆ degli studi di Padova, Universitˆ degli studi del
l'Insubria, CNRIFN Istituto di Fotonica e Nanotecnologia, Istituto d'Istruzione
Superiore Carlo Livi, Webbile, and Kaitos Games. We are especially thankful
for the recruitment efforts of the outreach departments at our institutions. 

We
acknowledge financial support from CONICET; ANPCyT (Argentina), Australian
Research Council Centre for Quantum Computation and Communication
Technology (CE110001027, CE170100012); Australian Research Council and
the University of Queensland (UQ) Centre for Engineered Quantum Systems
(CE110001013, CE170100009). A.G.W. acknowledges the University of
Queensland Vice-Chancellor's Research and Teaching Fellowship (Australia);
Austrian Academy of Sciences (OEAW), the Austrian Science Fund (FWF)
(SFB F40 (FoQuS); CoQuS No. W1210-N16), the Austrian Federal Ministry
of Education, Science and Research (BMBWF) and the University of Vienna
(project QUESS) (Austria), MEC; MCTIC (Brazil), Barcelona City Hall; Generalitat
de Catalunya (SGR 874 and 2014-SGR-1295; CERCA programme) (Catalonia),
PIA CONICYT (grants PFB0824 and PAI79160083); FONDECYT (grants
1140635, 1150101, 1160400, 3170596 and 11150325; Becas Chile); the
Millennium Institute for Research in Optics (MIRO); Becas CONICYT (Chile),
the National Fundamental Research Program (grant 2013CB336800); the
National Natural Science Foundation of China (grants 91121022, 61401441
and 61401443); the Chinese Academy of Science (Strategic Priority Research
Program (B) XDB04010200); the 1000 Youth Fellowship programme; the
National Natural Science Foundation of China (grant 11674193); the Science
and Technology Commission of Shanghai Municipality (grant 16JC1400402)
(China); ERC (grant agreements AQUMET 280169, 3DQUEST 307783,
OSYRIS 339106, ERIDIAN 713682, QITBOX, QUOLAPS, QuLIMA and
SuperQuNet 339871); ESA (contract number 4000112591/14/NL/US);
the European and Regional Development Fund (FEDER); H2020 (QUIC 641122); the Marie Sk\l{}odowska-Curie programme (grant agreement 748549); the FP7-ITN PICQUE project (grant agreement No 608062); the OPTIMAL project granted through FEDER; (European Commission); 
the Universit\'{e} C\^{o}te d'Azur (UCA, France) through its program Quantum@UCA; the Foundation Simone \& Cino Del Duca (Institut de France);  l'Agence Nationale de la Recherche (ANR, France) for the CONNEQT, SPOCQ and SITQOM projects (grants ANR-EMMA-002-01, ANR-14-CE32-0019, and ANR-15-CE24-0005, respectively); the iXCore Research Foundation (France); German Federal
Ministry of Education and Research (projects QuOReP and Q.com-Q)
(Germany); CONACyT graduate fellowship programme (Mexico), MINECO
(FIS2014-60843-P, FIS2014-62181-EXP, SEV-2015-0522, FIS2015-68039-P,
FIS2015-69535-R and FIS2016-79508-P; Ramon y Cajal fellowship
programme; TEC2016-75080-R); the ICFOnest + international postdoctoral
fellowship programme (Spain); the Knut and Alice Wallenberg Foundation
(project `Photonic Quantum Information') (Sweden); NIST (USA), AXA Chair
in Quantum Information Science; FQXi Fund; Fundaci— Privada CELLEX;
Fundaci— Privada MIR-PUIG; the CELLEX-ICFO-MPQ programme; Fundaci—
Catalunya-La Pedrera; and the International PhD-fellowship programme
`la Caixa'-Severo Ochoa.

\newcommand{\ContactSym}{\ddagger}
\newcommand{\ContactTag}{$^\ContactSym$}
\renewcommand{\ContactTag}{ (contact author)}

Author contributions: CA instigator, MWM\ContactTag~project leader. Coordination, gamification and networking (ICFO): 
SC general supervision, MM-P project management, MG-M, FAB, CA \stext{gamification design and execution} \btext{game design}, JT prediction engine,  
AH, MG-M, FAB Bellster recruitment and engagement strategy, design and execution, CA web infrastructure and 
networking, MWM\ContactTag, CA, JT main manuscript with input from all authors. 
Experiments: \GRIFFITH: GJP, RBP\ContactTag, FGJ, MMW, SS experiment design and execution. \EQUS: AGW, MR\ContactTag experiment design and execution. \USTC: J-WP\ContactTag, QZ\ContactTag~supervision, J-WP, QZ, XM, XY, YL experiment conception and design, ZW, LY, HL, WZ superconducting nanowire single-photon
detector (SNSPD) fabrication and characterization, JZ SNSPD maintenance, M-HL, CW, YL photon source design and characterization, J-YG, YL software design and deployment, XY protocol analysis, XY, YL data analysis, J-WP, QZ, CW, XY, YL manuscript, with input from all. \IQOQI: TS\ContactTag, AZ, RU supervision, conception and coordination, BL, JHa, DR experiment execution and analysis. \SAPIENZA: GC, LS, FA, MB, FS\ContactTag~experiment execution and analysis, RC theory support. \LMU: HW, WR\ContactTag, KR, RG, DB experiment design and execution. \ETHZ: 
JHe, PK, YS, CKA, AB, SK, PM, MO, TW, SG, CE, AW\ContactTag~experiment design and execution. \NICE: ST\ContactTag, TL, FK, GS, PV, OA, experiment design and execution. \ICFOONE: HdR\ContactTag, PF, GH experiment design and execution. \ICFOTWO: HdR\ContactTag, AS, AL, MM, DR, OJ, AM experiment design and execution, DC, A. Ac. theory support.  \UBA: MAL\ContactTag~coordination, server communication, LTK, IHLG, AGM experiment design and execution, CTS, AB input data formatting, LTK, IHLG, AGM, CTS, AB, MAL data analysis. \UC: GX\ContactTag~coordination, FT optical setup, PG, AAl, JF, A. Cuevas, GC optical setup support, JC electronics design and implementation, JC, FT, experiment execution, DM software, GL, PM, FS experimental support, ACa theory support, JC, ESG, J-{\AA}L data analysis. \NIST: LKS\ContactTag, SN, MS, OSM-L, TG, SG, PB, EK, RM experiment design, execution and analysis. 

Competing interests statement: The authors declare no competing financial interests.

Correspondence and material requests should be addressed to morgan.mitchell@icfo.eu. 
~ \\

\newcounter{MyID}
\setcounter{MyID}{192}
\newcommand{\ID}{\ding{192}\stepcounter{MyID}}

\clearpage
{\Large Methods}

\renewcommand{\figurename}{Extended Data Figure}


\section{Local realism, Bell parameters, Bell inequalities}
\label{Sec:LocalRealism}
In their 1935 article\mciteSI{EinsteinPR1935},  Einstein Podolsky and Rosen employed notions of locality (actions or observations in one location do not have immediate effects at other locations), and realism (observables have values even if we do not observe them) to argue that quantum theory was incomplete and could in principle be supplemented with information about which outcomes actually occur in any given run of an experiment. Bell formalized these notions by defining local hidden variable models, (LHVMs) a class of non-quantum theories that are simultaneously local and realistic.  We consider the simplest case, of two systems measured by two observers Alice and Bob. We write $x (y)$ to represent Alice's (Bob's) measurement setting,  $a (b)$ to represent their measurement outcomes, and $\lambda$ to represent the hidden variable, something we cannot measure, but which we include in the model to explain why $a$ and $b$ take on particular values.  The predictions of any such bi-partite LHVM are given by 
\begin{equation}
\label{Eq:LocalRealism1}
P(a,b|x,y) = \sum_\lambda P(a|x,\lambda) P(b|y,\lambda) P(\lambda) .
\end{equation}
where $P(\cdot|\cdot)$ indicates a conditional probability.  That is, the probability of getting outcome $a,b$ when Alice and Bob measure $x$ and $y$, respectively, is expressible in terms of the local conditional probabilities $P(a|x,\lambda)$ and $P(b|y,\lambda)$. $\lambda$ is averaged over, because of our ignorance of the value of this hidden variable. If the probabilities $P(a|x,\lambda)$ and $P(b|y,\lambda)$ are restricted to zero or one we have a deterministic LHVM, in which $\lambda, x$, and $y$ fully determine the outcomes $a$ and $b$. Such LHVMs are explicitly realistic in the 
EinsteinÐPodolskyÐRosen
sense. Locality is also explicit in the model: For example, $P(a|x,\lambda)$ depends on neither $b$ nor $y$, so that the events at Bob's station have no influence on Alice's measurement outcome $a$.  A mathematical notion of `freedom' is implicit in the LHVM: $x$ and $y$ are included as free parameters, and not, e.g., as functions of $\lambda$.   If $P(a|x,\lambda)$ and $P(b|y,\lambda)$ are allowed to take intermediate values, we speak of a non-deterministic LHVM.  Because the unknown $P(\lambda)$ can take on intermediate values, deterministic and non-deterministic LHVMs are equivalent, and from here on we drop the distinction. 

This class of models, which by construction embody the EPR assumptions, was shown by Bell to be incapable of reproducing the predictions of quantum mechanics.   For example, if Alice and Bob's local systems are spin-1/2 particles in a singlet state, then their measurements (assumed to be ideal) will 
agree, i.e., show that both are spin-up or both are spin-down, with probability $P({\rm agree}|\phi_a,\phi_b) \equiv P(\uparrow,\uparrow|\phi_a,\phi_b) +P(\downarrow,\downarrow|\phi_a,\phi_b)  =  \sin^2(\frac{\phi_b - \phi_a}{2})$, where $\phi_b - \phi_a$ is the angle between Alice's and Bob's analysis directions.  Eq.~(\ref{Eq:LocalRealism1}) cannot reproduce all the features of this distribution. No choice of $P(a|x,\lambda)$, $P(b|y,\lambda)$, and $P(\lambda)$ can give a probability $P(a,b|x,y)$ that simultaneously depends on the difference $\phi_b - \phi_a$,  has high-visibility (ranging from 0 to 1) and sinusoidal.  


This difference is efficiently captured by Bell inequalities.  A Bell parameter is a linear combination of conditional probabilities $P(a,b|x,y)$, and a Bell inequality indicates the bounds (within the class of LHVMs) of a Bell parameter.   Typically, the Bell inequalities of  interest are those that are not obeyed by quantum mechanics, i.e., those for which quantum correlations can be strong enough to violate the Bell inequality.  Bell's theorem shows that there are such inequalities, and thus that quantum mechanics cannot be `completed' with hidden variables. 

Bell inequalities also  make possible experimental tests of local realism: A Bell test is an experiment that makes many spatially-separated measurements with varied settings to obtain estimates of the $P(a,b|x,y)$ that appear in a Bell parameter.  If the observed Bell parameter violates the inequality, one can conclude that the measured systems were not governed by any LHVM.  It should be noted that this conclusion is always statistical, and typically takes the form of a hypothesis test, leading to a conclusion of the form `assuming nature is governed by local realism, the probability to produce the observed Bell inequality violation (or a stronger one) is $P({\rm observed~or~stronger} | {\rm local~realism}) \le p$.' This $p$-value is a key indicator of statistical significance in Bell tests.

\section{`Freedom' in Bell tests }
\label{Sec:Freedom}

The use of the term `free' to describe the choices in a Bell test derives more from mathematical usage than from its usage in philosophy, although the two are clearly related.  Bell \mcite{BellBook2004Ch7} (see Section \ref{Sec:JSB} below)  states that his use of `free will,' reflects the notion of `free variables,' that is, externally-given parameters in physical theories, as opposed to dynamical variables that are determined by the mathematical equations of the theory. 

As described above $x,y$ are the settings in a Bell test,  $a,b$ are the outcomes, and $\lambda$ is the hidden variable. Any local realistic hidden variable model is described by Eq.~(\ref{Eq:LocalRealism1}).
%
The mathematical requirements for the relevant `freedom' are made evident by a more general description in which the local realistic model includes also $x,y$, in which case it specifies the joint probability 
\begin{equation}
\label{Eq:LocalRealism2}
P(a,b,x,y) = \sum_\lambda P(a|x,\lambda) P(b|y,\lambda) P(x,y|\lambda) P(\lambda).
\end{equation}
Using the Kolmogorov definition of conditional probability $P(A,B) = P(A|B) P(B)$, we find that Eq.~(\ref{Eq:LocalRealism2}) reduces to Eq.~(\ref{Eq:LocalRealism1}), provided that $P(x,y|\lambda) = P(x,y)$, i.e., provided that the settings are statistically independent of the hidden variables.  By Bayes' theorem, this same condition can be written $P(\lambda|x,y) = P(\lambda)$ and $P(x,y,\lambda) = P(x,y)P(\lambda)$.  This condition is known in the literature as the freedom-of-choice assumption, although it implies more than just free choices.  A more accurate term  might be `measurement setting/hidden-variable independence.'  We note that {this condition}  does not require that $x$ be independent of $y$, nor does it require that $P(x,y)$ be unbiased.  Similar observations emerge from the more involved calculations required to assign p-values to observed data in Bell tests\pcite{ KoflerPRA2016}

The above clarifies the sense in which the basis choices should be `free.'  The desideratum is independence from the hidden variables that describe the particle behaviours, keeping in mind that the choices and measurements  could,  consistent with relativistic causality, be influenced by any  event in their backward light-cones .  Because the setting choices and the measurements will always have overlapping backward light-cones, it is impossible to rule out  the possibility of a common past influence through space-time considerations.  If human choices are free, however, such influences are excluded. It should also be noted that complete independence is not required, although the tolerance for interdependence can be low\pciteSI{HallPRA2011, BarrettPRL2011, PutzPRL2014} The theory that the entire experiment, including choices and outcomes, is pre-determined by initial conditions is known as superdeterminism. Superdeterminism cannot be tested\pciteSI{BellD1985}  

A very similar concept of `freedom' applies to the entangled systems measured in a Bell test.  A Bell inequality violation with free choice and under strict locality conditions implies either indeterminacy of the measurement outcomes or faster-than-light communications and thus closed time-like curves\pcite{ColbeckThesis2007,AcinN2016} If Bob's measurement outcome is predictable based on information available to him before the measurement, and if it also satisfies the condition for a Bell inequality violation, namely a strong correlation with Alice's measurement outcome that depends on his measurement choice, then Bob can influence the statistics of Alice's measurement outcome, and in this way communicate to her despite being space-like separated from her.  Considering, again, that Bob could in principle have information on any events in his backward light cone, this implies (assuming no closed time-like curves) that Bob's measurement outcome must be statistically independent of all prior events.

In this way, we see that `freedom,' understood as behaviour statistically independent of prior conditions, appears twice in a Bell test, first as a requirement on the setting choices, and second as a conclusion about the nature of measurement outcomes on entangled systems. These two are linked, in that the second can be demonstrated if the first is present.  

Prior tests using physical randomness generators to choose measurement settings thus demonstrate a relationship between physical processes, showing for example\mcite{HensenN2015, AbellanPRL2015} that if spontaneous emission is `free,' then the outcomes of measurements on entangled electrons are also `free.' By using humans to make the choices, we translate this to the human realm, showing, in the words of Conway and Kochen\cciteSI{ConwayFP2006}  ``if indeed there exist any experimenters with a modicum of free will, then elementary particles must have their own share of this valuable commodity.''   Here `experimenters' should be understood to refer to those who choose the settings, i.e., the Bellsters. See main text for a discussion of the locality loophole when using humans.

\section{John Stewart Bell on `free variables' }
\label{Sec:JSB}
A brief but informative source for Bell's positions on setting choices is an exchange of opinions with  Clauser, Horne and Shimony (CHS)\cciteSI{BellBook2004} in articles titled `The theory of local beables' and `Free variables and local causality.'  In the first of these articles Bell very briefly considers using humans to choose the measurement settings

\begin{quote}
It has been assumed [in deriving Bell's theorem] that the settings of instruments are in some sense free variables - say at the whim of experimenters - or in any case not determined in the overlap of the backward light cones.
\end{quote}
while the second article defends this choice of method and compares it against `mechanical,' i.e. physical, methods of choosing the settings.

\begin{quote}
Suppose that the instruments are set at the whim, not of experimental physicists, but of mechanical random number generators. Indeed it seems less impractical to envisage experiments of this kind...
\end{quote}
Bell proceeds to consider the strengths and weaknesses of physical random number generators in Bell tests, offering arguments why under `reasonable' assumptions physical random number generators might be trusted, but nonetheless concluding 
\begin{quote}
Of course it might be that these reasonable ideas about physical randomizers are just wrong - for the purpose at hand. A theory might appear in which such conspiracies inevitably occur, and these conspiracies may then seem more digestable than the non-localities of other theories.
\end{quote}

In sum, Bell distinguishes different levels of persuasiveness, noting that physical setting generators, while having the required independence in many local realistic theories, cannot be expected to do so in all such theories. In contemporary terminology, what he argues here is that  physical setting generators can only tighten, not close the FOCL.

Bell also defends his use of the concept of `free will' in a physics context, something that had been criticized by CHS. Bell writes

\begin{quote}
Here I would entertain the hypothesis that experimenters have free will \ldots it seems to me that in this matter I am just pursuing my profession of theoretical physics.  

\ldots A respectable class of theories, including contemporary quantum theory as it is practiced, have `free' `external' variables in addition to those internal to and conditioned by the theory. These variables É provide a point of leverage for `free willed experimenters', if reference to such hypothetical metaphysical entities is permitted. I am inclined to pay particular attention to theories of this kind, which seem to me most simply related to our everyday way of looking at the world.  

~ 

\noindent Of course there is an infamous ambiguity here, about just what and where the free elements are. The fields of Stern-Gerlach magnets could be treated as external.  Or such fields and magnets could be included in the quantum mechanical system, with external agents acting only on the external knobs and switches. Or the external agents could be located in the brain of the experimenter.  In the latter case the setting of the experiment is \textit{not} itself a free variable.  It is only more or less correlated with one, depending on how accurately the experimenter effects his intention. 
\end{quote}

It is clear from the last three sentences that Bell considers human intention -- that is, human free will -- to be a `free variable' in the sense he is discussing.  That is, he believes human intention fulfils the assumptions of Bell's theorem, as do experimental settings faithfully derived from human intention.

\section{Use of `freedom-of-choice loophole'  and `locality loophole' in this work}
\label{Sec:FoCLandLL}
As noted above, a statistical condition used to derive Bell's theorem is $P(x,y,\lambda) = P(x,y)P(\lambda)$, where $x$ and $y$ are choices and $\lambda$ describes the hidden variables.  This statistical condition, known as the freedom of choice assumption,  does not distinguish between three possible scenarios of influence: the condition could fail if the choices influence the hidden variables,  if the hidden variables influence the choices, or if a third factor influences both choices and hidden variables\pcite{ScheidlPNAS2010,LarssonJPA2014Official, KoflerPRA2016} According to Bayes' theorem, equivalent forms are $P(x,y|\lambda) = P(x,y)$, which expresses the fact that knowing $\lambda$ does not give information about $(x,y)$, and  $P(\lambda|x,y) = P(\lambda)$ which expresses the fact that knowing $(x,y)$ does not give information about $\lambda$.  The latter relationship makes clear that influence (in either direction) is incompatible with the freedom of choice assumption. The name for this condition should not be taken literally; the condition can be false even if the choices are fully free, in the sense of being independent of all prior conditions. This occurs for example if the choices are freely made but then influence the hidden variable.   

By long tradition, the `locality loophole' (LL) is the name given to the possibility of influence from Alice's (Bob's) choices or measurements to Bob's (Alice's) measurement outcomes.  The term `freedom-of-choice loophole'  was introduced in Scheidl et al.\ccite{ScheidlPNAS2010} to describe influence from hidden variables to choices.  The text of the definition was ``the possibility that the settings are not chosen independently from the properties of the particle pair.'' We note that this formulation centres on the act of choosing and its independence, which (assuming relativistic causality, an element of local realism) can only be violated by influences from past events, not future events.  These loophole definitions employ the concept of influence, which is directional, to explain how the non-directional relation of independence can be broken.  Similarly directional definitions have recently been applied to experiments using cosmic sources\pcite{HandsteinerPRL2017,WuPRL2017}

Our use of the term in this paper follows the definition of Scheidl et al.\mcite{ScheidlPNAS2010} described above:  FOCL refers to the possibility of influences on the choices from any combination of hidden variables and/or other factors within the backward light-cone of the choice, whereas the possibility of choices influencing the hidden variables, which necessarily occur in the forward light-cone of the choice, is included in the locality loophole. Such a division, in addition to fitting the common-sense notion of free choice, avoids counting a single possible channel of influence in both FOCL and the locality loophole. 


\section{Status of the freedom-of-choice loophole}

The FoCL remains unclosed after recent experiments simultaneous closing locality, detection efficiency, memory, timing, and other loopholes\pcite{HensenN2015, GiustinaPRL2015, ShalmPRL2015,RosenfeldPRL2017} Space-time considerations can eliminate the possibility of such influence from the particles to the choices\cciteSI{ScheidlPNAS2010,ErvenNPhoton2014,GiustinaPRL2015, ShalmPRL2015} or from other space-time regions to the choices\ccite{HensenN2015,RosenfeldPRL2017} but not the possibility of a sufficiently early prior influence on both choices and particles. To motivate freedom of choice in this scenario, well-characterized physical randomizers\mcite{AbellanPRL2015,FurstMOE2010}  have been used to choose settings. 

 In experiments\mcite{HensenN2015, GiustinaPRL2015, ShalmPRL2015} the physical assumption is that at least one of: spontaneous emission, thermal fluctuations, or classical chaos\mcite{AbellanPRL2015} is uninfluenced by prior events, and thus unpredictable even within local realistic theories.  In experiments\mciteSI{WeihsPRL1998, ScheidlPNAS2010, ErvenNPhoton2014, RosenfeldPRL2017} the physical assumption is that photodetection is similarly uninfluenced. While still requiring a physical assumption and thus not closing the freedom-of-choice loophole, this strategy tightens the loophole in various ways:  First, by using space-like separation to rule out influence from certain events, e.g. entangled pair creation, and from defined space-time regions.  Second, by using well-characterized randomness sources, for which the setting choice is known to faithfully derive from a given physical process, it avoids assumptions about the predictability of side-channel processes.  Third, in the case of\ccite{HensenN2015, GiustinaPRL2015, ShalmPRL2015, AbellanPRL2015} by using a physical variable that can be randomized by each of several processes, the required assumption is reduced from `x is uninfluenced' to `at least one of x, y and z is uninfluenced.'

%



\setcounter{figure}{0}

\begin{figure}[t]
\includegraphics[width=0.35\textwidth]{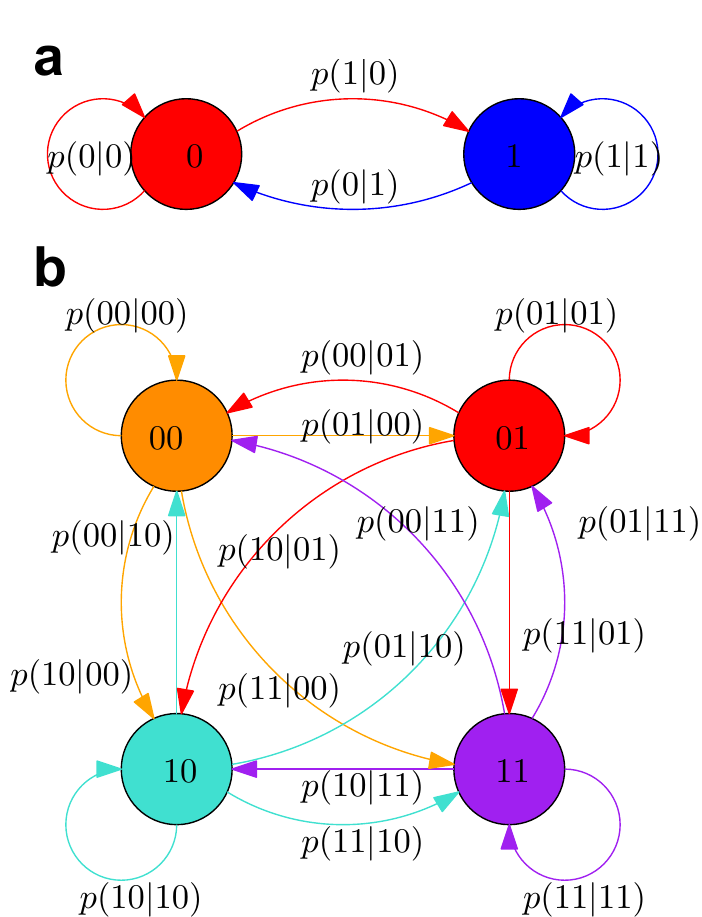}
\caption{\label{fig:ab}
{\bf Markov chains.} {\bf a}, Markov chain for $L = 1$. States
are represented by circles of different colours, and transitions between
states by arrows coloured as the initial state. The last input bit determines
the state of the predictor. The probability $p(a | b)$ of a transition from state
$b$ to state $a$ is estimated from the sequence $S_k$. {\bf b}, Markov chain for $L = 2$,
with four states determined by the last two input bits. The transitions
model the probability $p(a, b | c, d)$ that the user will input bits $ab$, given
that the last two bits were $cd$. The final prediction is based on the marginal
of the next single bit, which is extracted from these estimated probabilities.
}
\end{figure}

\section{Prediction engine}
\label{Sec:PredictionEngine}

Generation of random sequences by humans has been a subject of study in the field of psychology for decades\citeSI{Wagenaar72generationof,BarHillelAAM1991}. Early studies showed that humans perform poorly when asked to produce a random sequence, choosing in a biased manner and deviating from a uniform distribution.  It was shown in \citeSI{rapoport1992generation} that humans playing competitive, zero-sum games that reward uniform random choices tend to produce sequences with fewer identifiable biases. 
One such game is matching pennies: Players have to simultaneously choose between heads or tails; one player wins if the results are equal, the other wins if the results are different. This is a standard two-person game used in game theory \citeSI{gibbons1992game} (see also \citeSI{MOOKHERJEE199462}) with a mixed-strategy Nash equilibrium: As both players try to outguess the other, by behaving randomly they do not incentive either player to change their strategy.

{The BIG Bell Quest reproduces the coin-matching game, with a machine-learning algorithm (MLA) playing the part of the opponent. The MLA operates on simple principles that human players could employ: it maintains a model of the tendencies of the opponent, noting for example ``after choosing `0,' `0,' she usually choses `1' as her next bit.'' The MLA strategy operates with very little memory, mirroring the limited short-term memory of humans.    }

\newcommand{\concat}{^\frown}
\renewcommand{\concat}{||}

Formally, we write $x_i \in \{0,1\}$ for the $i$th input bit, $S_k \equiv \{x_1, \ldots, x_k\}$ for the sequence of $k$ input bits, and ${\bf x}_{j}^{(L)} \equiv \{x_j, x_{j+1}, \ldots , x_{j+L-1}\}$ for the length-$L$ sub-sequence of $S_k$ starting from bit $j$. Given $S_k$ as input, the algorithm predicts the value of $x_{k+1}\in \{0,1\}$ that maximizes 
\begin{equation}
\label{eq:preEstimator}
\max_{ L\in[1,L_{\rm max}]} f_L({x_{k+1}}|{\bf x}_{k-L+1}^{(L)})
\end{equation}
where $f$ estimates the probability of $x$ following  ${\bf x}$ in $S_k$:
\begin{equation}
\label{eq:estimator}
 f_L(x|{\bf x}) = \frac{\#\{ {\bf x}_i^{(L+1)}: {\bf x}_i^{(L+1)} = {\bf x} \concat x,\quad 1 \leq i \leq k-L\}}{\#\{{\bf x}_i^{(L)}: {\bf x}_i^{(L)} = {\bf x},\quad 1 \leq i \leq k-L\}},
\end{equation}
where $\concat$ indicates concatenation and \#A indicates the number of elements in set
A.  Equations (\ref{eq:preEstimator}) and (\ref{eq:estimator}) mean that the prediction algorithm identifies the most frequently input sequence, of length $L_{\rm max}+1$ or shorter, that the player can form when adding the bit $x_{k+1}$, and predicts the player will indeed produce the bit needed to complete that sequence.  $L_{\rm max}$ is chosen to be three, reflecting limited memory of a human opponent in the coin-matching game. 

In Eq.~(\ref{eq:estimator}), the estimator $f_L(x|{\bf x})$ of the probability that $x$ follows ${\bf x}$ in $S_k$ is based on modelling the user's input as a Markov process \citeSI{serfozo2009basics}. The MLA keeps a running estimate $T_L({\bf z}|{\bf y})$, updated with each new input bit, of the matrix describing probabilities of transitions among length-$L$ words, from word ${\bf y}$ to word ${\bf z}$. The estimates are simply the observed frequency of transitions in $S_k$. The MLA then obtains $f_L$ as a marginal probability distribution: the probability of the first bit of ${\bf z}$ being $x$, conditioned on the tail of $S_k$ being ${\bf y}$ (See \btext{Extended Data} Fig.~\ref{fig:ab}).

\section{Networking strategy and architecture}
\label{Sec:Networking}
The BBT required reliable, robust, and scalable operation of two linked networking tasks: providing the BIG Bell Quest video game experience, and live aggregation and streaming of user input to the running experiments.  From a networking perspective, the latter task resembles an instant messaging service, with the important asymmetry that messages from a large pool of senders (the Bellsters) are directed to a much smaller pool of recipients (the labs).  The network architecture is shown in Fig.~\ref{Fig:Architecture}{\bf c}, and was implemented using Amazon Web Services IaaS (Infrastructure as a Service) products. 

In the messaging component, we employed a two-layered architecture, shown in  Fig.~\ref{Fig:Architecture}{\bf c}. In the first layer \textit{Big Bell Test nodes} received input bits from the users and performed a real-time health check, described below, to block spamming by robot participants. The data were then sent to the second layer, a single instance \textit{Hub node} that concatenated all the bits from the first stage and distributed them to the labs. The communication between the two layers was implemented using a \textit{memcache} computation node to maximise speed and to simplify the synchronisation between the two layers. 

The gaming task was handled  by a single layer of \textit{Game nodes} and a database.  To protect the critical messaging task from possible attacks on the gaming components, we used separate instances to handle backend gaming tasks, such as user information and rankings, and to handle backend tasks in the messaging chain, such as data logging. Load balancers, networking devices that distribute incoming traffic to a scalable pool of servers, were used in both the gaming and messaging front ends to avoid overloading. This design pattern is known as horizontal scaling, and is a common practice in  scalable cloud systems. 

This specific architecture was not available as a standard service from web service providers, but was readily constructed from standard component services.  The architecture is not specific to the low-bitrate manual input collected for the BBT, and could straightforwardly be adapted to other data collectable by personal devices, e.g. audio or acceleration. The architecture was designed to solve a problem specific to time-limited projects with crowd-sourced input:  Due to the single-day nature of the BBT, the unknown number and geography of the participants, and the possibility of hackers/spammers, it was not practical to test the system under full-load conditions prior to the event itself. The two-layer architecture helped maintain all the critical servers isolated and independently operating, and helped us smoothly scale up the system when traffic increased. In the event, the traffic surpassed our initial estimates, and we deployed three additional BIG Bell Test nodes at 09:00 UTC  (when Europe was waking up) with no interruption of service.  Such scaling-up is expected to be critical for projects, \btext{e.g. ref.\mciteSI{HeckARX2017},} that combine laboratory experimentation, which tends to be time-limited due to stability and resource considerations, with crowd-sourcing, which usually entails unknown and fluctuating demand.

We now give more details on each computational resource.

\subsection{Big Bell Test nodes}

The first layer of computing resources received data from Bellsters, or more precisely from the BIG Bell Quest running in browsers on their computers and devices. A variable number of servers running the same software functionalities were placed behind a pre-warmed load balancer that was prepared to support up to 10,000 simultaneous connections. Users connected to the load balancer via a public URL end-point, and sent the data from their browsers using \textit{websocket} connections. {This} first layer of servers aggregated the data from {each} connection (i.e. from {each} user) in independent buffers during a $T=\SI{0.5}{\second}$ interval.  

A simple but important `health check' was performed to identify and block high-speed robotic participants.  If a given user contributed more than ten bits in a single half-second interval, corresponding to a rate of more than 20 keypresses per second, the user account was flagged as being non-human and all subsequent input from that user was removed from the data stream. No feedback was provided to the users in the event their account was flagged, to avoid leaking information on the blocking mechanism. This method could potentially ban honest users due to networking delays and other timing anomalies, but was necessary to prevent the greater risk that the data stream was flooded with robotic input.  

\subsection{Hub node}
The Hub node aggregated the data from all the Big Bell Test nodes and {also} handled the connection to the labs. In contrast to the Big Bell Test nodes, which had to service connections from an unknown and rapidly changing number of users, the Hub node was aggregating data from a small and relatively stable number of trusted instances. Overall, the two layer design simplified the networking task of delivering input from a large {and} variable {number} of {users to end points (the labs) receiving aggregated data streams at variable rates}. 

Laboratories connected to the Hub instance to receive random bits from the \Bellsters, which were distributed after aggregating four of the $T=\SI{0.5}{\second}$ batches from the Big Bell test nodes, i.e. in intervals of \SI{2}{\second}. At the end of each interval, bits were sent to each running experiment: if an experiment had requested $N$ bits, it was sent bits $\{x_0, x_1, \ldots , x_{N-1}\}$, i.e., the earliest bits to arrive in that interval. The same bits were in this way used simultaneously in many experiments.  This helps to tighten the locality loophole, since an influence from input bits to measurements would have to operate the same way in several independent experiments and in several locations. With the exception of \UC~and \NIST, the sent bits were used within the next \SI{2}{\second} interval. Experiments \UC~and \NIST, in order to run faster than the Bellster input rate, operated in a burst mode, accumulating bits for a time and then rapidly using them.  As with the Big Bell Test instances, these connections were established using websocket connections. When connecting to the Hub node, the labs specified their bitrate requirement, which could be dynamically changed. The Hub node then sent a stream of Bellster-generated bits at the requested rate.  In the event that insufficient Bellster-generated bits were arriving in real-time, archived bits from BBT participation prior to the day of the experiment were distributed to the labs in advance. In the event, the flux of live bits was sufficient and no experiments used these pre-distributed bits.

\subsection{Memcache node}
The interface between The Big Bell Test nodes and the Hub instance was implemented using a memcache node. While adding an extra computing resource slightly increased the complexity of the architecture, it added robustness and simplified operations. The memcache node, in contrast to the Big Bell Test and Hub nodes, had no internet-facing functionality, making its operation less dependent on external conditions.  For this reason, both the Big Bell Test nodes and the Hub node were registered and maintained on the memcache node, allowing the restart of any of these internet-facing instances without loss of records or synchronisation. 

In addition, and as detailed in the next section, there was an additional Monitor node in charge of (i) recording all the random bits that were being sent from the {\Bellsters} to the labs, and (ii) providing real-time feedback to the \Bellsters. This functionality was isolated from the operations of the Hub node. {Again}, by splitting the Monitor and Hub instances, a failure or attack in the public and non-critical real-time feedback functionality had no effect on the main, private, and critical random bit distribution task.

\subsection{Monitor node}
For analysis and auditing purposes, all of the bits passing through the first layer of servers were recorded in a database, together with metadata describing their origin (Monitor computing resource in Fig.~\ref{Fig:Architecture}{\bf c}). In particular, every bit was stored together with the username that created it and the origin timestamp. {The random bitstreams sent to the individual labs were similarly recorded bit-by-bit, allowing a full reconstruction of the input to the experiments.}

In post-event studies of the input data, we estimated the possible contribution from potentially machine-generated participations that were not blocked by the real-time blocking mechanism. We analysed participants whose contribution were significant, more than \SI{2}{\kilo\bit}  in total, and looked for anomalous timing behaviours such as improbably short time spent between missions and improbably large number of bits introduced per mission, both of which are limited by the dynamics of human reactions when playing the game. Flagging participants that contributed such anomalous participations as suspicious, and cross-referencing against the bits sent to the experiments, we find that no experiment received more than $0.1\%$ bits from the eleven suspicious participants.

\begin{figure*}[t]
\includegraphics[width=0.89\linewidth]{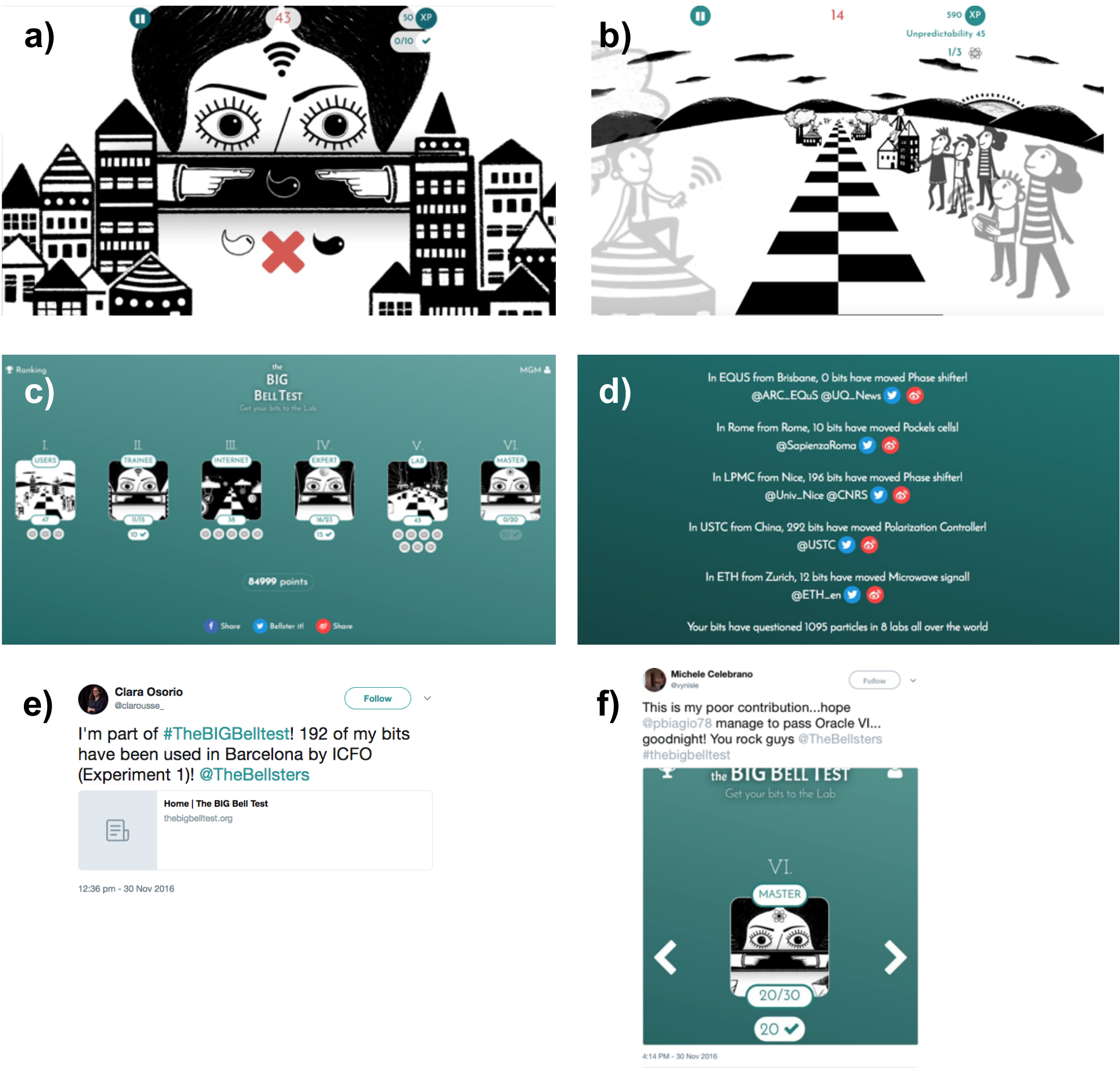}
\caption{\label{fig:Gamification} 
Screenshots from The BIG Bell Quest, illustrating various game elements. {\bf a)} the Oracle opponent uses a machine-learning algorithm to predict user input.  {\bf b)}  the `running' component of the game, in which participants are tasked to enter a minimum number of bits with a minimum fraction unpredicted in a limited time. {\bf c)} sequence of levels, of increasing difficulty, interspersed with oracle challenges {\bf d)} in-game feedback on the use of the user's input bits in running experiments. Blue and red buttons allow instant sharing on social networks Twitter and Weibo, respectively  {\bf e)} \newbtext{a social media post for direct sharing of participant results} {\bf f)} \newbtext{a social media post by a participant} who completed the very difficult last Oracle level. \newbtext{BIG Bell Quest artwork by Maria Pascual (Kaitos Games).} See also Methods~\ref{Sec:Gamification}.  
}
\end{figure*}

In the Monitor computing resource, in addition to being used to store in a database all the information that was streamed to the labs, we also implemented a real-time feedback mechanism to improve the \Bellsters' participation experience.  After accomplishing each mission, users were shown a report on the use of their bits in each of the labs running at that moment, as illustrated in \btext{Extended Data}  Fig.~\ref{fig:Gamification}{\bf d}.
 The numbers shown were calculated as a  binomial random process $B(n, p_i)$ with parameters $n=N$ and $p_i = R_i / R$, where  $N$ is the number of bits introduced by a user in his/her last mission, $R_i$ is the number of bits sent to lab $i$, and $R$ is the total number of bits entered in the last $T=\SI{0.5}{\second}$ interval.  

\section{Gamification}
\label{Sec:Gamification}
The BBT required a large number of human-generated random bits in a short time, thus requiring many participants (\Bellsters), rapid input, and sustained participation. The gamification strategy was designed to maximize all of these factors.  The game itself The BIG Bell Quest can be viewed and played at \href{https://museum.thebigbelltest.org/quest/}{https://museum.thebigbelltest.org/quest/}. Relevant screenshots are shown in \btext{Extended Data}  Fig.~\ref{fig:Gamification}. 

While still adhering to common conventions of video games (levels, power-ups, boss battles, animations, sound effects, etc.), the intended appeal of the BIG Bell Quest is less its entertainment value than the opportunity to contribute to the BBT experiments, and to test one's unpredictability against a computer opponent.  The \stext{gamification} \btext{game design} incorporated internationalization, connection to social networking, community-building features, and a feedback system to inform users about their contribution to the experiment, all considered essential to attract the necessary tens of thousands of participants. 

The game has a classic challenge and reward incentive structure.  The challenge is to produce random bits, avoiding being predicted by the Oracle, (see \btext{Extended Data}  Fig.~\ref{fig:Gamification}{\bf a}). This reproduces the `penny-matching' game studied in psychology\citeSI{gibbons1992game, MOOKHERJEE199462}, and resembles the well-known `rock-paper-scissors,' thus requiring little explanation.  The Oracle is a machine-learning algorithm that predicts player behaviour based on patterns in past input, described in Section \ref{Sec:PredictionEngine}.   Most player time was spent in a rapid `speed game' (see \btext{Extended Data}  Fig.~\ref{fig:Gamification}{\bf b}), in which the Bellster moves along a road by hitting 0s and 1s.  This part of the game requires rapid bit generation (a few bps) in order to complete the level in time.  Every 20 bits an indicator shows the player's `unpredictability,' i.e., the percentage of unpredicted bits entered thus far, and the final score reflects the number of unpredicted bits, with a power-up multiplier for bits entered during a particular time window. 

The rewards are multiple: at the individual level, the player is given a score for each level (to encourage a high fraction of unguessed input), and a cumulative score (to encourage repeated play).  At the community level, a sharing platform offers rankings and a tool to create groups, so that \Bellsters can compare their performance among friends and colleagues.  Players can also post their scores to social networks (Facebook, Twitter or Weibo) at the press of a button, see \btext{Extended Data}  Fig.~\ref{fig:Gamification}{\bf e} and {\bf f}.  At the scientific level, the game provides a report on which laboratories have used how many of the player's bits, and for what purpose (see \btext{Extended Data}  Fig.~\ref{fig:Gamification}{\bf c}). Finally, the player is occasionally rewarded with a short video pre-recorded in one of the laboratories, in which experimentalists explain a part of their experiment.  User feedback, see \btext{Extended Data}  Fig.~\ref{fig:Gamification}{\bf e}, suggests this approach succeeded in making \Bellsters feel meaningfully involved in the project, with a positive effect on retention and propagation.

This Speed Game + Oracle structure was repeated through three levels of difficulty, or `worlds:'  `Users', `Internet', and `Laboratory', illustrating the travel of the bits from the fingers of the \Bellsters, through the Internet, to the labs of the scientists (see \btext{Extended Data}  Fig.~\ref{fig:Gamification}{\bf c}). 
The times, speeds, and required unpredicted fraction in each of the levels were adjusted with the help of beta testers to avoid offering levels of trivial or impossible difficulty.  The final Oracle level was objectively difficult even for an experienced player.  To pass, it required $n\ge 20$ unguessed bits in a time allowing at most 30 bits to be entered. Even for a sequence of 30 ideal random input bits, the condition $n\ge 20$  occurs less than 5\% of the time by binomial statistics. For 30 bits predictable with probability 0.6, the chance of success drops below $0.003$. Nevertheless, several players persisted and completed the game (see \btext{Extended Data}  Fig.~\ref{fig:Gamification}{\bf f}). 

\btext{
\section{Data Availability}
Experimental data are available upon reasonable request from the contact author of each experiment, as indicated in the author contributions. Other project data are available upon reasonable request from the corresponding author.  
}




\clearpage

%
%
%
%


\clearpage

\onecolumngrid

\newcommand{\BBTcite}[1]{\citeBBT{#1}}
\newcommand{\BBTNodeTitle}[1]{\noindent{\large \NODE~#1}\\}
\newcommand{\BBTNodeAuthorList}[1]{\noindent{Authors: #1}}
\newcommand{\NODE}{NODE}


%
%

\renewcommand{\figurename}{Supplementary Figure}
\newcommand{\TextFig}{Suppl.~Fig.}

\renewcommand{\BBTcite}[1]{\citeGU{#1}}
\renewcommand{\NODE}{\GRIFFITH}
\graphicspath{{./PartnerContributions/GUBBT/}}


\BBTNodeTitle{Quantum steering using human randomness }

\BBTNodeAuthorList{Raj B. Patel, Farzad Ghafari Jouneghani, 
Morgan M. Weston, Sergei Slussarenko, and Geoff J. Pryde}

\begin{figure*}[tbh]
{\centering
	\includegraphics[width=1\textwidth]{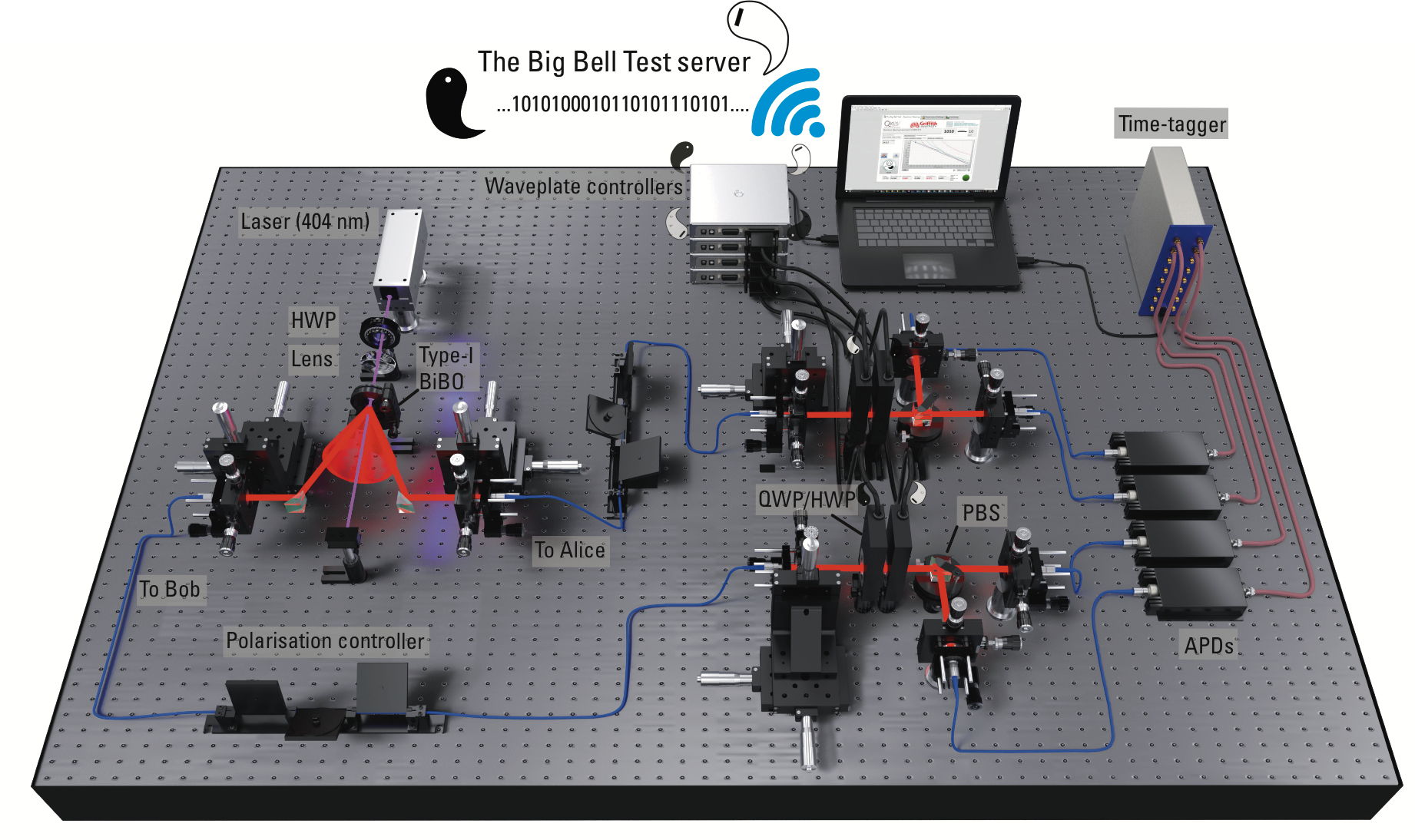}}
\caption{Experimental setup. Alice and Bob's measurements stages are set according to the random stream of bits acquired from the Big Bell Test server. Entangled photon pairs are generated by pumping a sandwiched Type-I BiBO crystal with a continuous wave laser diode at 404 nm. Each photon from the down-converted pair is sent along an optical fiber to polarisation analysis stages, one for Alice and one for Bob. Photons are detected using silicon avalanche photodiode detectors and the electrical signals for each event \btext{are} time-tagged. A computer is used to calculate the steering parameter $S_{16}$ in real-time. As additional random measurement settings are retrieved from the server, the experiment is iterated for improved statistics.}
\label{fig:SetupGU}
\end{figure*}
Schr\"{o}dinger first coined the term `steering'\BBTcite{schrodinger} as a generalisation of the EPR-paradox. With the advent of quantum technologies, steering has been recognised as being well suited to certain quantum communication tasks. Here we report a demonstration of EPR-steering using polarisation entangled photons, where Alice and Bob's measurement settings are chosen based on data randomly generated by humans. We use a 404 nm UV continuous wave laser diode to pump \btext{a} pair of sandwiched type-I nonlinear bismuth triborate (BiBO) crystals to generate entangled photon pairs at 808 nm via spontaneous parametric down-conversion. The generated state is the singlet state $|\psi\rangle=\frac{1}{\sqrt{2}}(|H_{A}V_{B}\rangle-|V_{A}H_{B}\rangle)$. The generated photon pairs are sent to two separate measurement stages consisting of polarisation analysers and single-photon avalanche photodiode detectors. {The stages,} designated Alice and Bob, were located 50 cm apart from one another. Single photons are measured shot-by-shot. That is, for each random measurement setting, a short burst of detection events are collected and time-tagged. From this set of detections, only the very first joint detection is kept and the others are discarded.

During the Big Bell test event, bits were acquired at a rate of \SI{4}{bps} for 24 hours. A random four bit sequence represents one of $n=16$ measurement settings per side.  After performing all sixteen measurements, the following steering inequality was calculated, ${S_{16}} = \frac{1}{n}\sum\limits_{k = 1}^n {\left\langle {{A_k}\sigma _k^B} \right\rangle }\le C_n$ (refs\BBTcite{dylan,adamprx}). Here ${S_{16}}$ is referred to as the steering parameter whilst $A_k\in\{-1,1\}$ and $\sigma^{B}_{k}$ is Alice's measurement outcome and the Pauli operator corresponding to Bob's measurement setting, respectively. The correlation function is bounded by +1 (maximal correlations) and -1 (maximal anti-correlations) with a value of 0 representing no correlation at all.  It should be noted that fair sampling of all the detected photons is assumed.  Given these parameters we obtain  ${S_{16}} = 0.965 \pm 0.008$ which beats the bound of $C_{16}=0.511$ by 57 standard deviations. This is first demonstration of quantum steering with human-derived randomness.

\clearpage
\renewcommand{\BBTcite}[1]{\citeEQUS{#1}}
\renewcommand{\NODE}{\EQUS}
\graphicspath{{./PartnerContributions/EQUSBBT/}}

\BBTNodeTitle{\btext{Quantum Correlations} in Time}

\BBTNodeAuthorList{Martin Ringbauer and Andrew White}

\begin{figure*}[h]
\centering
	\includegraphics[width=0.95\columnwidth]{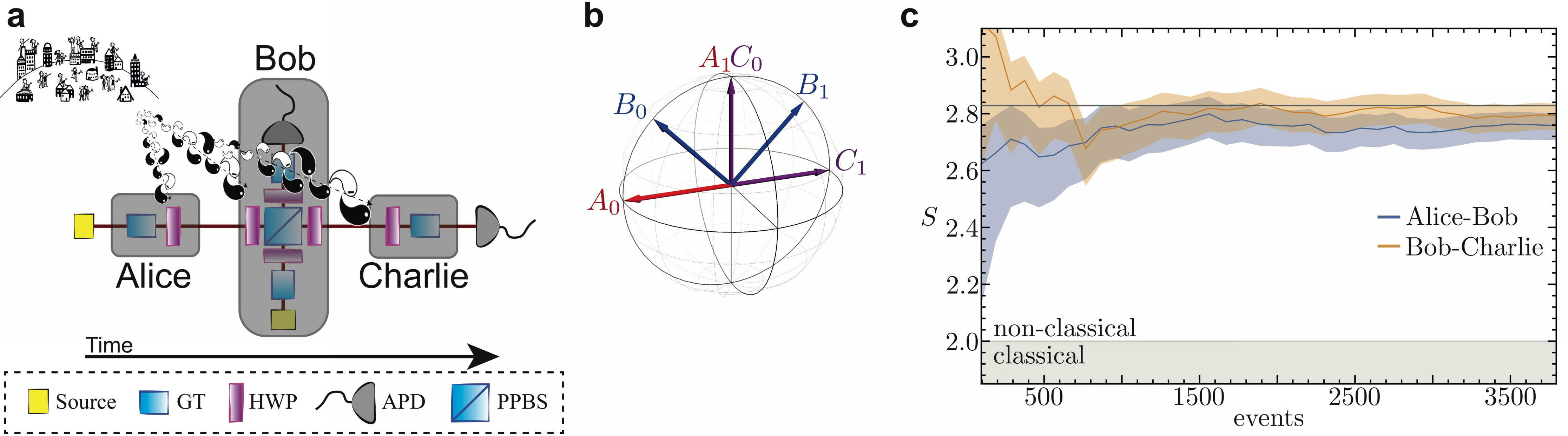}
\caption{\textbf{(a) The experimental setup used to test \btext{temporal quantum correlations}.}  A single photon, produced by spontaneous parametric down-conversion (not shown), is first measured by Alice, then by Bob and finally by Charlie. Alice and Charlie use a half-wave plate (HWP) and a Glan-Taylor polarizer (GT) for their measurement, while Bob, being in the middle, measures the quantum system indirectly by entangling it to another photon (using a partially polarizing beam splitter, PPBS) and detecting that photon in an avalanche photo diode (APD). All three observers choose their measurement settings by using the human-generated random bits supplied by the Bellsters to set the angles of their waveplates.
\textbf{(b) Measurement settings.} Visualization of the measurement settings for Alice ($A_0, A_1$), Bob ($B_0, B_1$), and Charlie ($C_0, C_1$) on the single-qubit Bloch-sphere.
\textbf{(c) Experimental results.} Shown is the cumulative violation of the CHSH inequality in time between the observers Alice and Bob (blue), and between Bob and Charlie (orange) versus the number of observed detection events. The data shows a strong violation of both inequalities and thus a violation of entanglement monogamy. The shaded regions correspond to $1\sigma$ statistical confidence intervals.}
\label{fig:SetupEQUS}
\end{figure*}

\newcommand{\myemph}[1]{{#1}}

{
In the scenario originally considered by Bell~\BBTcite{Bell1964}, pairs of entangled particles are shared between spacelike-separated observers, Alice and Bob, who perform local measurements on their particles. Under the conditions of realism, measurement independence, and local causality, the correlations between Alice's and Bob's measurement outcomes must then satisfy a set of Bell inequalities. The correlations of entangled quantum systems, on the other hand, violate these inequalities and are thus said to be in conflict with the above assumptions. \btext{These correlations}, however, \btext{are} not limited to spacelike-separated scenarios, which \btext{are} in fact quite challenging to achieve in practice. 

In the present experiment, we thus do not probe traditional Bell-nonlocality, but instead consider correlations between measurements performed on the \myemph{same} quantum system at \myemph{different} points in time. Such experiments are expected to reveal \btext{correlations that resemble those observed in entangled quantum systems, and just like in the spatial case, we can test for these correlations using a Bell-inequality}~\BBTcite{Brukner2004} 
\begin{equation}
\label{eq:CHSH}
S_{\textsc{ab}}=\langle A_0B_0\rangle + \langle A_0B_1\rangle + \langle A_1B_0\rangle - \langle A_1B_1 \rangle \leq 2 ,
\end{equation}
where $A_0$ and $A_1$ ($B_0$ and $B_1$) are Alice's (Bob's) measurement settings. This inequality is derived under equivalent assumptions as the spatial case, with local causality replaced by an analogous temporal no-fine-tuning assumption, see Ref.~\BBTcite{Wood2015CausalDiscovery,TemporalFramework} for details.

Experimentally we test \btext{temporal quantum correlations} using the setup in \TextFig~\ref{fig:SetupEQUS}a, where a single photon is subject to a sequence of three polarization measurements, first by Alice, then Bob, and finally Charlie. Pairs of single photons at a wavelength of $820$~nm are produced via spontaneous parametric downconversion in a $\beta$-Barium borate (BBO) crystal, pumped by a femtosecond-pulsed Ti:Sapphire laser at a wavelength of 410~nm and a repetition rate of $76$~MHz. One of these photons acts as the system which is subject to a series of projective measurements by Alice, then Bob, and finally Charlie. The second photon is used as an ancilla for Bob's measurement, which is implemented in a non-destructive fashion. Specifically, Bob entangles the system to the ancilla (the meter) using a non-deterministic controlled-NOT gate based on non-classical interference on a partially polarizing beamsplitter, and then measures the meter photon in the computational basis. Operating the experiment at low pump power to suppress higher-order emissions with a $g^{(2)}(0)\sim 10^{-2}$, the gate achieved a relative Hong-Ou-Mandel interference visibility of $1.00\pm0.01$. Through local rotations of the system before and after the entangling operation, this design thus allowed for arbitrary non-destructive measurements on the system with a fidelity of $\overline{\mathcal{F}} = 0.98\pm0.02$ with the ideal projective measurement, and an average purity of $\overline{\mathcal{P}}=0.97\pm0.03$, as determined using quantum process tomography. The remaining imperfections can be attributed to imperfect mode overlap in the entangling gate, limiting the entanglement generation between Bob and Charlie to a concurrence of $\mathcal{C}=0.970\pm0.006${.}


The measurement settings for each party, shown on the Bloch-sphere in \TextFig~\ref{fig:SetupEQUS}b, are set through waveplate rotations driven by the human-generated random numbers supplied by the Bellsters. Due to the sequential nature of the experiment a gated detection scheme was chosen, where for each waveplate setting a pair of avalanche photo diodes (APDs) recorded data for 100ms. This resulted in an average event-rate of $1.60\pm0.07$ detected photons per gate window for all combinations of measurement settings. {In contrast to the other experiments in the BBT, this sometimes produced multiple events from the same human-generated bits.} It should be noted that the experiment in its current form is subject to a detection loophole, and fair sampling is thus assumed for all observed events. Closing the detection loophole would require not just more efficient detectors, but also an event-ready version of the intermediate non-destructive measurement, instead of the current non-deterministic design. On the other hand, since the experiment is explicitly timelike separated, it is not subject to a locality loophole in the usual sense. There is, however, the related assumption that there is no hidden (i.e. fine-tuned) communication channel between the time steps~\BBTcite{Ringbauer2017TemporalCM}. 

\TextFig~\ref{fig:SetupEQUS}c shows the cumulative statistics for inequality~\eqref{eq:CHSH} between Alice and Bob, and between Bob and Charlie as a function of the number of recorded events. The observed values of $S_{\textsc{ab}}=2.75\pm0.05$ and $S_{\textsc{bc}}=2.79\pm0.05$ not only demonstrate \btext{a violation of Eq.~(\ref{eq:CHSH})}, but also constitute the first experimental demonstration of a key difference between spatial and temporal \btext{quantum correlations}. In the spatial case entanglement between Alice and Bob precludes either of them from being entangled with a third party, which is known as \myemph{monogamy of entanglement}~\BBTcite{Coffman2000}. In contrast, our results show that polygamy is possible in the temporal case: Bob can \btext{violate a CHSH-inequality with} Alice, and, at the same time, \btext{with} Charlie. 
This experiment is thus a first step towards exploring the rich structure of multipartite quantum correlations in time, which remains widely unexplored. Understanding the relationship between temporal and spatial correlations might be important for the design of future distributed quantum networks, where both kinds of correlations are expected to play a role. Temporal correlations may also be useful for temporal quantum communication~\BBTcite{Brukner2004} and computation~\BBTcite{Markiewicz2014} schemes. In light of these potential applications it would be of particular interest to explore more complex scenarios involving genuine multipartite temporal correlations and correlations arising from generalized measurements. 
}



\clearpage
\renewcommand{\BBTcite}[1]{\citeUSTC{#1}}
\renewcommand{\NODE}{\USTC}
\graphicspath{{./PartnerContributions/USTCBBT/}}
\BBTNodeTitle{Bell tests with imperfectly random human input}

\BBTNodeAuthorList{
Yang Liu, Xiao Yuan, Cheng Wu, Weijun Zhang, Jian-Yu Guan, Jiaqiang Zhong, Hao Li, Ming-Han Li, Sheng-Cai Shi, Lixing You, Zhen Wang, Xiongfeng Ma, Qiang Zhang and Jian-Wei Pan}

\begin{figure*}[tbh]
\centering
\includegraphics[width=0.85\textwidth]{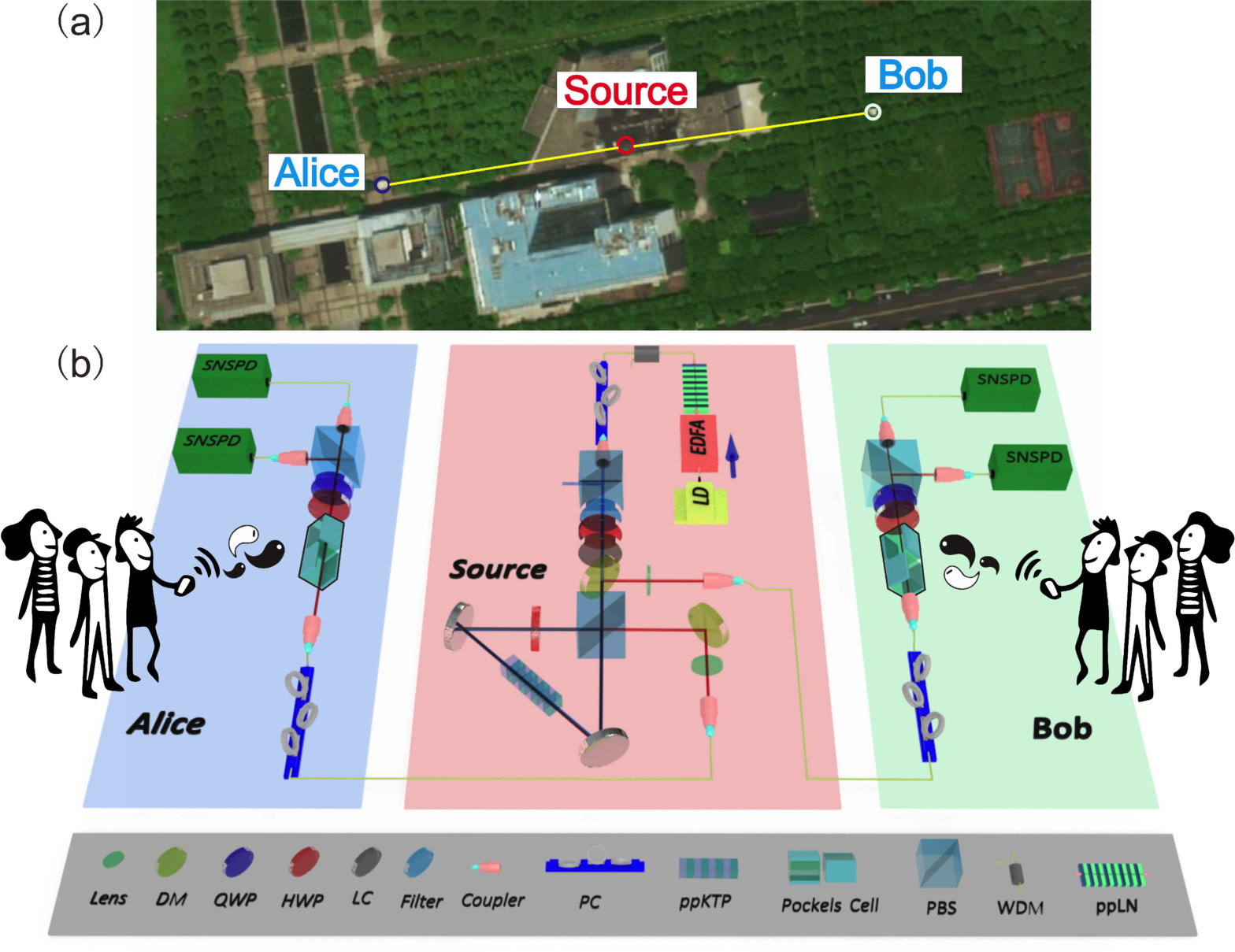}
\caption{Bell test using imperfect input randomness. (a) A bird's-eye view of the entanglement source, Alice's detection and Bob's detection. The distances from the source to Alice and Bob are $87\pm2$ m and $88\pm2$ m, respectively. (b) Schematic setup of the Bell test. A distributed feedback (DFB) laser diode (LD) at $\lambda=1560$ nm is modulated to produce pulses with a repetition rate of \SI{100}{\kilo\hertz} and a pulse width of \SI{10}{\nano\second}. The pulses are amplified with an erbium-doped fiber amplifier (EDFA) and then are up-converted to 780 nm via second-harmonic generation (SHG) in an in-line periodically poled lithium niobate (PPLN) waveguide. The residual 1560 nm light is filtered with a wavelength-division multiplexer (WDM) and a shortpass filter. After adjusting the polarization using a half-wave plate (HWP) and a liquid crystal (LC), the 780 nm pump light is focused to a periodically poled potassium titanyl phosphate (PPKTP) crystal in a ``Sagnac'' geometry to generate entangled photon pairs. A series of dichroic mirrors (DMs) are used to remove the residual pump light at 780 nm and fluorescence before the entangled pairs are collected. In Alice's and Bob's detection stations, a polarization controller (PC), a quarter-wave plate (QWP), a HWP and a polarizing beam splitter (PBS) are used to set the measurement angle. Random numbers control the Pockels cells to dynamically select the bases. Superconducting nanowire single-photon detectors (SNSPD) are used to detect the photons after the PBS.}
\label{fig:SetupUSTC}
\end{figure*}

Humans are not perfectly random, and tend to produce patterns that make their choices somewhat predictable. For example, we ran the NIST statistical test {suite}\BBTcite{RukhinNIST2010} on the human-generated random numbers from the BBT and of the 14 different tests for uniformity, the human random numbers only passed 2. Nevertheless, human randomness is very attractive for Bell tests because of the element of human free will; if this freedom exists, the human choices are not controlled by hidden variables.  Remarkably, it is possible to say how well the hidden variable would have to control the human choices to explain a Bell inequality within local realism. If $a, b$ and $x, y$ denote the binary outputs and inputs, respectively, then the imperfection of the input randomness can be characterized by a  bound on the conditional setting probability $P(xy|\lambda)$
\begin{equation}
\begin{aligned}
	l &= \min_{xy\lambda}P(xy|\lambda).
\end{aligned}
\end{equation}
where $l \in[0,1/4]$ and a smaller value of $l$ indicates more imperfection of the input randomness.
We report two Bell tests, one using the well-known Clauser-Horne-Shimony-Holt (CHSH) \BBTcite{CHSH} inequality and the other using the measurement dependent local (MDL) \BBTcite{putz14} inequalities. The MDL is designed to be more robust against influence on the settings.  We observe large violations of the two Bell inequalities and give the $l$ required of the input human randomness to rule out local realism. 

Our experimental setup is shown in shown in \TextFig~\ref{fig:SetupUSTC}(b). A 1560 nm seed laser with frequency $f=\SI{100}{\kilo\hertz}$, width $\Delta t=\SI{10}{\nano\second}$ is amplified and up-converted to \SI{780}{\nano\meter} via second-harmonic generation (SHG). Pumped with this laser, entangled pairs are generated in the Sagnac based setup, and then collected into single mode fibers for detection. The measurement devices in Alice's and Bob's detection station are around 90 meters from the source, as shown in \TextFig~\ref{fig:SetupUSTC}(a). {The spatial separation makes sure the measurement in Alice's detection station will not affect that in Bob's detection station, and vice versa.} A field-programmable gate array (FPGA) board is used to generate synchronizing signals and to distribute the random numbers in real time. ICFO provided us the human generated random numbers, which we re-distribute to modulate the basis immediately. Note that the system only works when there is a batch of random numbers comes in. The random numbers control the Pockels cell by applying a zero or a half-wave voltage, setting the basis to  $A_0/A_1$ for Alice and $B_0/B_1$ for Bob. Passing through the PBS, the photons are finally detected with a superconducting nanowire single-photon detectors (SNSPD), and recorded by a time-digital convertor (TDC) for off-line data analysis.

First, the maximum entangled state $\ket{\Psi^+}=(\ket{HV}+\ket{VH})/\sqrt{2}$ is generated and tested. We measure the visibility in the horizontal/vertical basis as 99.2\% and in the diagonal/anti-diagonal basis as 98\%. State tomography shows the fidelity to the ideal state is around 97.5\%. We test the standard CHSH inequality with quantum random numbers, calculate the value $S=2.804$, with $E(A_1,B_1)=-0.751$, $E(A_1,B_2)=0.651$, $E(A_2,B_1)=0.657$, $E(A_2,B_2)=0.745$ where $E(A_x, B_y) \equiv \sum (-1)^{a+b+xy}p(ab|xy)$. Considering MDL correlation with the randomness measure $l$, that is $P(xy|\lambda)\ge l, \forall xy\lambda$, the maximal achievable CHSH value with MDL correlation is $4(1-2l)$ \BBTcite{putz14,  Yuan15b}. To achieve the experimentally obtained Bell value with MDL correlation, we thus need $4(1-2l) < S = 2.804$ and thus $l<0.1495$. In other words, if the input human randomness is high enough, i.e., if $l>0.1495$, the experimentally observed data cannot be explained with any MDL correlation.

Next, the human random numbers are used to test the MDL inequality \BBTcite{putz14} which is defined by
$lP(0000)-(1-3l)[P(0101) + P(1010) + P(0011)] \le 0. $
Interestingly, as long as $P(xy|\lambda)\ge l$, such an inequality cannot be violated with any MDL correlation. In this case, observing the violation of the MDL inequality indicates the existence of non-classical correlation even in the presence of measurement dependence.

The non-maximal entangled state $\cos(69.1^\circ)\ket{HV}+\sin(69.1^\circ)\ket{VH}$ is generated. State tomography shows the fidelity to the ideal state is 98.7\%. Using the setup described above, the locality loophole is closed by considering the spatial distance and the randomness loophole is closed with human free will. The experiment was performed during 30 Nov. and 1 Dec. when the public helps us to generate random numbers. The coincidence values were tested with human-generated bits determining the basis choices. The experimental data are divided into to several one-hour periods for statistical analysis.

As this MDL inequality cannot be violated with MDL correlation satisfying $P(xy|\lambda) \ge l, \forall xy\lambda$, we calculate \btext{$l_0$,} the smallest possible value of $l$ \btext{that satisfies the inequality} \stext{such that the equal sign is saturated}. In experiment, we obtain $l_0=0.10\pm0.05$ for MDL inequality by using human random numbers. As a comparison, we also use quantum random numbers to choose the basis. With the rest of the setup remaining the same, we obtain $l_0=0.106\pm 0.007$ which gives a similar $l$ value with less fluctuation. This is because the total amount of data using quantum random numbers is much larger than using human random numbers. The input of human random numbers was at most a few thousand per second whereas for quantum random numbers we had a steady \SI{100}{kbps} for controlling the basis. In total, we accepted and used around \SI{80}{Mb} of human random numbers for the MDL inequality test in two days. For comparison, we tested the inequality with quantum random numbers in around one and half hours, using more than \SI{500}{Mbits}.



\clearpage
\renewcommand{\BBTcite}[1]{\citeIQOQI{#1}}
\graphicspath{{./PartnerContributions/IQOQIBBT/}}
\renewcommand{\NODE}{\IQOQI}

\BBTNodeTitle{Violation of a Bell Inequality using Entangled Photons and Human Random Numbers}

\BBTNodeAuthorList{Bo Liu, Johannes Handsteiner, Dominik Rauch, Rupert Ursin, Thomas Scheidl and Anton Zeilinger}

\begin{figure*}[tbh]
\centering
	\includegraphics[width=0.75\textwidth]{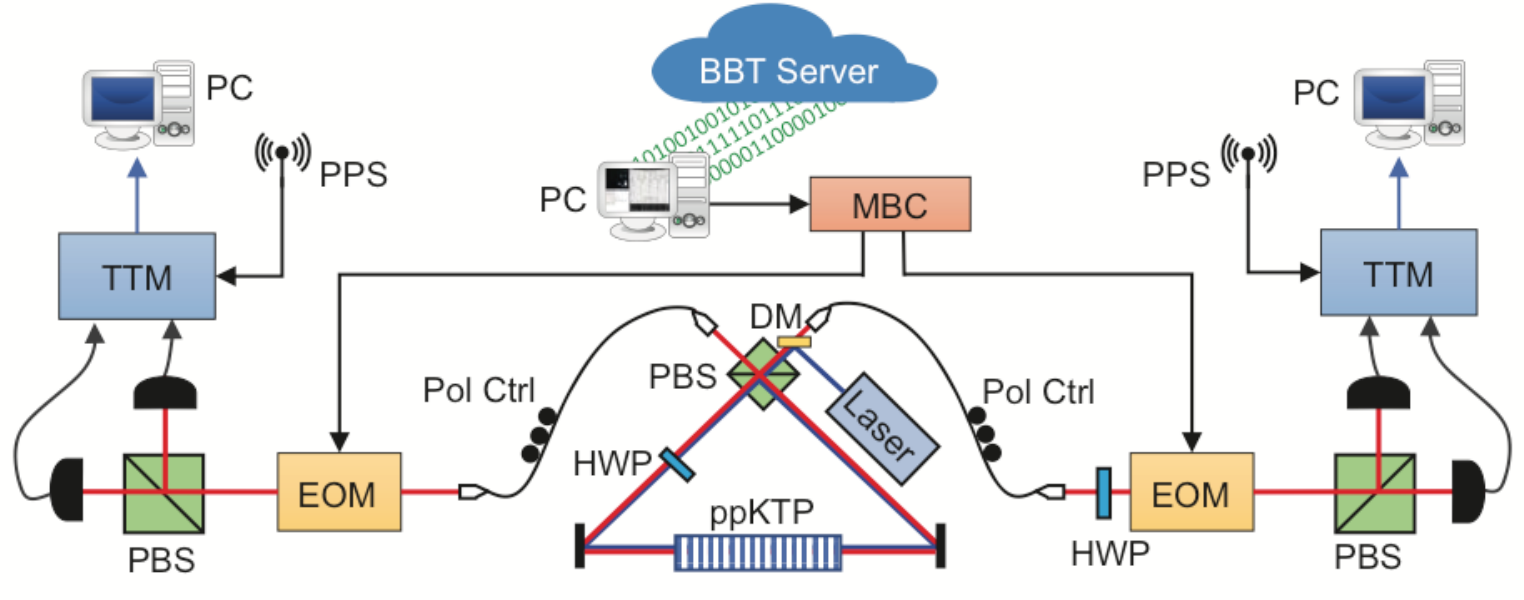}
\caption{Experimental setup. The wavelength of the laser is 405nm. MBC: measurement basis controller; Pol Ctrl: polarization controller; DM: dichroic mirror; HWP: half-wave plate; EOM: electro-optical modulator; PBS: polarized beam splitter; TTM: time tagging module; PPS: pulse per second.}
\label{fig:SetupIQOQI}
\end{figure*}

In the Big Bell test (BBT) experiment, we tested the CHSH form of the Bell inequality with polarization-entangled photon pairs. As shown in \TextFig~\ref{fig:SetupIQOQI}, the entangled photon source was based on a Sagnac interferometer generating polarization-entangled photon pairs in the maximally entangled state $|\Psi^-\rangle=\frac{1}{\sqrt{2}}\left(|H_AV_B\rangle-|V_AH_B\rangle\right)$. Each photon pair was guided to two receivers, called Alice and Bob, via single mode fibers. At each receiver, the entangled photons were coupled out of the fiber and guided to an electro-optical modulator (EOM), which allowed for fast switching between complementary measurement bases. The EOM was followed by a polarizing beam splitter (PBS) with a single-photon avalanche diode (SPAD) detector in each output port. The EOM settings were driven by the measurement basis controller (MBC), which read the random bits from the BBT server and converted them in real-time to the corresponding EOM control signals. In this way, real-time measurements in the following linear polarization bases have been implemented for Alice (Bob): $0^\circ/90^\circ$ ($22.5^\circ/112.5^\circ$) if the random number was '0' and $45^\circ/135^\circ$ ($67.5^\circ/157.5^\circ$) if the random number was '1'. Finally, using a time tagging module (TTM) and a computer, Alice and Bob generated and recorded time stamps of all SPAD detection events, the EOM settings and an additional shared pulse per second (PPS) signal.

\begin{figure*}[tbh]
\centering
	\includegraphics[width=0.95\textwidth]{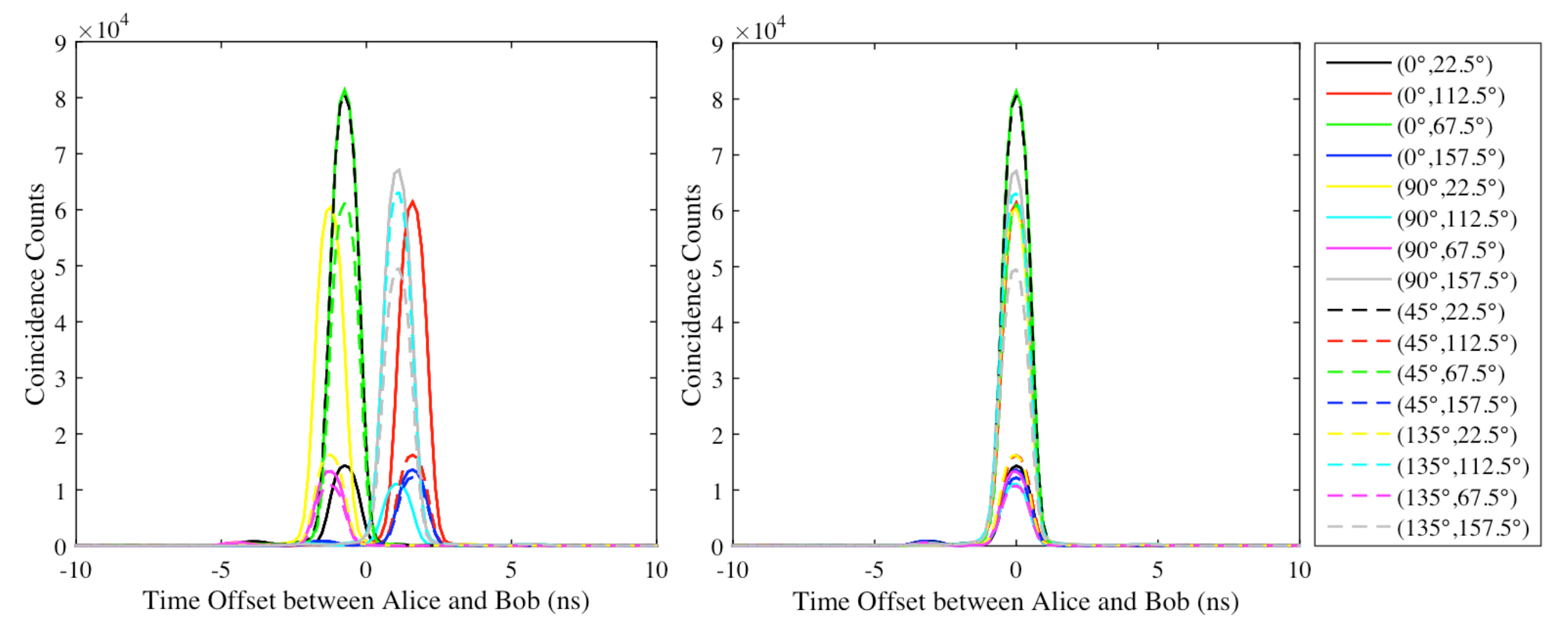}
\caption{Coincidence counts between Alice and Bob for the experiment with human random numbers (HRN) mixed with quantum random numbers  (QRN).}
\label{fig:histogram}
\end{figure*}

On the BBT day, we performed measurements using all three different types of random numbers provided via the BBT server, i.e. real-time human random numbers (HRN), quantum random numbers (QRN) and human random numbers from the database (DB). In the first experiment, we used a mix of HRN and QRN for a total measurement time of 200 seconds. During post-processing, we separated the results obtained with HRN from the results obtained with QRN. For the HRN (QNR) data we obtained a Bell value of 2.6387 (2.6434) with a 81 (115) sigma violation of the local-realistic bound of 2. The second experiment was performed with a mix of HRN and DB for a total measurment time of 210 seconds. For the HRN (DB) data we obtained a Bell value of 2.6403 (2.6371) with a 63 (122) sigma violation of the local-realistic bound of 2. Errors are calculated assuming Poissonian photon statistics. In our experiments, the time delay between Alice's and Bob's time-tags slightly varied for the different detector combinations as can be seen from the coincidence peaks shown in \TextFig~\ref{fig:histogram}. However, this could easily be compensated during data post-processing, such that we could use a coincidence window of 1.25ns. All results are summarized in detail in Table~\ref{tab:DetailedRes}.

\begin{table}[!htb]

\caption{Detailed measurement Results.}
\label{tab:DetailedRes}
\begin{tabular}{|l|l|l|l|l|l|l|l|l|l|l|l|}

\hline
\bf Exp \# &  \bf Time (s) & \bf Source & \bf RN rate (bps) & \bf Total CCs  & \bf P1 & \bf P2 & \bf P3 & \bf P4 & \bf S Value &\bf Xigma & \bf Visibility
\\
\hline
1  & 200 & HRN & 1612.88 & 160178 & 0.6520 & -0.6947	& -0.6592 & -0.6329 & 2.6387 & 81.0265 & 0.9329
\\
 & & QRN & 2727.68 & 294465 & 0.6560 & -0.6938 & -0.6544 & -0.6391 & 2.6434 & 115.8174 & 0.9346
\\
\hline
2 & 210 & HRN & 910.21 & 95024 & 0.6500 & -0.7009 & -0.6539 & -0.6355 & 2.6403 & 63.0984 & 0.9335
\\
 & & DB & 3404.49 & 337463 & 0.6498 & -0.6984 & -0.6589 & -0.6300 & 2.6371 & 121.7927 & 0.9324
\\
\hline

\end{tabular}
\end{table}


\bibliographystyleIQOQI{../../biblio/BBT}
\bibliographyIQOQI{./PartnerContributions/IQOQIBBT/IQOQI}

\clearpage
\renewcommand{\BBTcite}[1]{\citeROMA{#1}}
\graphicspath{{./PartnerContributions/ROMABBT/}}
\renewcommand{\NODE}{\SAPIENZA}
\BBTNodeTitle{Experimental bilocality violation with human randomness}

\BBTNodeAuthorList{Luca Santodonato, Gonzalo Carvacho, Francesco Andreoli, Marco Bentivegna, Rafael Chaves and Fabio Sciarrino}

\begin{figure*}[tbh]
\centering
	\includegraphics[width=0.85\textwidth]{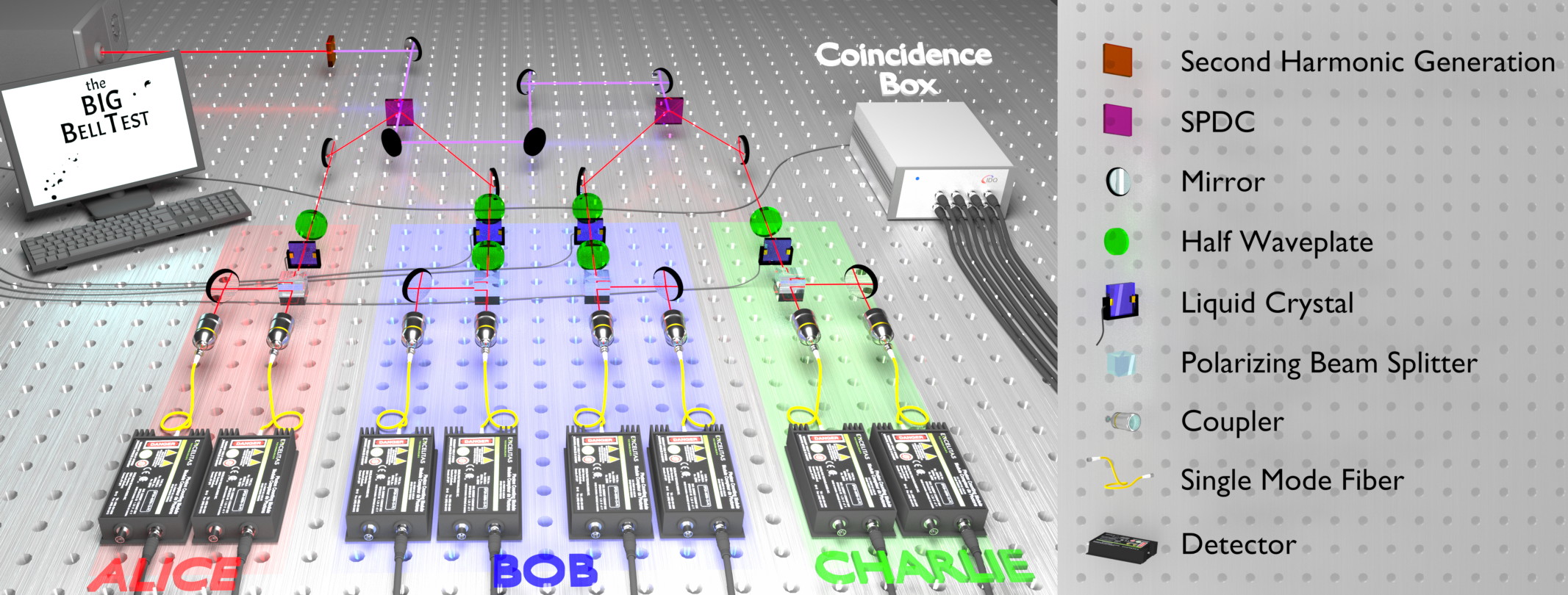}
\caption{\textbf{Experimental setup for bilocality test exploiting human randomness.} Two couples of photons are independently generated in the singlet state by a pulsed pump laser. For each pair, one photon is sent to one of the furthest stations (A or C, respectively), while the other one reaches the central node B. All parties perform single qubit polarization analysis, exploiting Half wave-plates (HWPs) and Liquid Crystals (LC). The measurement choice is controlled by the random bits provided by the ICFO BBT cloud infrastructure, which drives the voltage applied on the LC depending on the value of three random bits.}
\label{fig:SetupROMA1}
\end{figure*}

Nowadays, complex networks are attracting a large interest for the development of future quantum technologies but also for the understanding of generalized Bell-like inequalities.
{Quantum information processing will certainly rely on quantum networks which until now have been exploited for studies ranging from generalized Bell non-locality \BBTcite{Branciard2010,Branciard2012,polinomial,fritz2012,takavoli2014,rosset,takavoli2016,Gisin2017,AndreoliNJP2017}, distribution and control of entanglement \BBTcite{distribution,control,Cavalcanti2011}, non-Markovian dynamics \BBTcite{markovian} to quantum network of clocks with GHZ states \BBTcite{clocks}}.
Here by considering a tripartite scenario where two parties (Alice and Charlie) share entangled states (independently generated) with a third one (Bob), we are able to witness a new kind of non-classical behavior called quantum non-bilocality \BBTcite{Branciard2010,Branciard2012}, where the results are incompatible with {any possible model described by local realism, two independent hidden variables and the free-will assumption}.
Alice, Bob and Charlie perform dichotomic measurements respectively labeled by the variables x, y and z, and they return some binary outcomes described by the variables a, b and c. Since we are considering two physically different and independent entangled sources we must consider also two different sets of hidden variables $\lambda_{1}$ and $\lambda_{2}$.
Any correlation allowed by this model can be written as $p(a,b,c \vert x,y,z)= \sum_{\lambda_1,\lambda_2}p(\lambda_1)p(\lambda_2)p(a\vert x,\lambda_1)p(b\vert y,\lambda_1,\lambda_2) p(c\vert z,\lambda_2).$
The set of correlations compatible with the described model is constrained by the so-called bilocality inequality: $B=\sqrt{|I|}+\sqrt{|J|}\leq 1$ (with $I=\frac{1}{4} \sum_{x,z} \langle A_{x}B_{0}C_{z} \rangle$, $J=\frac{1}{4}  \sum_{x,z} (-1)^{x+z} \langle A_{x}B_{1}C_{z} \rangle,$ (summing on $x,z=0,1$) and where: $\langle A_{x}B_{y}C_{z} \rangle=  \displaystyle \sum_{a,b,c=0,1} (-1)^{a+b+c} p(a,b,c|x,y,z)$), which can be violated by quantum mechanics \BBTcite{bilocality_nostro,saunders,expandreoli}.
In our experiment we implement a photonic network consisting of three parties and two independent sources of quantum states (see \TextFig~\ref{fig:SetupROMA1}) and we evaluate the bilocality parameter $\mathcal{B}$, exploiting random measurements settings driven by the bits provided by the Big Bell Test experiment. 

\begin{figure*}[h]
\centering
	\includegraphics[width=0.75\textwidth]{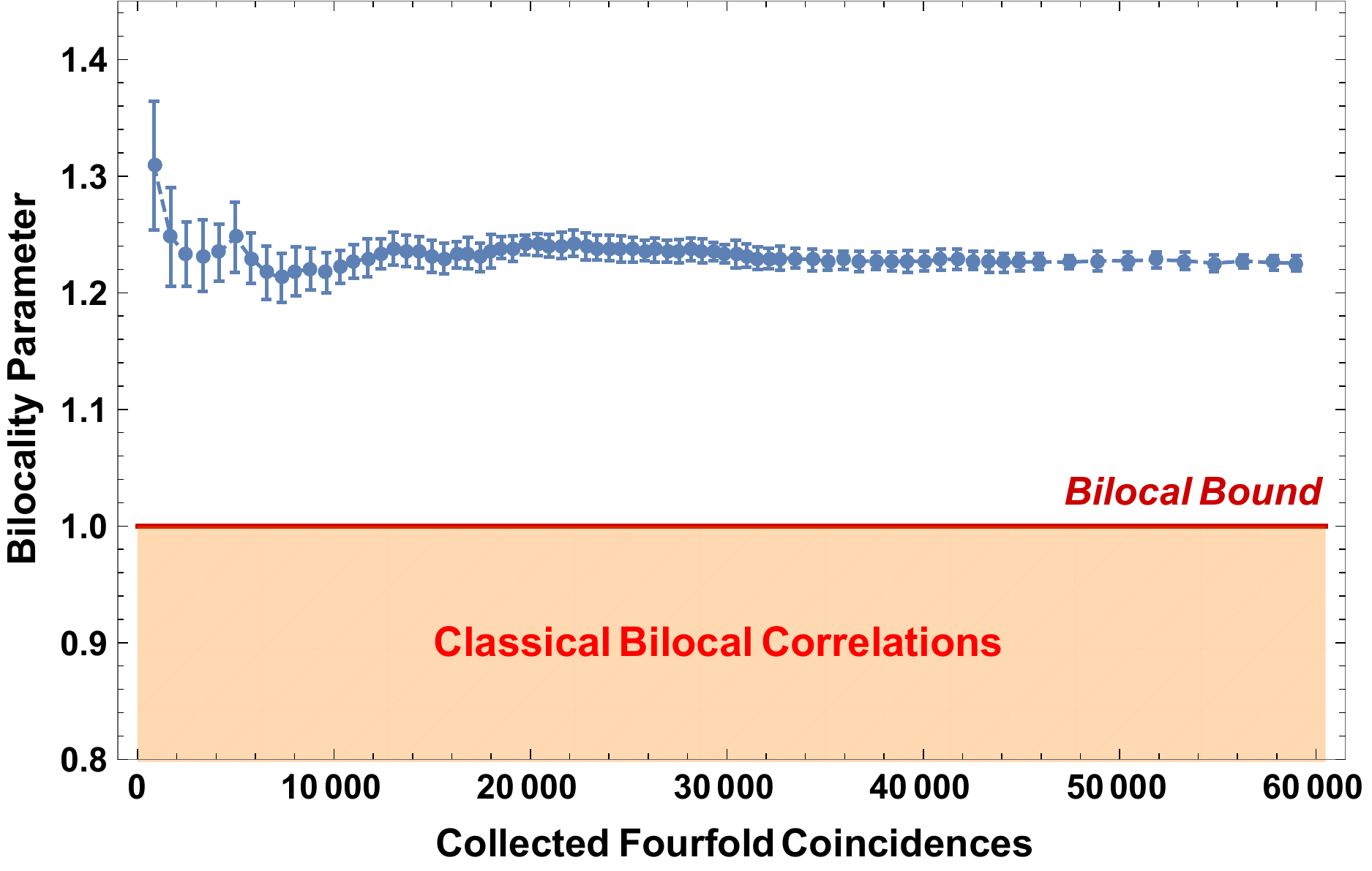}
\caption{\textbf{Experimental bilocality violation with human randomness.}  We cumulatively evaluated the bilocality parameter $\mathcal{B}$ and the total number of fourfold coincidences collected. A red line shows the bound of the bilocality inequality, which delimits the classical bilocal region depicted with a shaded red area.}
\label{fig:SetupROMA2}
\end{figure*}

The choice of measurement settings in our stations was physically implemented by the use of liquid crystals, driven in real time by the human random numbers provided by web users during the BBT day. We kept each setting for 5 s, recorded the coincidences in that time and then received fresh random bits for the next setting. This allowed us to witness a strong violation of the bilocality inequality of $\mathcal{B}=1.2251 \pm 0.0066$ by cumulatively evaluation of the $\mathcal{B}$ parameter for a growing number of collected coincidences (see \TextFig~\ref{fig:SetupROMA2}) observing a violation of almost $34$ standard deviations. {The obtained violation provides clear evidence that, up to the loopholes present in our implementation (low detection efficiencies and the locality loophole), the correlations generated in our experiment cannot be explained by any classical model based on the assumptions of local realism, free will of the participants and the independence of two physically different source of states. That is, the real-time (without post-processing of the data) violation of the bilocality inequality shows that bilocal classical models augmented with free-will are incompatible with our experimental data. This experiment provides a proof-of-principle for generalizations of Bell's theorem to more complex causal networks (involving more than one source of states) and thus can pave the way towards the implementation of new quantum information protocols exploiting non-bilocal quantum correlations as a novel resource.}

\clearpage
\renewcommand{\BBTcite}[1]{\citeLMU{#1}}
\graphicspath{{./PartnerContributions/LMUBBT/}}
\renewcommand{\NODE}{\LMU}
\BBTNodeTitle{Violation of Bell's inequality with a single atom and
single photon entangled over a distance of $400\,\mathrm{m}$}

\BBTNodeAuthorList{Kai Redeker, Robert Garthoff, Daniel Burchardt,
Harald Weinfurter, Wenjamin Rosenfeld}

\begin{figure*}[tbh]
\centering{}\includegraphics{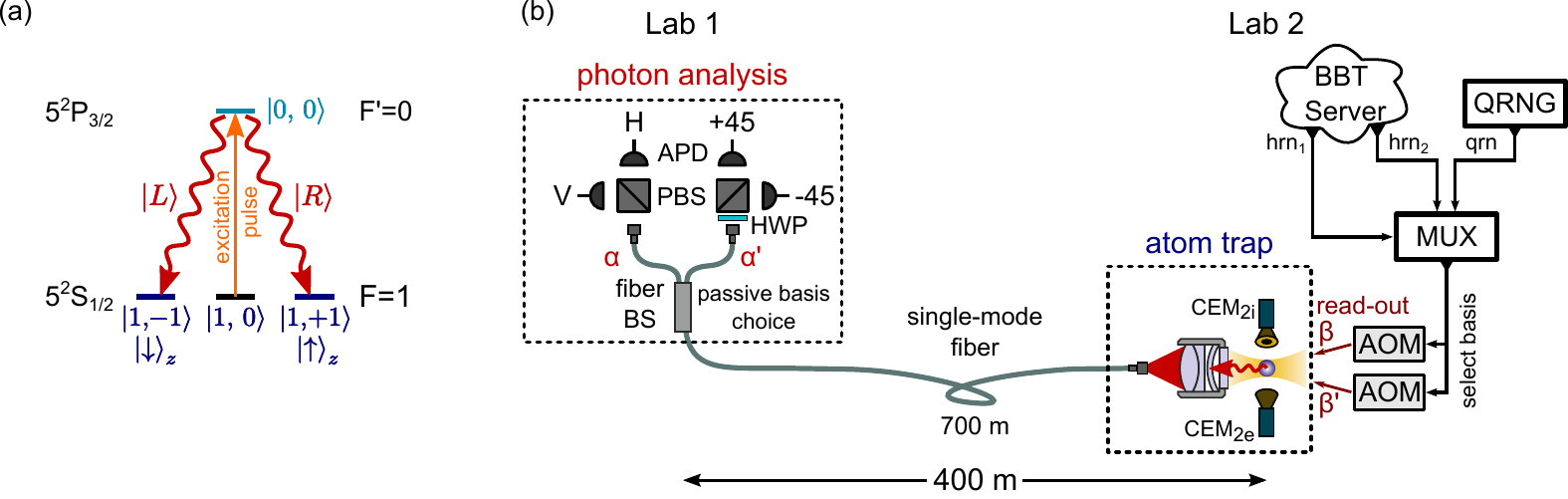} 
\caption{\label{fig:ExperimentalScheme}(a) Scheme of the atomic levels involved
in the entanglement between the spin state of the atom and polarization
of the photon. The atomic qubit consists of Zeeman spin states $\left|m_{F}=\pm1\right\rangle $
of $5^{2}S_{1/2}$, $F=1$ hyperfine ground level. (b) Overview of
the experiment. A single atom, trapped in Lab 2, is entangled with
a single photon. The photon undergoes a polarization analysis in Lab
1 where the measurement direction $a\in\{\alpha,\alpha'\}$ is passively
selected at a beam-splitter (fiber BS). The measurement direction
on the atomic spin $b\in\{\beta,\beta'\}$ is actively selected using
input from the BBT server and a local quantum random number generator
(QRNG).}
\end{figure*}

We demonstrate a violation of a CHSH Bell inequality with a spin of
a single trapped $^{87}\mathrm{Rb}$ atom which is entangled with
the polarization state of a single photon. The random numbers generated
by human contributors were used here to decide which measurement direction
had to be applied to the atomic spin. Our experiment employs 3 different
kinds of randomness: randomness in detecting a photon in a certain
output of an optical beam-splitter, a quantum random number generator
and human-generated random numbers.

In the experiment the generation of entanglement between the spin
of the atom and polarization of the photon \BBTcite{Volz2006} is
performed by optical excitation as shown in \TextFig~\ref{fig:ExperimentalScheme}(a).
The subsequent spontaneous decay leads to the atom-photon state $\left|\Psi\right\rangle _{AP}=\frac{1}{\sqrt{2}}\left(\left|\downarrow\right\rangle _{z}\left|L\right\rangle +\left|\uparrow\right\rangle _{z}\left|R\right\rangle \right)$
where $\left|\uparrow\right\rangle _{z}$ and $\left|\downarrow\right\rangle _{z}$
are the atomic spin states and $\left|L\right\rangle $, $\left|R\right\rangle $
denote the left- and right-circular polarization states of the photon
(corresponding to $\left|\downarrow\right\rangle _{z}$, $\left|\uparrow\right\rangle _{z}$
states of the photonic spin, respectively). The emitted photon is
coupled into a single-mode fiber and guided to another laboratory
which is located at a distance of $400\,\mathrm{m}$, \TextFig~\ref{fig:ExperimentalScheme}(b).
There the photon passes a polarization-independent $50:50$ beam-splitter
(BS) where the (passive) choice is made which basis will be used for
the photon measurement according to probabilistic rules of quantum
mechanics. In one output of the BS the measurement is performed in
the $H/V$ basis by means of a polarizing beam-splitter (PBS) and
two avalanche photo-diodes (APDs). In the other output of the BS the
measurement is performed in the $\pm45^{\circ}$ basis with the help
of an additional half-wave retarder plate (HWP) oriented at $22.5^{\circ}$.

After the state of the photon was measured the read-out of the atomic
state is performed. Here, for each measurement round, two random numbers
are retrieved from the server. The first one ($hrn_{1}$) decides
whether in the current round the measurement direction shall be selected
by a local quantum random number generator (QRNG) \BBTcite{Fuerst2010}
or a human contributor. For these means the QRNG output ($qrn$) and
the second random number ($hrn_{2}$) from the server are fed into
a multiplexer (MUX). Its output activates one of the two acousto-optical
modulators (AOMs) thereby determining the polarization of an optical
read-out pulse applied to the atom. This polarization defines the
basis for the analysis of the atomic spin (some details on the read-out
process can be found in \BBTcite{Rosenfeld2017}). The outcomes of
the atom and photon measurements are combined into the CHSH Bell parameter
$S=\left|\left\langle \sigma_{\alpha}\sigma_{\beta}\right\rangle -\left\langle \sigma_{\alpha}\sigma_{\beta'}\right\rangle \right|+\left|\left\langle \sigma_{\alpha'}\sigma_{\beta'}\right\rangle +\left\langle \sigma_{\alpha'}\sigma_{\beta'}\right\rangle \right|$
in terms of correlators $\left\langle \sigma_{a}\sigma_{b}\right\rangle =\frac{1}{N_{a,b}}(N_{a,b}^{\uparrow\uparrow}+N_{a,b}^{\downarrow\downarrow}-N_{a,b}^{\uparrow\downarrow}-N_{a,b}^{\downarrow\uparrow})$,
$N_{a,b}^{A,B}$ being the number of events with the respective outcomes
$A$, $B$ for measurement directions $a\in\{\alpha,\alpha'\}$, $b\in\{\beta,\beta'\}$.

During the experimental run of about $13$ hours $39614$ events were
collected, $19716$ where atomic measurement direction was chosen
by a human random number and $19898$ events where this choice was
done according to the local QRNG. The results are $S=2.427\pm0.0223$
for the first set and $S=2.413\pm0.0223$ for the second, respectively.
Both sets show a strong violation of Bell's inequality by more than
$18$ standard deviations. No statistically significant difference
in the results for the two sets can be observed. 

\begin{figure}
\begin{centering}
\includegraphics[width=0.6\columnwidth]{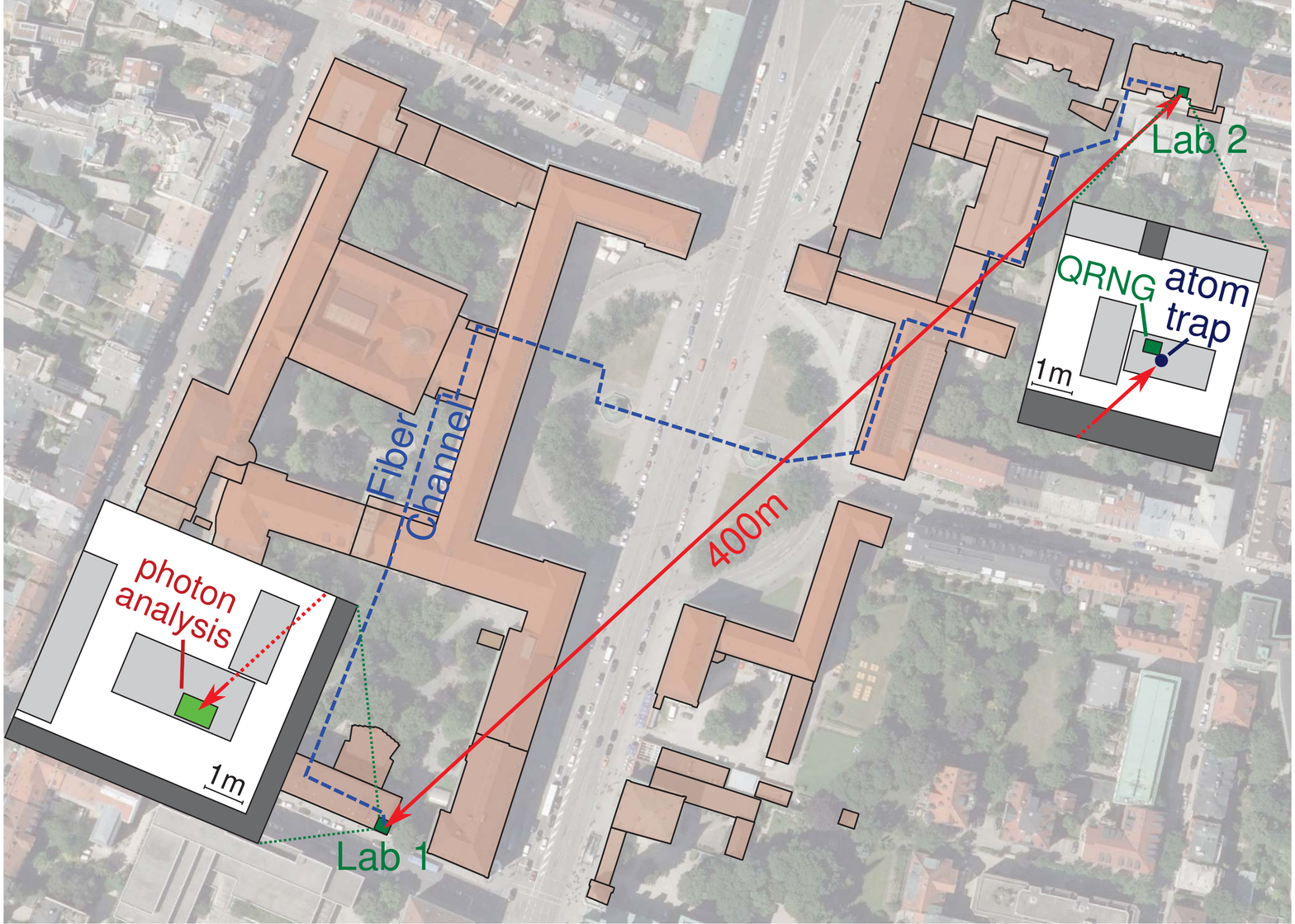}
\par\end{centering}
\caption[~]{Top view of the main campus of LMU. The photon analysis arrangement is located in the of the faculty of physics (Lab~1) while the atomic trap in the basement of the department of economics (Lab~2). Map data was provided by \BBTcite{vermessungsamt}
.}
\end{figure}



\clearpage
\renewcommand{\BBTcite}[1]{\citeETHZ{#1}}
\graphicspath{{./PartnerContributions/ETHZBBT/}}
\renewcommand{\NODE}{\ETHZ}



\newcommand{\newjh}[1]{{#1}}
\newcommand{\cphase}{\texttt{c}\textsc{-phase}}			



\BBTNodeTitle{Violation of a Bell inequality using superconducting qubits}

\BBTNodeAuthorList{Johannes Heinsoo, Philipp Kurpiers, Yves Salath{\'e}, Christian Kraglund Andersen, Adrian Beckert, Sebastian Krinner, Paul Magnard, Markus Oppliger, Theodore Walter, Simone Gasparinetti, Christopher Eichler, Andreas Wallraff}

\begin{figure*}[h]
\centering
	\includegraphics{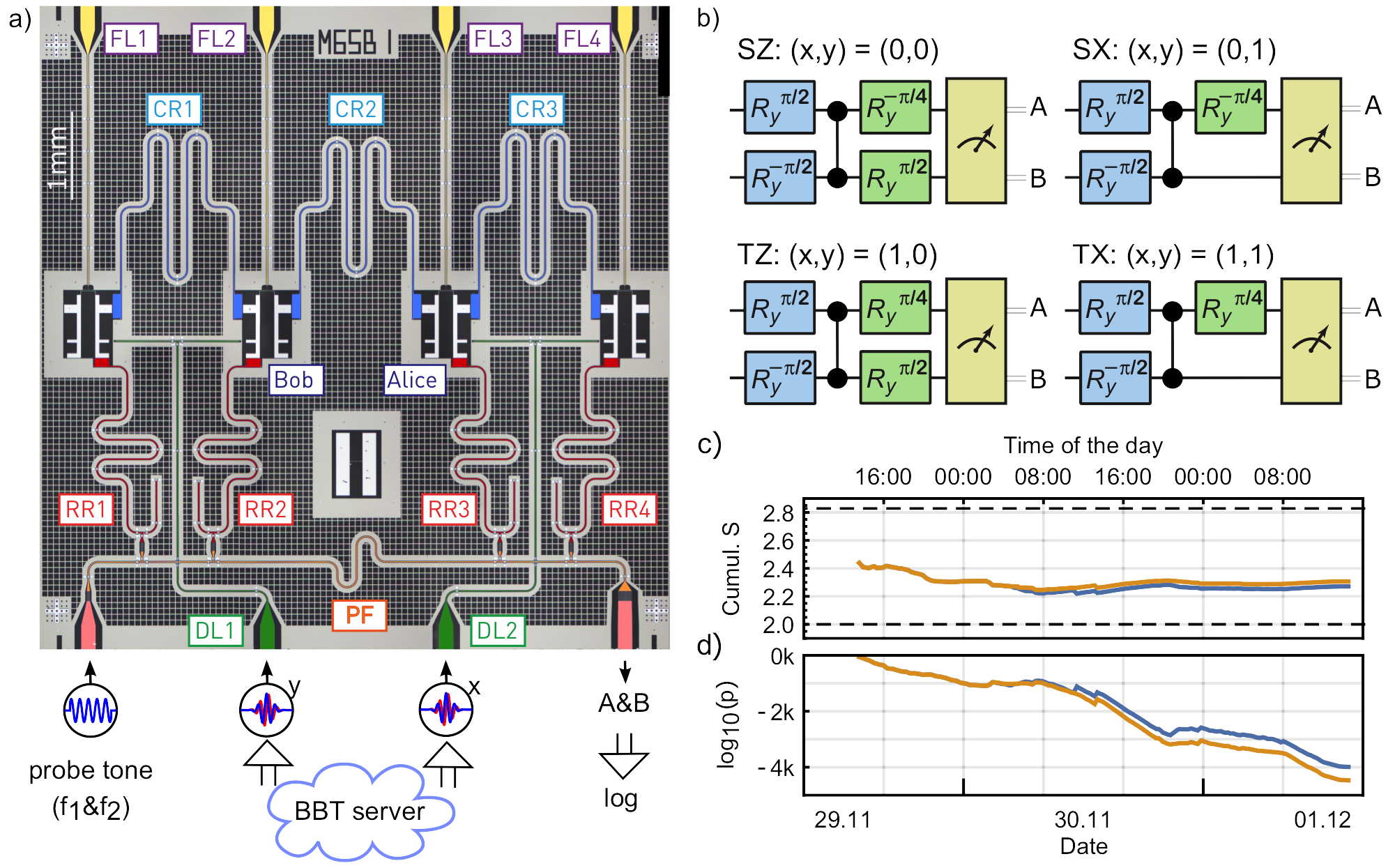}
\caption{(a) False coloured photograph of the superconducting chip used for the Bell test experiment. Each of the four qubits, of which the two central ones were used, has its own flux line (FL) and readout resonator (RR). Neighbouring  qubits are coupled through the coupling resonators (CR) with flux controllable effective interaction strength. To perform single qubit gates microwave pulses are applied via drive lines (DL). Probe tones at frequencies $f_1$ and $f_2$ are applied to the Purcell filter (PF) to measure the qubit states. (b) Depending on the bit string (x,y) received from the BBT server we \newjh{execute one of the four quantum gate sequences labeled $SZ$, $SX$, $TZ$ and $TX$. Single qubit gates $R_y^{\phi}$ rotate the state vector around the $y$-axes of the Bloch sphere by an angle $\phi$. The state preparation single qubit gates (blue) and the \cphase{} gate (connected black dots) remain unchanged,} while the single qubit gates (green) applied prior to the correlation sensitive readout (yellow) depend on the input bits from the server. (c) The cumulative $S$ value and (d) $p$ value as a function of time covering the entire period of the BBT, including all data (blue) and considering only data with successful calibration results (orange). The classical and the {non-signalling} thresholds are indicated as dashed horizontal lines.
}
\label{fig:SetupETHZ}
\end{figure*}

The Bell test in this work is performed using two superconducting qubits~\BBTcite{Bouchiat1998,Koch2007,Houck2008}, referred to as Alice and Bob, which are located about~\SI{1}{\milli\meter} apart from each other as shown in \TextFig~\ref{fig:SetupETHZ}a. The qubit transitions frequencies are set to $\omega_A / 2\pi = \SI{5.963}{\giga\hertz}$ and $\omega_B / 2\pi = \SI{5.464}{\giga\hertz}$, respectively. Single qubit gates are performed by applying \SI{580}{\mega\hertz} wide DRAG pulses resonant with the respective qubit transition frequencies~\BBTcite{Motzoi2009,Gambetta2011a}. The \cphase{} two qubit gate is realized by non-adiabatically tuning the state $\ket{11}$ into resonance with the $\ket{02}$ state for half the oscillation period~\BBTcite{Strauch2003}.

For each pair of bits received from the Big Bell test (BBT) server we perform one trial of the experiment by executing one of the four quantum circuits
shown in \TextFig~\ref{fig:SetupETHZ}b, each of which corresponds to a specific pulse sequence played from an arbitrary waveform generator (AWG). Each of the four sequences comprises the deterministic generation of an entangled state $\frac{1}{\sqrt{2}}\left(\ket{0+}-\ket{1-}\right)$ followed by a set of single qubit gates, which rotates the different measurement axes into the computational basis prior to readout. The readout of the qubit state is achieved by measuring the qubit state dependent transmission through the Purcell filter (see \TextFig~\ref{fig:SetupETHZ}a) recorded and processed with an analog to digital converter (ADC) and field programmable gate array (FPGA). A single shot fidelity of \SI{96.6}{\percent} and \SI{93.2}{\percent} for Alice and Bob respectively is achieved by employing a near-quantum limited \stext{travelling wave amplifier} \btext{parametric amplifier}~\BBTcite{Macklin2015}. Between two sequential runs of the experiment, a wait time of $\SI{50}{\micro\second} \approx 5 T_1$ allows the qubits to decay back to their ground state. The successful initialization of the qubits using this passive method is verified by reading out the qubit state prior to each experimental run. The \stext{datasets} \btext{data sets} with failed state preparation were discarded. About every $\SI{20}{\minute}$ both the qubit transition frequencies and two qubit gate parameters were recalibrated. Moreover, Bell state tomography was used to benchmark each round of calibration.





During the 48 hours of \stext{continiusly} \btext{continuously} running experiments, \num{16.34} million human generated random numbers were used to perform \num{8.17} million individual Bell measurements of which \num{7.69} had succeful state initialization and calibration. By counting the different measured two qubit states for different basis choices we directly evaluate the CHSH inequality~\BBTcite{CHSH1969}. The resulting S-value is shown in \TextFig~\ref{fig:SetupETHZ}c and converges to a final value of \num{2.271(1)} in case all data is included and to \num{2.307(1)} if \stext{datasets} \btext{data sets} with failed calibration are ignored. The observed S value is mostly limited by qubit readout fidelity and qubit decay during the $\SI{300}{\nano\second}$ long pulse sequence.

The confidence of which the experiment violates the CHSH inequality is quantified by estimating the $p$-values for the data under the hypothesis of \newjh{a local hidden variable (LHV) model~\BBTcite{Larsson2014}. The estimation based on the Bentkus' inequality~\BBTcite{bentkus2004} is shown to be tight and accounts for all possible memory effects and input bias~\BBTcite{Elkouss2016}}. Note that the latter is crucial for reasonable $p$-value estimation as BBT participants tended to input $(1,0)$ and $(0,1)$ pairs twice more often than $(1,1)$. We observed an exponentially decreasing $p$-value with increasing number of trials as shown in \TextFig~\ref{fig:SetupETHZ}d.
The exponentially \stext{descreasing} \btext{decreasing} p-value is expected as long as the performance of the experiment is not degraded over time, which is ensured by our automated recalibrations. The final $p$-value of about $\sim10^{-4000}$ indicates that with high confidence locality is violated either by the incompleteness of a classical description of reality under the additional assumption of free will of the participants or by a locality loophole in our specific experimental setup.

\clearpage
\renewcommand{\BBTcite}[1]{\citeNICE{#1}}
\renewcommand{\NODE}{\NICE}
\graphicspath{{./PartnerContributions/NICEBBT/}}


\BBTNodeTitle{Telecom-compliant source of polarisation entangled photons for the violation of Bell inequalities driven by human-generated random numbers}

\BBTNodeAuthorList{Florian Kaiser, Tommaso Lunghi, Gr\'egory Sauder, Panagiotis Vergyris, Olivier Alibart, and S\'ebastien Tanzilli}

\begin{figure*}[h]
\centering
	\includegraphics[width=0.98\textwidth]{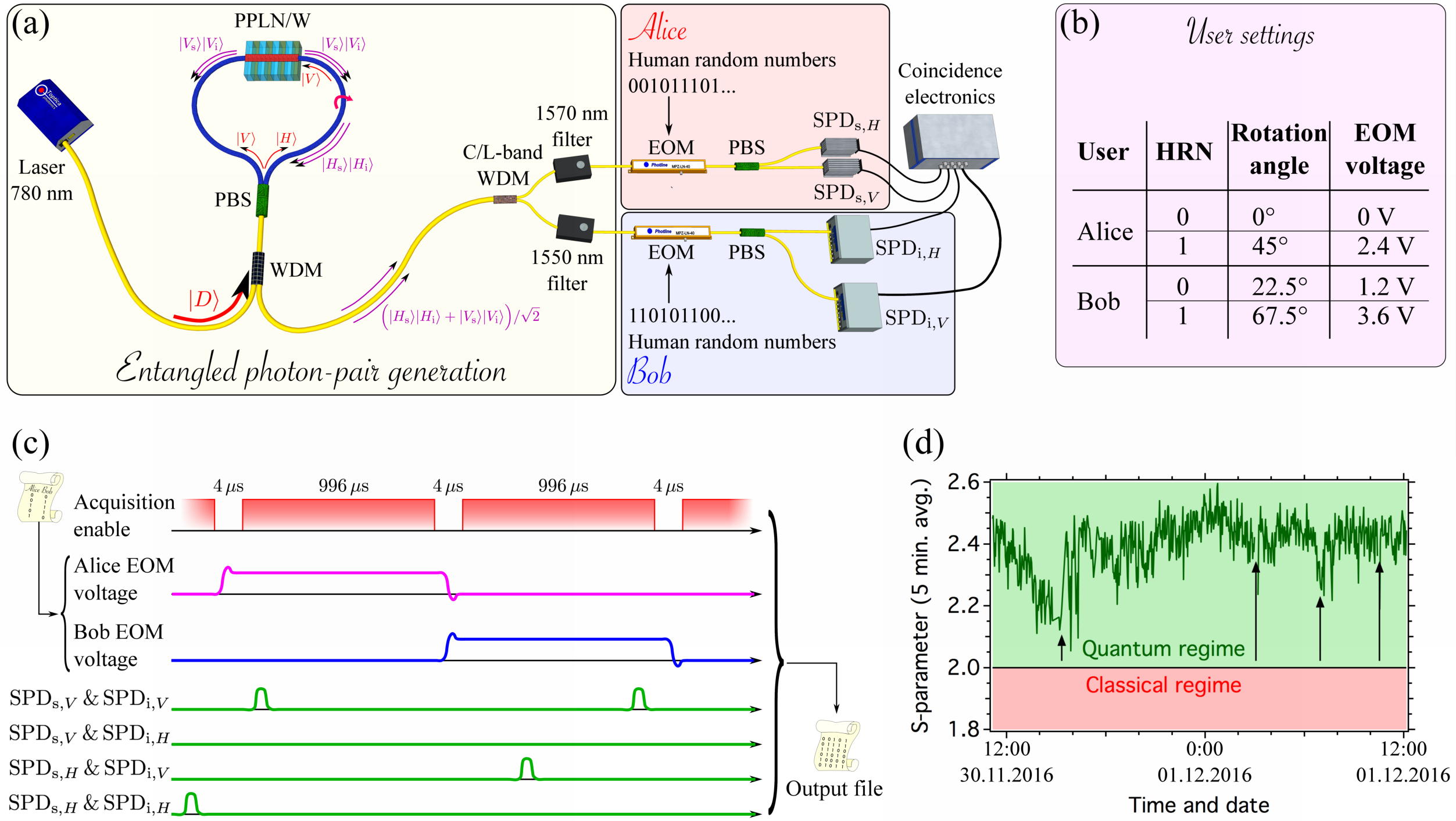}
\caption{(a) Experimental setup. A 780\,nm continuous-wave laser pumps a PPLN/W placed in a Sagnac interferometer. This way, polarisation entangled photon pairs are generated in the state $|\psi \rangle = \Big( |H_{s} \rangle |H_{i} \rangle+ |V_{s} \rangle |V_{i} \rangle \Big) / \sqrt{2}$. After being deterministically separated using a C/L-band WDM, the photons are directed to Alice and Bob. Both rotate, accordingly to the HRN, the polarisation state of their respective photon with an EOM. Finally, the polarisation state is analysed using a PBS, followed by two SPDs. (b) User settings that are applied at Alice's and Bob's stations accordingly to the HRN inputs, provided by the Bellsters. (c) Schematic sketch of the data acquisition strategy. (d) Development of the 5-minute averaged $S$-parameter as a function of time. The four vertical arrows indicate the times at which unexpected incidents occurred to the experiment.}
\label{SetupNICE}
\end{figure*}

We violate the Bell inequalities by more than 143 standard deviations using polarisation entangled photon pairs~\BBTcite{Bell_EPR_1964,CHSH_1969,The_Quantum_Challenge_Book,Nielsen_book_2011}.
The related experimental setup is shown in \TextFig~\ref{SetupNICE}(a).
We developed a photon pair source which is entirely based on guided-wave photonics in order to guarantee the highest possible stability~\BBTcite{VergyrisQST2017}.

Diagonally polarised light, $|D \rangle$, from a fibre-coupled continuous-wave 780\,nm pump laser passes through a wavelength division multiplexer (WDM), which is the fibre optics equivalent of a bulk dichroic mirror. The subsequent fibre polarising beam-splitter (PBS) defines the input and output of a fibre Sagnac loop~\BBTcite{Lim_Sagnac_2010,VergyrisQST2017}.
Inside the loop, vertically (horizontally) polarised pump light propagates clockwise (anti-clockwise) in polarisation maintaining fibres (indicated by blue lines). By rotating the right-hand side fibre by $90^{\circ}$, the horizontally polarised pump light is effectively turned to be vertical.
The pump light is injected in a periodically poled lithium niobate waveguide (PPLN/W) in which vertically polarised pump photons are converted to vertically polarised signal and idler photon pair contributions through type-0 spontaneous parametric downconversion, \textit{i.e.} $|V \rangle \rightarrow |V_s \rangle |V_i \rangle$. Here, the subscripts $s$ and $i$ denote signal and idler photons at wavelengths of $\lambda_{\rm s}=1570\,\rm nm$ and $\lambda_{\rm i}=1550\,\rm nm$, respectively.
The photon pair contribution obtained in the clockwise sense is rotated in the polarisation maintaining fibre, and therefore becomes $|H_s \rangle |H_i \rangle$.
Both contributions, \textit{i.e.} $|V_s \rangle |V_i \rangle$ and $|H_s \rangle |H_i \rangle$, are then coherently combined at the PBS and separated from the pump light using the WDM.
It is important to note that we choose the pump laser power to be low enough such that the source generates only one photon pair contribution at a time, \textit{i.e.} the probability for generating both contributions $|V_s \rangle |V_i \rangle$ and $|H_s \rangle |H_i \rangle$ simultaneously is negligible.
Therefore, the following maximally polarisation entangled photon pair state is generated: $
|\psi \rangle = \Big( |H_{s} \rangle |H_{i} \rangle+ |V_{s} \rangle |V_{i} \rangle \Big) / \sqrt{2}.$
Signal and idler photons are deterministically separated using a standard telecom C/L-band WDM, filtered down to $\sim$1\,nm bandwidth and further directed to Alice's and Bob's stations, which are separated by 35\,m	of optical fibre, corresponding to a physical separation of 5\,m.
For entanglement analysis, both users are equipped with fibre electro-optic phase modulators (EOM), allowing them to rotate the polarisation state of their respective photon on demand. {For entanglement analysis, both users are equipped with fibre electro-optic phase modulators (EOM), allowing them to rotate the polarisation state of their respective photon on demand. EOM phases are set \SI{4}{\micro\second} prior to the detection window. As such, there is not space-like separation of the measurement setting from the remote detection. Nevertheless, the \SI{800}{m} fibre length employed would allow this kind of space-like separation if the fibre were linearly deployed.}
\TextFig~\ref{SetupNICE}(b) shows the operations that Alice and Bob apply depending on the human random numbers (HRN) that are generated by the Bellsters and provided by the central server at the ICFO~\BBTcite{TBBT_website}. The settings we use correspond to the standard ones for violating the Bell inequality using polarisation entangled photons~\BBTcite{Bell_EPR_1964,CHSH_1969,The_Quantum_Challenge_Book,Nielsen_book_2011}.
Finally, each user analyses the polarisation state of their photon using a PBS, followed by two Indium Gallium Arsenide avalanche single photon detectors (SPD).
A coincidence electronic circuit is used to account only for events in which both Alice and Bob receive a photon from the same pair.

A schematic of our data acquisition scheme is shown in \TextFig~\ref{SetupNICE}(c).
Every two seconds, a table of $2 \times 1000$ random bits (Alice and Bob) is downloaded from the ICFO server, and the following sequence is repeated 1000 times:
\begin{enumerate}
\item A $4\rm\,\mu s$ time window is reserved in order to adjust Alice's and Bob's EOM settings accordingly to the HRNs. During this time, no coincidences are recorded. The following information is written into the raw data output file: a time stamp at the end of the $4\rm\,\mu s$ window, and Alice's and Bob's EOM settings.
\item Two-photon coincidences are recorded for $996\rm\,\mu s$. Time stamps and information about the corresponding pairs of detectors are stored in the raw data output file.
\item Thereafter, the experiment is halted for one second to give the computer time to analyse the data and store them on the hard-disk drive.
\end{enumerate}
We note that during the second step in the above-mentioned procedure it might occur that no or multiple coincidence events are recorded.
This is why we construct a new data file from the raw data file.
Here, no-coincidence events are deleted, and for very rarely occurring multi-coincidence cases only the event belonging to the first time stamp is kept. Single coincidence events are not modified and stored as is.

From the new data file, the $S$-parameter is constructed in real-time following the standard approach~\BBTcite{Bell_EPR_1964,CHSH_1969,The_Quantum_Challenge_Book,Nielsen_book_2011}.
In \TextFig~\ref{SetupNICE}(d), we show the 25-hour time line of the measured $S$-parameter with a 5-minute running average. Within the first four hours, we observe a drop due to a strong temperature change in our laboratory which caused a slight misalignment of the PPLN/W.
At $\sim$3:20\,pm, the photon pair coupling from the PPLN/W to PMFs was re-optimised, and the resulting $S$-parameter increased. At $\sim$3:00\,am the experiment was halted for about 10 minutes due to an electric power failure. A second realignment procedure was required at $\sim$7:00\,am due to a fast temperature change in the laboratory induced by the building air conditioning. Finally, the experiment was halted a second time at $\sim$10:30\,am due to a SPD overheating problem.

Note that none of the above-mentioned issues is actually related to the optical design of our fully guided-wave photon pair source, but exclusively due to infrastructural influences that could not be controlled.
Nevertheless, the $S$-parameter is maintained always above 2 which exceeds the classical-quantum boundary, therefore proving the quantumness of our results.

The $S$-parameter, averaged over the full experiment, which started on 30.11.2016 at 11:09\,am and ended on 01.12.2016 at 12:10\,pm is calculated to be $S_{\rm full} = 2.431$ with a standard error of $\sigma_{\rm full} = \pm 0.003$, corresponding to a  violation of the Bell inequalities by more than 143 standard deviations.
Throughout the course of the experiment, the Bellsters provided us with $2 \times 19.5 \cdot 10^6$ HRNs, with which we performed $2.9 \cdot 10^6$ successful measurements (coincidences).



\clearpage
\renewcommand{\BBTcite}[1]{\citeICFOONE{#1}}
\graphicspath{{./PartnerContributions/ICFO1BBT/}}
\renewcommand{\NODE}{\ICFOONE}

\BBTNodeTitle{Bell test using entanglement between a photon and a collective atomic excitation, driven by human randomness }

\BBTNodeAuthorList{Pau Farrera, Georg Heinze, Hugues de Riedmatten}

\begin{figure}[h]
\includegraphics[width=.85\textwidth]{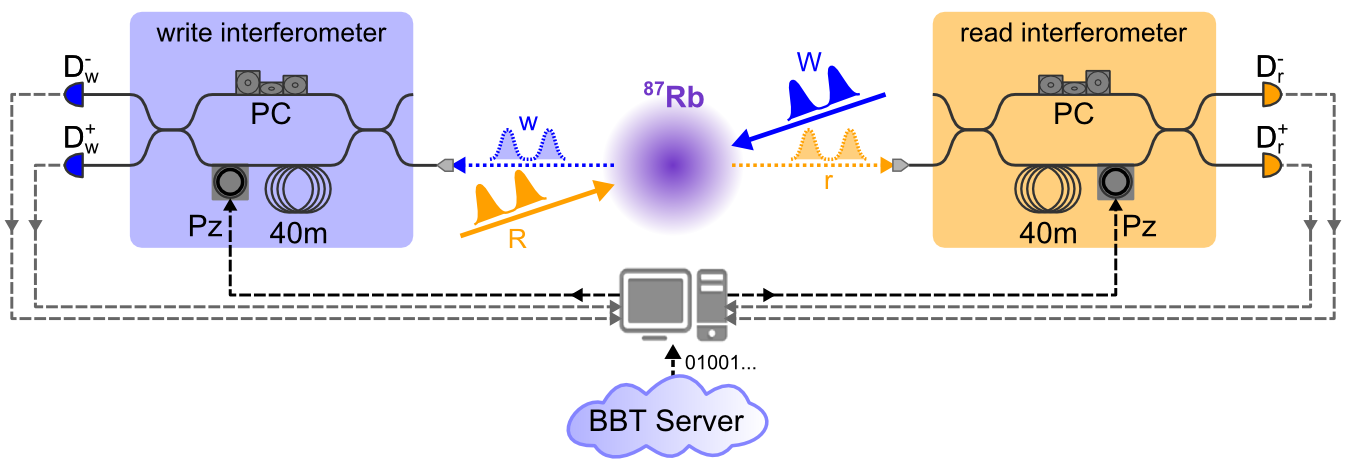}
\caption{Schematic drawing of the experiment. Entangled atom-photon time-bin qubits are generated in a cold $^{87}\mathrm{Rb}$ ensemble ($T\approx 100\mathrm{\mu K}$) and analyzed via two imbalanced Mach-Zehnder interferometers. Abbreviations: W write pulse; w write photon; R read pulse; r read photon; PC polarization controller; Pz piezo-elecric fiber stretcher; 
$D^{+(-)}_{w(r)}$ single photon detector.}
\label{Figure1ICFO1}
\end{figure}

Our experiment within the Big Bell Test involves the generation and analysis of entanglement between a photon and a collective atomic excitation in a cloud of laser-cooled $^{87}\mathrm{Rubidium}$ atoms \BBTcite{Matsukevich2005}. Herefore, we generate photon-atom entangled qubits, which are encoded in the time-bin degree of freedom, as proposed in \BBTcite{FarreraPRL2018}.

As shown in \TextFig~\ref{Figure1ICFO1}, we follow the DLCZ (Duan-Lukin-Cirac-Zoller) protocol \BBTcite{Duan2001} sending a double-peaked write laser pulse (W) to probabilistically generate a write photon (w) which is paired with an excitation in the atomic cloud. The write pulse interacts with many atoms, leading to an excitation that is delocalized over all the atoms, i.e., a collective atomic spin excitation (``spin-wave''). Since this process can happen in any of the two time-bins, the photon and the atomic excitation are entangled in the time-bin degree of freedom. The entangled state can thus be written as $\left|\Psi_{w,a}\right\rangle=\frac{1}{\sqrt{2}}\left(\left|E_wE_a\right\rangle+e^{i\varphi}\left|L_wL_a\right\rangle\right)$, where $E_{w(a)}$ denotes a write photon (atomic excitation) generated in the early time-bin, $L_{w(a)}$ denotes a write photon (atomic excitation) generated in the late time-bin, and the phase $\varphi$ can be controlled by changing the phase between the early and late time-bins of the write and read pulses. In order to temporally distinguish the atomic excitations created in the two different time-bins, we apply an homogeneous magnetic field to induce collective de- and re-phasings of the stored spin-waves \BBTcite{Albrecht2015}. Subsequently, the atomic qubit can be converted into a photonic time-bin qubit (r) using a resonant double-peaked read laser pulse (R). This converts the photon-atom entangled state into a photon-photon entangled state, which is more suited for the entanglement analysis. The analysis is done with two imbalanced Mach-Zehnder interferometers and single photon detectors, which allow us to perform qubit projective measurements in any basis that lies on the equator of the Bloch sphere \BBTcite{Marcikic2004}. The bases in which the write and read qubits are projected depend on the phase delays $\phi_w$ and $\phi_r$ between the two arms of each interferometer. The quantum correlations of the two photons is then assessed by the CHSH Bell parameter \BBTcite{Clauser1969}
\begin{equation}
S=\left|E(\phi_w,\phi_r)+E(\phi_w,\phi'_r)+E(\phi'_w,\phi_r)-E(\phi'_w,\phi'_r)\right|
\end{equation}
where $E(\phi_w,\phi_r)=\left[p_{++}(\phi_w,\phi_r)-p_{+-}(\phi_w,\phi_r)-p_{-+}(\phi_w,\phi_r)+p_{--}(\phi_w,\phi_r)\right]/p(\phi_w,\phi_r)$ are the correlation coefficients, $p_{ij}(\phi_w,\phi_r)$ are the probabilities to detect coincidences between the write and read photons at detectors $D_w^i$ and $D_r^j$, and $p(\phi_w,\phi_r)=\sum_{i,j=+,-}p_{ij}(\phi_w,\phi_r)$.

In order to change the phase of the interferometers to the desired values of $\phi_w$, $\phi'_w$, $\phi_r$, and $\phi'_r$, a short part of the fiber of the $40\,\mathrm{m}$ long interferometer arms is rolled around a piezo-electric ceramic cylinder. Applying a voltage $U_w$ or $U_r$ to the corresponding piezo cylinder stretches the fiber and therefore changes the phase delay between the two arms of the write or the read interferometer. The random numbers generated by the participants of the Big Bell Test were used to decide which voltage is applied to each interferometer, hence controlling the measurement bases.
Trapped atom clouds were created at a rate of 59 Hz. For each cloud, we performed 608 entanglement trials. The bases were changed  in between each cloud and not between each single trial because of the limited bandwidth of the piezo fiber stretcher. However, the typical write-read photon coincidence detection probability  per trial $p(\phi_w,\phi_r)$ was between $10^{-6}$ and $10^{-5}$. We can therefore say that for each detected coincidence event, the bases were chosen randomly.

\begin{figure}[h]
\includegraphics[width=.45\textwidth]{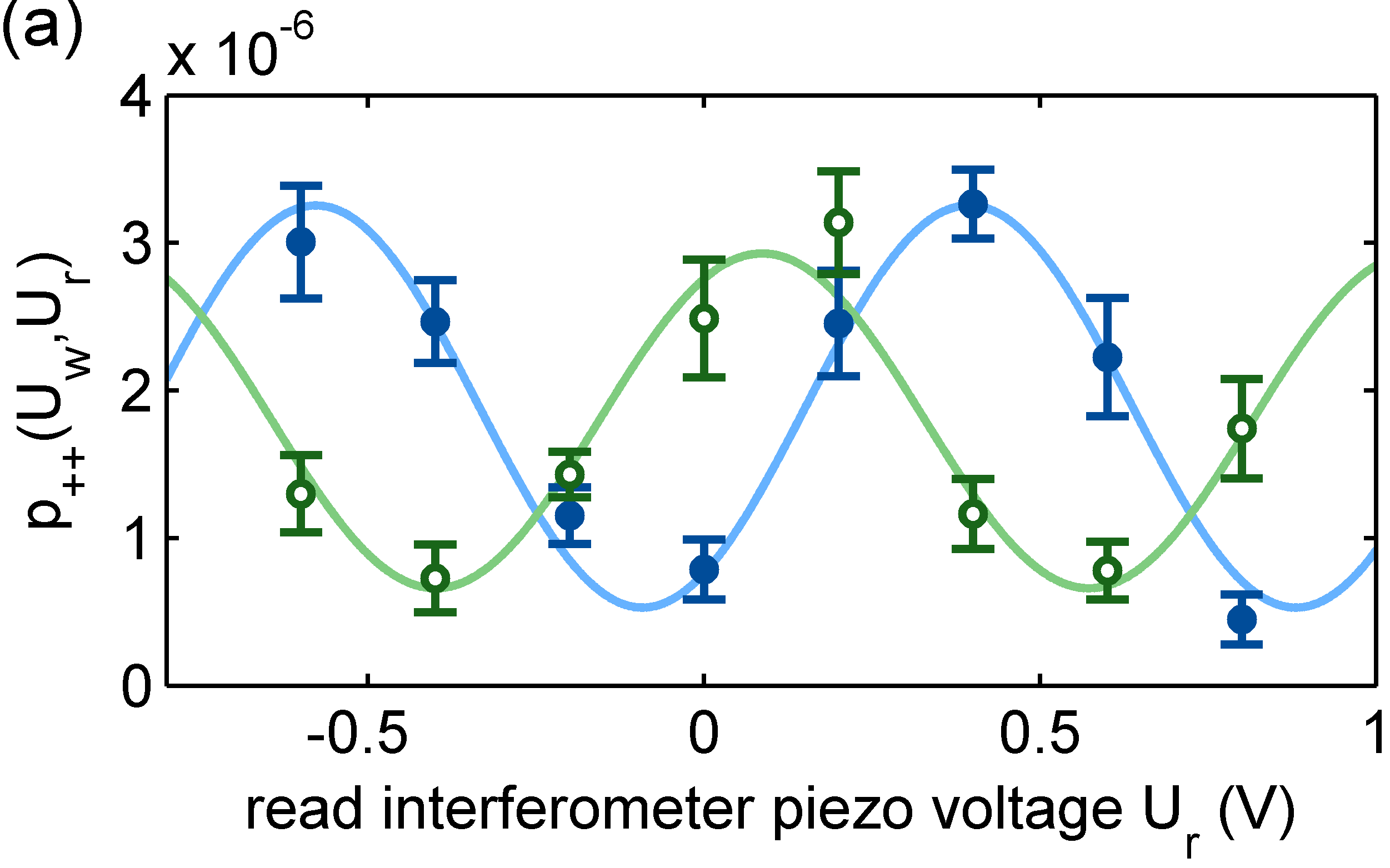}
\hspace{1cm}
\includegraphics[width=.45\textwidth]{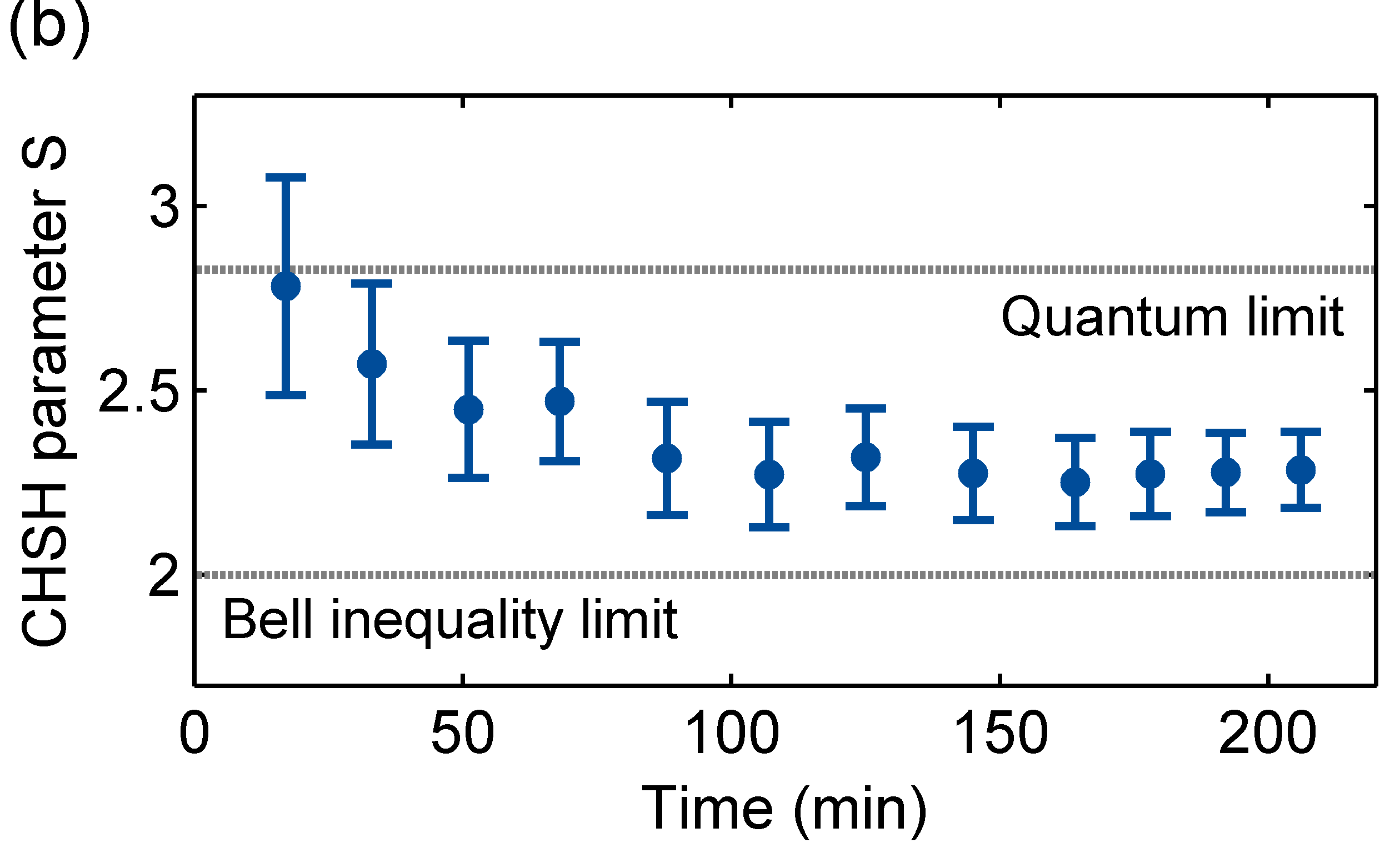}
\caption{(a) Photon coincidence detection probability between detectors $D_w^+$ and $D_r^+$ as a function of the voltage applied to the piezo of the read photon interferometer. The two different curves correspond to two different voltages applied to the piezo of the write photon interferometer: $U_w=0\,\mathrm{V}$ (blue dots) and $U_w=0.296\,\mathrm{V}$ (green circles).
(b) Accumulated CHSH Bell parameter $S$ as a function of data acquisition time for the measurement with stored human random numbers.}
\label{Figure2ICFO1}
\end{figure}

During the time window of the Big Bell Test we used the live human random numbers to randomly choose between predefined phases in the two analysing interferometers to take the data shown in \TextFig~\ref{Figure2ICFO1}(a). Here, $p_{++}$ is plotted as a function of $U_r\propto\phi_r$ for two different values of $U_w\propto\phi_w$. The two fringes are shifted by $114(9)$ degrees exhibiting visibilities of $V_1=0.72(0.08)$ (blue dots) and $V_2=0.63(0.10)$ (green circles). These visibilities are sufficient to prove entanglement between both time-bin qubits, as follows from the Peres separability criterion ($V>1/3$) under the assumption of equally distributed noise for all possible outcomes \BBTcite{Peres1996}. However, the values didn't surpass the threshold of $V>1/\sqrt{2}\approx0.707$ to guarantee Bell-type non-local correlations. We obtained similar visibilities without the use of human random numbers, which confirmed that the measurements were affected by experimental instabilities during the required relatively long integration times of several hours.

After improving the long-term stability of the experiment, we redid a Bell test at a later stage with human random numbers received and stored during the day of the Big Bell Test. Here, the human random numbers were used to switch randomly between the four settings required for the measurement of the CHSH inequality. The experimental results of that Bell Test are shown in \TextFig~\ref{Figure2ICFO1}(b). The data acquisition lasted 3 hours and 26 minutes, during which we performed $364800000$ experimental trials. In this time we recorded 1100 photon coincidence events which led to a final CHSH Bell parameter of $S=2.29\pm0.10$. This measurement shows a violation of the Bell inequality $|S|\leq2$ by approximately three standard deviations.

\clearpage
\renewcommand{\BBTcite}[1]{\citeICFOTWO{#1}}
\graphicspath{{./PartnerContributions/ICFO2BBT/}}
\renewcommand{\NODE}{\ICFOTWO}

\BBTNodeTitle{Violation of a Bell inequality using high-dimensional frequency-bin entangled photons }

\BBTNodeAuthorList{Andreas Lenhard, Alessandro Seri, Daniel Riel\"{a}nder, Osvaldo Jimenez, Alejandro M\'{a}ttar, Daniel Cavalcanti, Margherita Mazzera, Antonio Ac\'{i}n, and Hugues de Riedmatten}

\begin{figure*}[tbh]
\centering
	\includegraphics[width=0.75\textwidth]{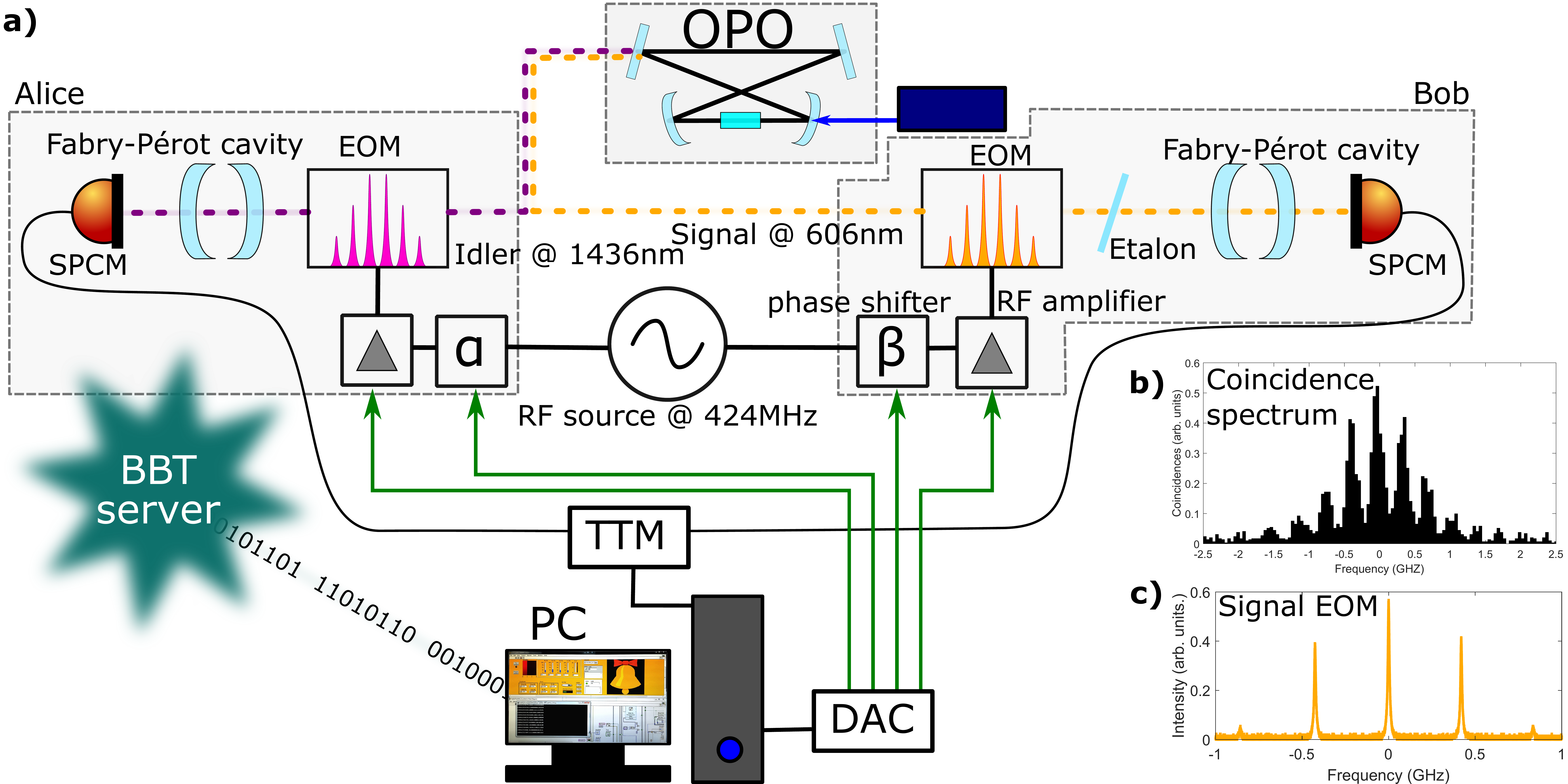}
\caption{a) Experimental setup. Photon pairs are generated via cavity enhanced SPDC in multiple frequency modes. The pairs are split and the Idler photons sent to Alice and Signal photons to Bob. At both ends there are EOMs, controlled by a RF phase shifter and an amplifier while the RF source is shared. Before detecting coincidences, single frequency bins are selected by cavity filters. The basis settings are chosen by HRN and a computer generates analog voltages (DAC) controlling the RF phases and powers accordingly. b) Spectrum of the photons, illustrating frequency-bins. c) Modulation spectrum of Signal EOM.}
\label{fig:SetupICFOTWO}
\end{figure*}

\btext{
\begin{figure}[htb]
\centering
\includegraphics[width=0.70\textwidth]{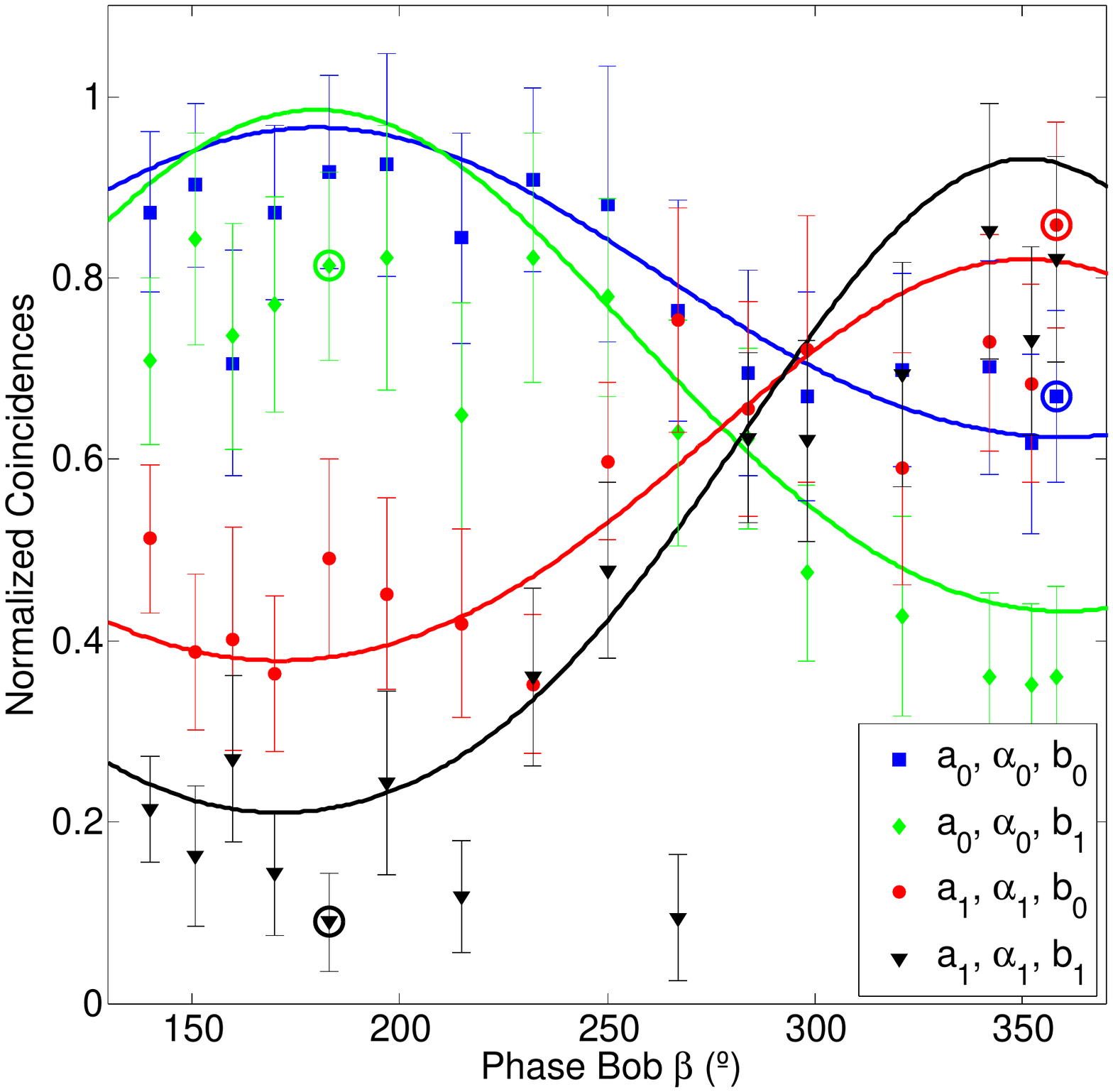}
\caption{Two photon interference fringes obtained by using HRN to choose the setting of modulation amplitude and phase shift for Alice and Bob's EOMs. While the phase of Bob's EOM was chosen among 16 possible values (corresponding to the x-axis), the other values were randomly selected between two possible values: $a_0 = 0.19$, $a_1 = 0.57$, $\alpha_0 = 0\,^\circ$, $\alpha_1 = 180\, ^\circ$, $b_0 = 0.63$, $b_1 =1.06$.}
\label{fig:ICFO2Fringes}
\end{figure}
}

We report a Bell inequality violation with high-dimensional frequency-bin entangled photons. Due to energy conservation, photon pairs generated by spontaneous parametric down-conversion (SPDC) are naturally correlated in frequency, such that the sum of the signal and idler frequencies is always equal to the frequency of the pump beam. One can then define frequency-bins in the spectrum and find correlations between different spectral bins of the two photons, leading to so-called frequency bin entanglement \BBTcite{Olislager2012}. In our experiment, we generate narrowband photon pairs in discrete frequency modes by SPDC in an optical parametric oscillator (OPO) operated below threshold \BBTcite{Rielander2016}. One photon is at visible wavelength (606 nm) while the other is at the telecommunication wavelength (1436 nm). The photon spectrum consists of around 8 equally spaced modes {of different amplitudes}, separated by the free spectral range of the cavity (FSR=424~MHz, see \TextFig~\ref{fig:SetupICFOTWO}b).
The two photons are separated deterministically and sent to Alice and Bob for analysis. The measurement process uses electro-optic phase modulators (EOMs) driven at a radio frequency (RF) corresponding to the FSR of the OPO (see \TextFig~\ref{fig:SetupICFOTWO}c) to mix neighboring modes and introduce a phase shift. Each measurement setting for Alice and Bob corresponds to a given modulation depth ($a$ and $b$, set by the RF power) and to a given phase ($\alpha$ and $\beta$, set by the RF phase). We select single spectral modes of the signal and idler photons via filter cavities and detect the photons with single photon counting modules (SPCM). The full setup is illustrated in \TextFig~\ref{fig:SetupICFOTWO}.

We observe two-photon interference fringes depending on the phase difference between signal and idler EOMs. We develop a theoretical model for our setup from which we estimate the optimal experimental configurations leading to a violation of the CH-inequality\BBTcite{Clauser1974} $S_{CH}\leq2$.  
During the BBT day we used the human random numbers (HRN) to choose in real time one out of four settings for ($a_i\alpha_ib_j$; $i,j=0,1$) while the second phase setting ($\beta_k$; $k=0..15$) was chosen out of 16 possible phases. This allowed us to record {the full interference fringes shown in \TextFig~\ref{fig:ICFO2Fringes}.} We kept each setting for 10~s, recorded the coincidences in that time and then received fresh HRN for the next setting. {Selecting the data points circled in the fringes of \TextFig~\ref{fig:ICFO2Fringes},} our experiment reaches $S=2.25(8)$, corresponding to a violation of the Bell inequality by three standard deviations. After the BBT, new measurements with stored HRN have been made, considering only the four best settings leading to the maximal violation. In this way data can be collected much faster, resulting in a violation of 8.5 standard deviations \BBTcite{RielanderQST2018}.



\clearpage
\renewcommand{\BBTcite}[1]{\citeBA{#1}}
\graphicspath{{./PartnerContributions/BABBT/}}
\renewcommand{\NODE}{\UBA}

\BBTNodeTitle{Violation of the CHSH inequality using polarization-entangled photons}

\BBTNodeAuthorList{Laura T. Knoll, Ignacio H. L\'opez Grande, Agustina G. Magnoni, Christian T. Schmiegelow, Ariel Bendersky, and Miguel A. Larotonda}

\begin{figure*}[tbh]
\centering
	\includegraphics[width=0.85\textwidth]{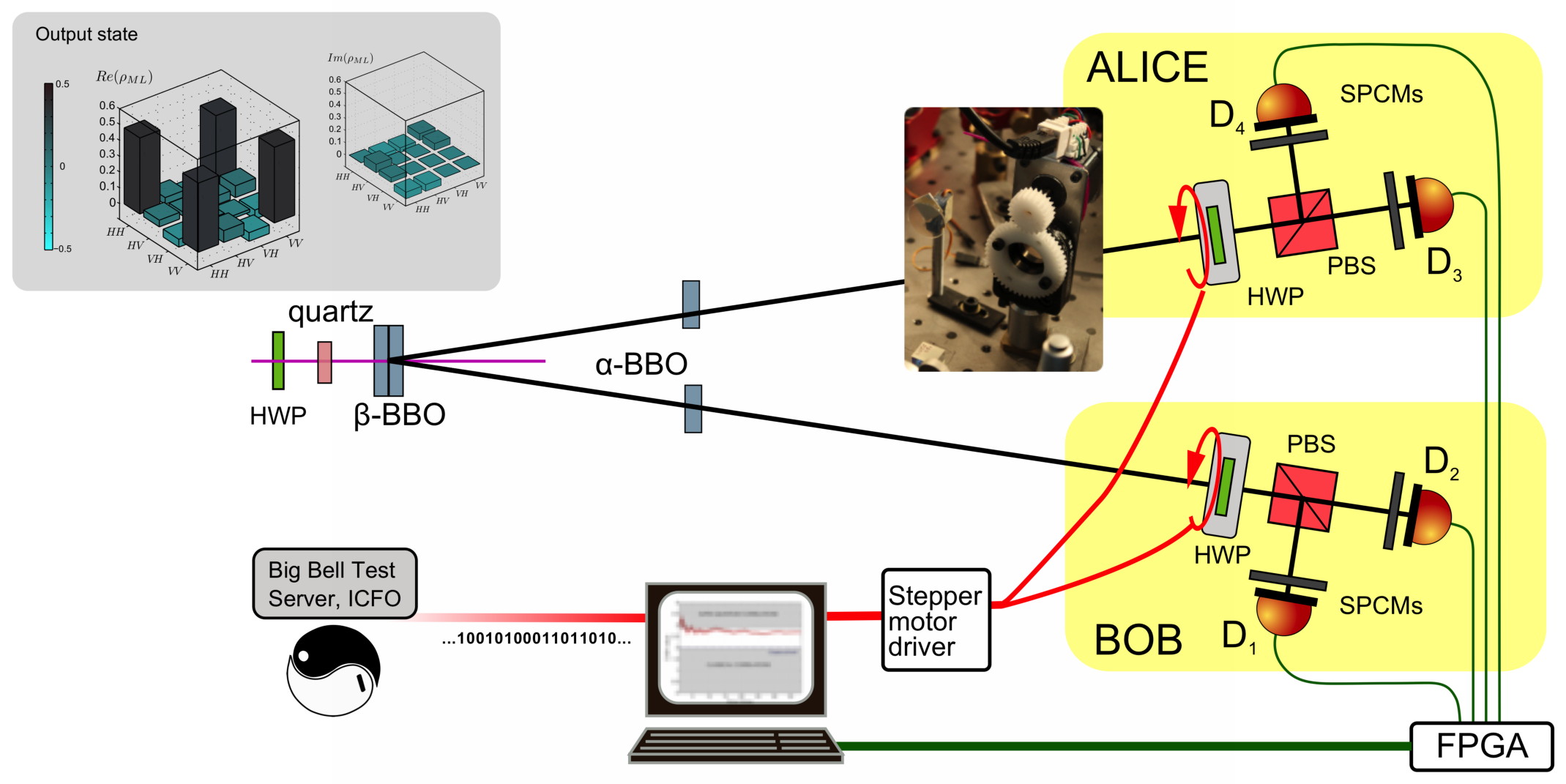}
\caption{Experimental setup. A 405 nm diode laser generates degenerate photon pairs by spontaneous parametric downconversion at 810 nm. A series of birefringent crystals placed at the pump beam and at the downconverted paths compensate temporal and spatial walk-off  ensuring a high degree of 
indistinguishability; state tomography characterization of the produced state is shown in the inset
. Random bit strings are retrieved from the BBT server and fed to the stepper motor drivers that control the projective measurements via rotations of the waveplates. The photo shows an actual motorized waveplate mount. 
Coincidences are registered with an FPGA-based detector and read by the control computer.
}
\label{fig:SetupBA}
\end{figure*}

We show the results of a two-channel experiment to test a Clauser-Horne-Shimony-Holt (CHSH)-like inequality, using polarization entangled photons. Photon pairs are generated by spontaneous parametric downconversion in a BBO type-I nonlinear crystal arrangement, pumped by a 405~nm cw laser diode polarized 45 degrees with respect to both crystals. Both photons are generated at 810~nm \BBTcite{knoll2014remote}. A series of birefringence compensating crystals included in the photon pair source guarantee a high 
level of entanglement: a tomographic reconstruction of the output state of the source is shown at the top left inset on \TextFig~\ref{fig:SetupBA})
\BBTcite{kwiat1999ultrabright}. Polarization projections are performed with motorized waveplates, which are fed with the random strings generated at the BBT cloud infrastructure. {Photons at both sides of the experiment are collected and spatially filtered by means of single mode fibers. Taking advantage of the angular distribution of the parametric fluorescence spectrum, the bandwidth of the correlated photon pairs is filtered down to \SI{10}{\nano\meter} using the same fiber coupling optics. An additional set of longpass filters is used to remove residual scattered light. Two pairs of single-photon counting devices detect the incoming photons and send the detections to an FPGA-based coincidence counter; correlated photons are detected using \SI{8}{\nano\second} coincidence windows.} 

The experiment was fully controlled using a single PC. Each experimental run consisted in the following sequence: a pair of random bits were gathered from the BBT server data and fed into the waveplate stepper motor drivers. Once the polarization projections were set, an FPGA-based coincidence counter started a repetition of 20 measurements on the four coincidence detections D$_1$ $\land$ D$_3$, D$_2$ $\land$ D$_3$, D$_1$ $\land$ D$_4$ and D$_2$ $\land$ $D_4$. The full setup is shown in \TextFig~\ref{fig:SetupBA}.
The brightness of our source and a detection window of 200~$\mu$s for each measurement gave a single-channel coincidence probability below 0.1. In particular, the mean D$_1$ $\land$ D$_3$ coincidence number per measurement was $\mu$=0.091 for one particular waveplate setting. This results on a total mean coincidence number per single experimental run (i.e. the expected value of detecting a coincidence on any channel) of $\nu$=0.227. Statistics of occurrence of either of the above situations follow a Poisson distribution. This condition minimizes the occurrence of multiple coincidences. Repetitions of the measurement increased its chance of success: we used the first single-channel coincidence from the list of 20 measurements as the result from each run, therefore increasing the "bit usage efficiency" of the experiment up to 97.5\%. This rate also takes into account discarded runs with simultaneous coincidence detections. The total duration for each run was close to 2~s, which accounted 
for the mechanical rotation of the waveplates and the measuring time. Overall, almost 34~kbit were required from the BBT server at the Buenos Aires Node; some of them were used for initial and mid-term alignment of the optical setup. 

\begin{figure*}[tbh]
\centering
	\includegraphics[width=0.75\textwidth]{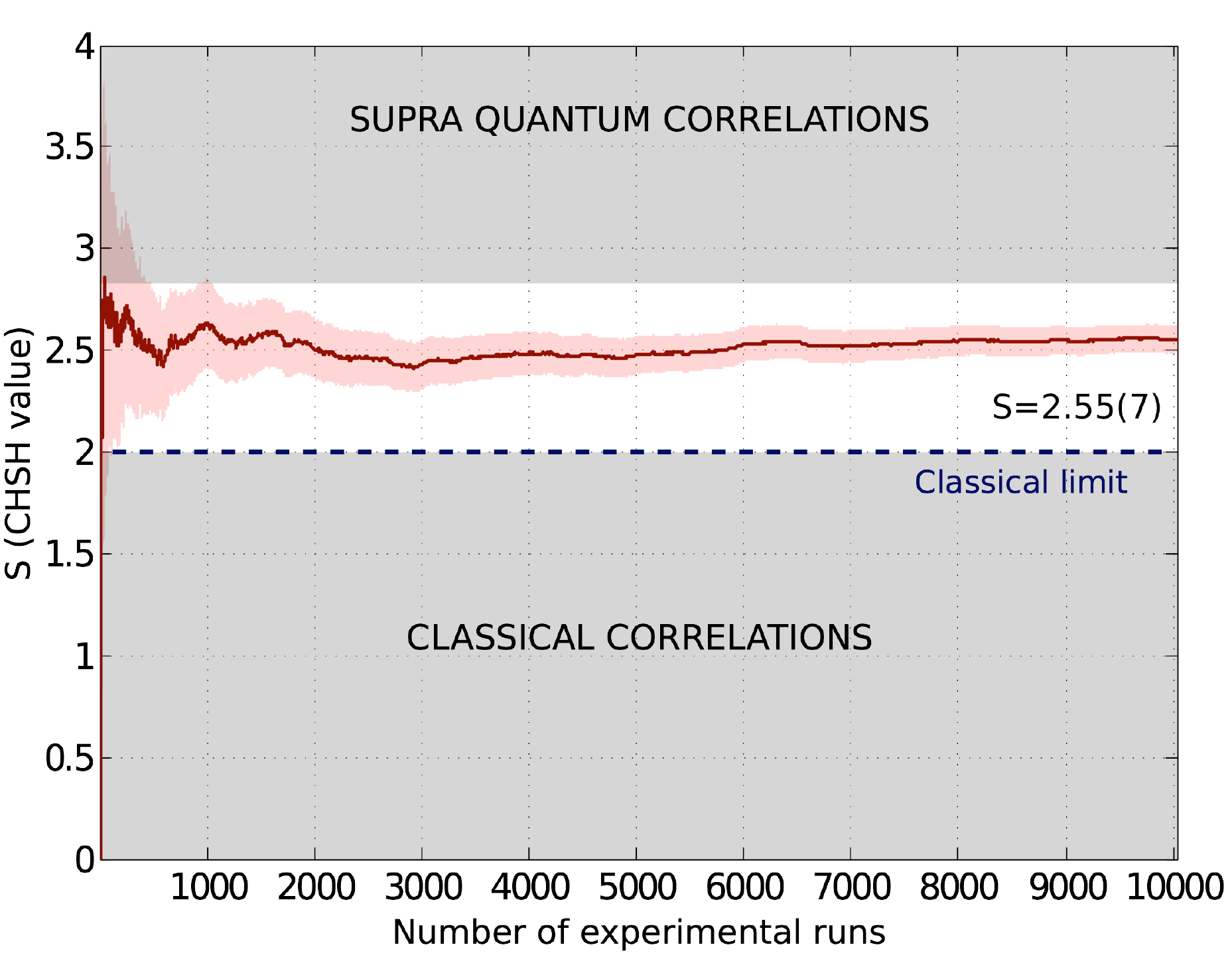}
\caption{Evolution of the CHSH test statistic ``S'' throughout the BBT day. Approximately 2000 experimental runs are sufficient to obtain a violation of the inequality with 95\% confidence level}
\label{fig:databsas}
\end{figure*}

The complete experiment carried out during the BBT day built statistics from 10033 polarization settings measured, and resulted in a measurement of the S-parameter value of S=$2.55(7)$, which implies a violation of the CHSH-Bell inequality that bounds 
local correlations by 7 standard deviations. {Since the choice of measurement of both Alice and Bob was obtained through the BBT game, we can be certain about the freedom of choice of both Alice and Bob. We can thus conclude that the assumption of free will was satisfied, therefore the free will loophole was closed. On the other hand, the locality loophole remains open in our setup.} \TextFig~\ref{fig:databsas} shows the evolution of the S-value obtained throughout the day. The red-shaded area represents the statistical uncertainty derived from the measured CHSH test.

\clearpage
\renewcommand{\BBTcite}[1]{\citeUdeC{#1}}
\graphicspath{{./PartnerContributions/UdeCBBT/}}
\renewcommand{\NODE}{\UC}
\BBTNodeTitle{Post-selection loophole-free  energy-time Bell test fed with human-generated inputs}

\BBTNodeAuthorList{Jaime Cari\~ne, Felipe Toledo, Pablo Gonz\'alez, \'Alvaro Alarc\'on, Daniel Martinez, Jorge Fuenzalida, Jean Cort\'es, \'Alvaro Cuevas, Gonzalo Carvacho, Aldo Delgado, Fabio Sciarrino, Paolo Mataloni, Jan-\AA ke~Larsson, Ad\'an Cabello, Esteban S. G\'omez, Gustavo Lima and Guilherme B. Xavier}

\begin{figure*}[tbh]
\centering
	\includegraphics[width=0.85\textwidth]{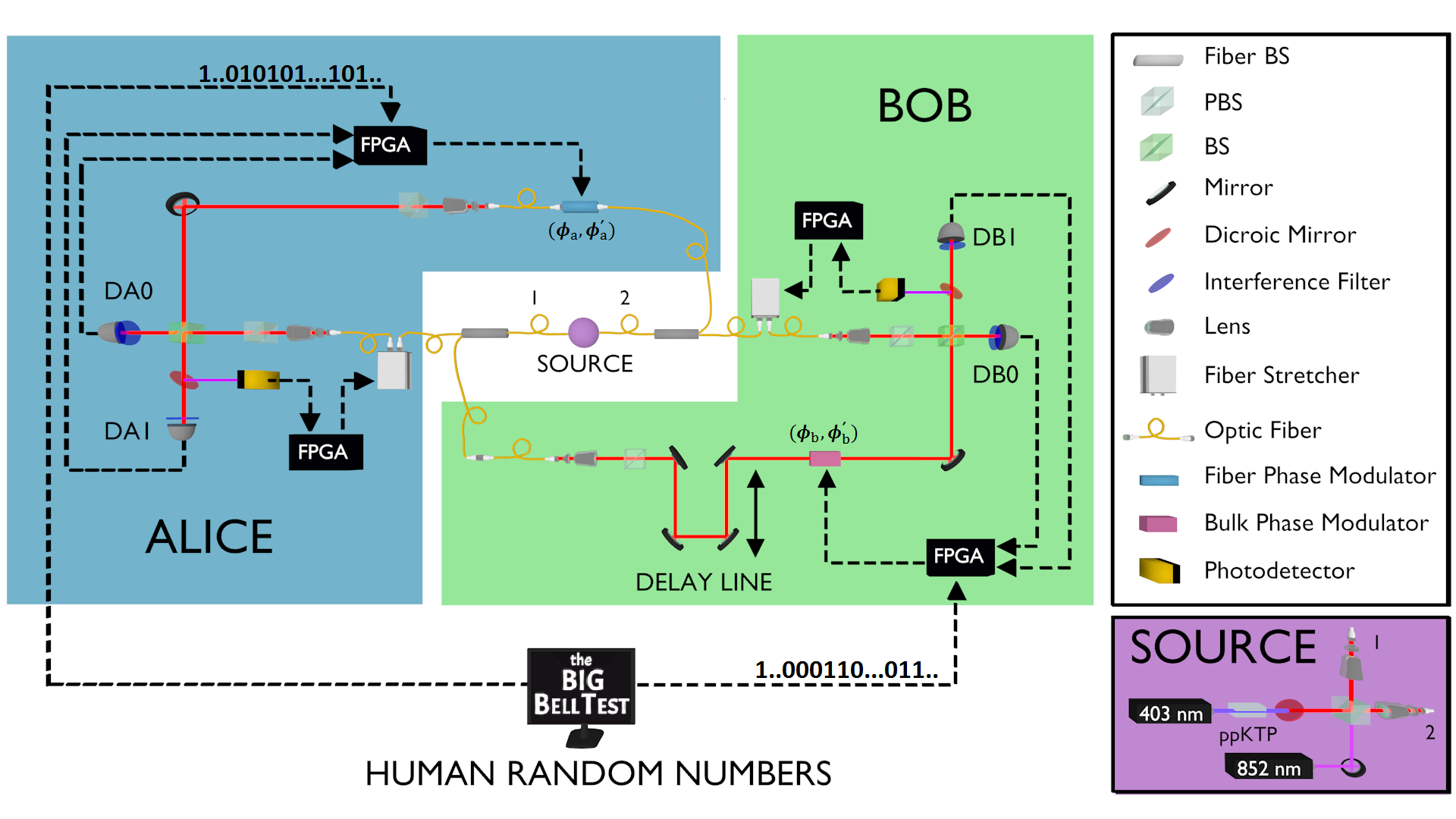}
\caption{\sffamily 
Experimental setup. Dashed and red lines represent electrical connections and free-space beams, respectively. The propagation distance between the source to Alice is 15~m and from the source to Bob is 45~m, determined mainly by the length of the piezoelectric spooled fiber-optical stretchers (10~m long for Alice's interferometer and 40~m for Bob's). Please see text for further details.
}
\label{SetupUdeC}
\end{figure*}

Energy-time Bell tests based on the standard and widely employed Franson configuration \BBTcite{Franson89} are affected by the post-selection loophole \BBTcite{Aerts99}. The only experimentally demonstrated solution to this problem is the so-called ``hug'' configuration \BBTcite{Adan09, Cuevas13, Cuevas15}. {Here we perform the first post-selection loophole-free energy-time Bell test where the measurement settings are randomly chosen by humans. The human random numbers were provided by the Big Bell Test (BBT) server and were used to choose the settings for each detected event. Several optoelectronic upgrades compared to our previous efforts were implemented to support the randomly and dynamically changing measurement settings.} The experimental setup is shown in \TextFig~\ref{SetupUdeC}, where the source generates energy-time entangled photon pairs through the process of spontaneous parametric down-conversion. A periodically poled potassium titanyl phosphate (ppKTP) 20 mm long waveguide crystal emits degenerate type-II (orthogonally polarized) photon pairs at 806 nm, when pumped with a single-mode longitudinal continuous wave laser at 403 nm. The single-photons are deterministically split with a polarizing beam splitter (PBS), with each photon being coupled to a single-mode fiber and then guided to one of the two bidirectional 50:50 fiber couplers whose outputs are then cross-sent to Alice and Bob. The main characteristic of the ``hug'' configuration is the presence of two crossed (or ``hugging'') Mach-Zehnder interferometers (MZI) serving as the communication channel between the source and the communicating parties. Alice and Bob place single-photon detectors at the outputs of each interferometer. The single-photon detectors are free-running silicon-based single-photon avalanche detectors, D$_{A1}$ and D$_{A2}$ for Alice and D$_{B1}$ and D$_{B2}$ for Bob. The interferometers are asymmetrical, as in the standard Franson scheme, with each consisting of a short (S) and a long (L) arm.

Each interferometer needs to be actively stabilized against environmental phase drifts. In order to do so, a reference laser beam (852 nm) is superposed to the path of the single-photons in the source, before coupling to the single-mode fibers. The intensity of this reference signal is detected by p-i-n photodiodes after both interferometers (split from the single-photons with dichroic mirrors) and are used to close individual electronic control loops based on field programmable gate array (FPGA) electronics. Each control loop employs a piezoelectric fiber-optical stretcher, capable of dynamically changing the length of the fiber \BBTcite{Xavier11}. With each interferometer independently stabilized, electro-optical phase modulators (PMs) are used to apply a relative phase difference in a short time window (5 $\mu$s), compared to the control response time. With such a technique the control in itself is not disturbed, and thus independent settings may be applied while the two interferometers remain stable \BBTcite{Xavier11}. 

A motorized delay line is used to adjust the long-short path difference in both interferometers to be within the coherence length of the photon pair. Once this is achieved the following state is generated without the need for temporal post-selection: $|\Phi\rangle=1/\sqrt{2}(|SS\rangle+e^{i(\phi_a+\phi_b)}|LL\rangle)$, where $\phi_a$ and $\phi_b$ are phase shifts within the long arm of Alice and Bob's respective MZIs. We test the well-known Clauser-Horne-Shimony-Holt (CHSH) inequality $S = E(\phi_{\text{a}},\phi_{\text{b}}) + E(\phi^{\prime}_{\text{a}},\phi_{\text{b}}) + E(\phi_{\text{a}},\phi^{\prime}_{\text{b}}) - E(\phi^{\prime}_{\text{a}},\phi^{\prime}_{\text{b}})$ \BBTcite{CHSH69}, where $E(\phi_{\text{a}}, \phi_{\text{b}}) = P_{11}(\phi_{\text{a}}, \phi_{\text{b}}) + P_{22}(\phi_{\text{a}}, \phi_{\text{b}}) - P_{12}(\phi_{\text{a}}, \phi_{\text{b}}) - P_{21}(\phi_{\text{a}}, \phi_{\text{b}})$, with $P_{\text{ij}}(\phi_{\text{a}}, \phi_{\text{b}})$ corresponding to the probability of a coincident detection at Alice and Bob's detectors $i$ and $j$ respectively, while the relative phases $\phi_{\text{a}}$ and $\phi_{\text{b}}$ are applied to Alice and Bob's interferometers. For the maximum violation of the Bell CHSH inequality the phase settings are $\phi_{\text{a}} = \pi/4$, $\phi_{\text{b}} = 0$, $\phi_{\text{a}}^{\prime} = -\pi/4$ and $\phi_{\text{b}}^{\prime} = \pi/2$. 

\begin{figure*}[tbh]
\centering
	\includegraphics[width=0.95\textwidth]{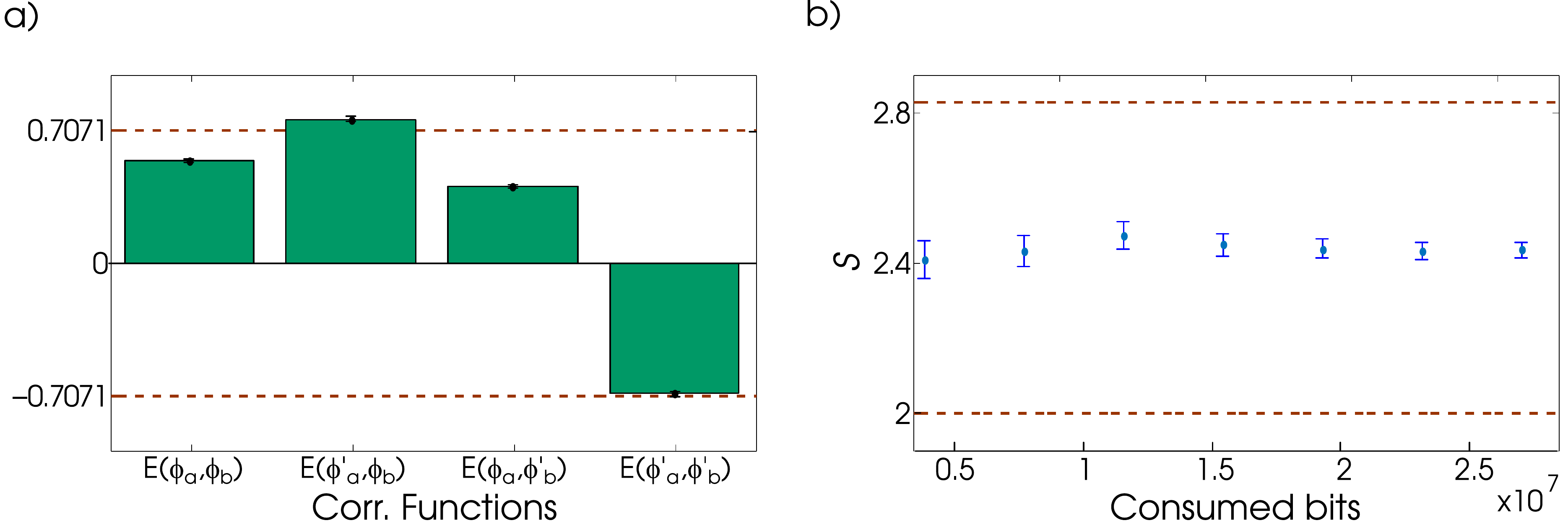}
\caption{\sffamily Experimental results. a) Correlation functions $E$ as a function of the different settings for the CHSH inequality. Dashed lines correspond to the value of $E$ for a maximal violation of the CHSH inequality. b) Cumulative CHSH violation (S parameter) as a function of the amount of consumed bits. Dashed red lines represent the local bound (2) and the quantum bound ($2\sqrt{2}$). Error bars in both subfigures represent the standard deviation.}
\label{fig2UdeC}
\end{figure*}

For each measurement round two bits are used from the BBT server to choose one out of the four possible setting combinations (two settings for Alice and two for Bob). The detection electronics (also FPGA-based) register if a coincident detection occurred, and record the values for the employed phases. This is done at a rate of 52 kbit/s while consuming approximately 27 million bits from the BBT database. The average expected values for the entire run are shown in \TextFig~\ref{fig2UdeC}a), with the cumulative S parameter in \TextFig~\ref{fig2UdeC}b), whose final value is $2.4331 \pm 0.0218$, corresponding to a violation of 19.87 standard deviations for a total of 20676 detection events. 

\clearpage
\renewcommand{\BBTcite}[1]{\citeNIST{#1}}
\graphicspath{{./PartnerContributions/NISTBBT/}}
\renewcommand{\NODE}{\NIST}
\BBTNodeTitle{Using human generated randomness to violate a Bell inequality without detection or locality loopholes}

\BBTNodeAuthorList{Lynden Krister Shalm, Martin Stevens, Omar S. Maga\~{n}a-Loaiza, Thomas Gerrits, Scott Glancy, Peter Bierhorst, Emanuel Knill, Richard Mirin, Sae Woo Nam}

\begin{figure*}[tbh]
\centering
\includegraphics[width=1.\textwidth]{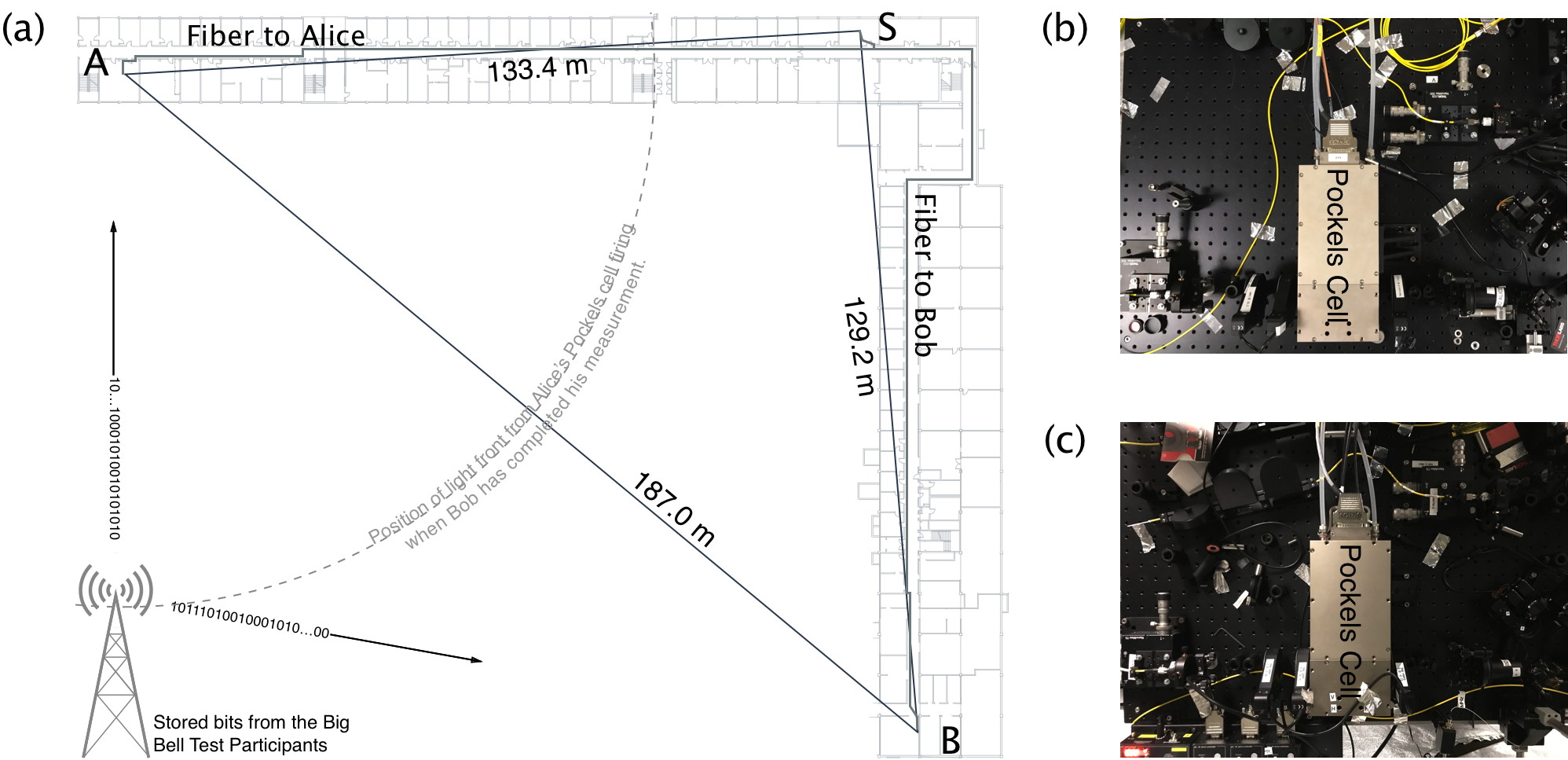}
\caption{Experimental setup. (a) The locations of the Source (S), Alice (A), and Bob (B). Each trial, the source lab produces a pair of photons in the non-maximally polarization-entangled state $\left|\psi \right\rangle \approx 0.982 \left|HH \right\rangle + 0.191 \left|VV \right\rangle$,
where $H$ ($V$) denotes horizontal (vertical) polarization. One photon is sent to Alice's lab while the other is sent to Bob's lab to be measured. Alice's computed optimal polarization measurement angles, relative to a vertical polarizer, are $\{a =-3.7^o, a' = 23.6^o\}$ while Bob's are $\{b = 3.7^o, b' = -23.6^o\}$. Each trial, Alice and Bob each use a bit from the Bellsters to choose which polarization measurement to make. Panel (b) shows the Pockels cell that Alice uses, and panel (c) shows Bob's Pockels cell. Alice and Bob are ($187 \pm 1$) m apart from one another. At this distance, in any given trial when Bob has completed his measurement, information traveling at the speed of light from Alice's Pockels cell firing is still ($88.25 \pm 1.1$) m away from his measurement setup (represented by the light front). When Alice completes her measurement, Bob's light front from his Pockels cell firing is ($127.2 \pm 1.1$) m away from her measurement apparatus.}
\label{fig:Setup}
\end{figure*}

We report a violation of a Bell inequality free of fair-sampling and locality assumptions. We use polarization-entangled photons generated by a nonlinear crystal pumped by a pulsed, picosecond laser. The laser repetition rate is 79.3 MHz, and each pulse that enters the crystal has a probability of $\approx 0.002$ of creating an entangled photon pair. The two entangled photons from each pair are then separated, with each one being sent to one of two measurement stations ($187 \pm 1$) m apart, named Alice and Bob. At Alice and Bob, a Pockels cell and polarizer combine to allow the rapid measurement of the polarization state of the incoming photons. Alice's and Bob's decisions on how to set their Pockels cells are dictated by the bits supplied by the Big Bell Test participants (Bellsters). The photons are then detected using fiber-coupled superconducting single-photon nanowire detectors, each operating at over $90 \%$ efficiency \BBTcite{marsili2013detecting}.

Our experimental setup, shown in \TextFig \ref{fig:Setup}, is nearly identical to the experimental setup used in the fully loophole-free Bell Test conducted in 2015 \BBTcite{PhysRevLett.115.250402}. The primary difference is that random numbers used to set Alice's and Bob's polarization settings are provided by the Bellsters instead of random number generators. Like the 2015 experiment, we consider the measurement of Alice's or Bob's photon detection to be complete when the amplified electrical signal from Alice's or Bob's detector reaches the input of the time tagger that records the outcomes. The polarization measurement is carried out using a high-voltage Pockels cell. When this device switches states, it is possible for an electro-magnetic pulse to be emitted. If these electro-magnetic signals travel at the speed of light, then Bob completes his measurement $(294.4 \pm 3.7)$ ns before any signal from Alice's Pockels cell could arrive at his station. Similarly, Alice completes her measurement $(424.2 \pm 3.7)$ ns before a switching signal from Bob's Pockels cell could arrive at her location. {Under these conditions the implementation of Alice's setting and Bob's measurement outcome are space-like separated from one another and vice-versa, so it is impossible for one station's setting to influence the other station's measurement.}

Testing the hypothesis of local realism requires repeating the experiment many times, each of which is called a trial.  We then compare the trials' statistics to those that local realistic theories predict. In our setup we perform 100,000 trials per second. This is limited by the speed of our Pockels cells at Alice and Bob. Because the probability of generating a photon pair with each pump pulse is small, Alice and Bob look for photons generated by 15 successive pulses during each trial. This increases our probability of creating a photon pair per trial by more than an order of magnitude, allowing us to achieve a stronger violation using fewer trials. Each trial then consists of 15 different time bins in which Alice and Bob may register photon detections. In our experiment we collected data for just under 7 minutes and performed 40,559,990 trials. Each trial, Alice and Bob each used one bit provided by the Bellsters to set their polarization measurement choice (consuming a total of 81,119,980 bits over the course of the experiment). Immediately after collecting data using bits provided by the Bellsters, we repeated the experiment using random bits provided solely by computer-based random number generators located at both Alice and Bob. In this second data run we performed a total of 74,400,000 trials.

Each trial, Alice uses a bit from the Bellsters to choose between two settings ${a, a'}$ while Bob uses a bit from the Bellsters to choose between two different settings ${b, b'}$. Because we include 15 different time bins each trial, when certain settings are chosen there is a significant probability in a given trial of Alice and Bob both detecting photons, but in different time bins. Using conventional Bell inequalities, these detections would normally be considered a coincidence event even though they involve unentangled photons from different photon pairs. To help limit the effect of these unwanted coincidences, we use a modified Clauser-Horne (CH) inequality \BBTcite{PhysRevD.10.526} that accounts for the 15 individual time bins that a detection can take place in at either Alice or Bob. During a trial if Alice (Bob) detects a photon in time bin $i$, we label the event $i_{A}$ ($i_{B}$). If no detection is observed, then the outcome is labelled with a $0_{A}$ for Alice or $0_{B}$ for Bob. If two separate photon events are detected by either Alice or Bob in the same trial, only the first bin is counted. The modified inequality is:
\begin{equation}
\label{E:CH}
K = \sum_{i=1}^nP(i_A,i_B|ab)
 - \sum_{i=1}^nP(i_A,i_B|a'b') -\sum_{i=1}^nP(0_A,i_B|a'b)
 -\sum_{i=1}^n\sum_{j\ne i}P(j_A,i_B|a'b)-\sum_{i=1}^nP(i_A,0_B|ab')-\sum_{i=1}^n\sum_{j\ne i}P(i_A,j_B|ab') \leq 0
\end{equation}

Here $n=15$ for our experiment. The term $P(i_A,i_B|ab)$ represents the probability that Alice and Bob both detect photons in time bin $i$, given that Alice chooses setting $a$ and Bob chooses setting $b$. The term $ P(i_A,i_B|a'b')$ is the same, but for settings $a'$ and $b'$. This eliminates from the inequality the high probability of accidental coincidences where Alice and Bob, with joint settings $a'$ and $b'$, detect photons in different time bins ($i\neq j$) during the same trial. The terms $P(0_A,i_B|a'b)$ and $P(i_A,0_B|ab')$ correspond to cases where only one party detects a photon in a given trial. Finally, the $P(j_A,i_B|a'b)$ and $P(i_A,j_B|ab')$ terms correspond to trials where Alice and Bob both detect photons, but not in the same time bins.

To see that equation \ref{E:CH} is a valid Bell inequality, we note that if local realism holds then each term can be written as an integral over hidden variables, $\lambda$. Then, because Alice and Bob are far enough apart, Alice's measurement setting ($S_A$) and outcome ($O_A$) can be factored from Bob's measurement setting ($S_B$) and outcome ($O_B$):
\begin{equation}
\label{E:BellArgument}
P(O_a,O_b|S_aS_b)=\int P(O_a,O_b|S_aS_b\lambda)\rho(\lambda)d\lambda=\int P(O_a|S_a\lambda)P(O_b|S_b\lambda)\rho(\lambda)d\lambda.
\end{equation}

Since the probability of some event, $E$, occurring is $P(E) = 1- P(\bar{E})$, where $P(\bar{E})$ is the probability the event does not occur, the two terms with double sums can be re-written, as
\begin{equation}
\label{E:doubleFactor1}
\begin{split}
\sum_{i=1}^n\sum_{j\ne i}P(j_A,i_B|a'b) &= \sum_{i=1}^n\int P(i_B|b\lambda)[1-P(i_A|a'\lambda)-P(0_A|a'\lambda)]\rho(\lambda)d\lambda, \\
\sum_{i=1}^n\sum_{j\ne i}P(i_A,j_B|ab') &= \sum_{i=1}^n\int P(i_A|a\lambda)[1-P(i_B|b'\lambda)-P(0_B|b'\lambda)]\rho(\lambda)d\lambda.
\end{split}
\end{equation}

Now, since everything is just a sum over $i$, we can first pull the sum and then the integral out of all terms. A number of terms cancel, leaving us with:
\begin{equation}
\label{E:CHLocality}
K = \sum_{i=1}^n\int \big[P(i_A|a\lambda) P(i_B|b\lambda)-P(i_A|a'\lambda)P(i_B|b'\lambda) -   P(i_B|b\lambda)
+ P(i_B|b\lambda)P(i_A|a'\lambda) - P(i_A|a\lambda)
+P(i_A|a\lambda)P(i_B|b'\lambda)	\big]\rho(\lambda)d\lambda.
\end{equation}
The fact that the four remaining probabilities, $P(i_A|a\lambda)$, $P(i_B|b\lambda)$, $P(i_A|a'\lambda)$, and $P(i_B|b'\lambda)$ are all bounded between 0 and 1 implies that the maximum value of each integrand is 0, following the arguments of \BBTcite{PhysRevD.10.526}.

 Using this inequality, we measure a Bell parameter of $K = (1.70 \pm 0.20) x 10^{-4}$, corresponding to an 8.7-$\sigma$ violation of a Bell inequality (under the assumptions of a normal distribution for the Bell parameter, and of each trial being independent and identical). Table \ref{T:NISTData} contains the number of photons detected and the number of trials for each term in Equation \ref{E:CH}. We also analyzed the data from the experiment that used computer-based randomness, and obtained a Bell parameter of $K = (1.55 \pm 0.14)x 10^{-4}$ corresponding to an 11.2-$\sigma$ violation a Bell inequality. The violation is larger because we performed nearly twice as many trials with the computer-based random number generator, allowing us to obtain better statistics. Both of these results constitute a strong rejection of local realism. A full record of the data, along with the code used to analyze it, \stext{can be found at ** 
 } \btext{is available on request}.

 \begin{table}[]
 \centering
 \caption{The number of photons detected for each of the terms in equation \ref{E:CH}. Due to the large bias towards zeros in the bits received from the Bellsters, it is necessary to normalize the counts by the number of trials conducted at each setting. A total of 40,559,990 trials were carried out over the course of the experiment. A second experiment 74,400,000 trials long that used only unbiased random bits from a computer was run immediately afterwards.}
 \label{T:NISTData}
 \begin{tabular}{ccccccc}
 \cline{2-7}
 \multicolumn{1}{c|}{}        & \multicolumn{1}{c|}{$\sum_{i=1}^{15}P(i_A,i_B|ab)$}       & \multicolumn{1}{c|}{$\sum_{i=1}^{15}P(i_A,i_B|a'b')$}   & \multicolumn{1}{c|}{$\sum_{i=1}^{15}P(0_A,i_B|a'b)$}    & \multicolumn{1}{c|}{$\sum_{i=1}^{15}\sum_{j\ne i}P(j_A,i_B|a'b)$}      & \multicolumn{1}{c|}{$\sum_{i=1}^{15}P(i_A,0_B|ab')$}       & \multicolumn{1}{c|}{$\sum_{i=1}^{15}\sum_{j\ne i}P(i_A,j_B|ab')$}       \\ \hline

\multicolumn{7}{|c|}{Random bits from Bellsters}  \\ \hline

 \multicolumn{1}{|c|}{Events} & \multicolumn{1}{c|}{20265}    & \multicolumn{1}{c|}{712}     & \multicolumn{1}{c|}{7475}     & \multicolumn{1}{c|}{143}      & \multicolumn{1}{c|}{8181}     & \multicolumn{1}{c|}{123}      \\ \hline

 \multicolumn{1}{|c|}{Trials} & \multicolumn{1}{c|}{11126350} & \multicolumn{1}{c|}{9202147} & \multicolumn{1}{c|}{10125716} & \multicolumn{1}{c|}{10125716} & \multicolumn{1}{c|}{10105777} & \multicolumn{1}{c|}{10105777} \\ \hline

 \multicolumn{7}{|c|}{Random bits from computers}  \\ \hline

 \multicolumn{1}{|c|}{Events} & \multicolumn{1}{c|}{33543}    & \multicolumn{1}{c|}{1677}     & \multicolumn{1}{c|}{13911}     & \multicolumn{1}{c|}{249}      & \multicolumn{1}{c|}{14589}     & \multicolumn{1}{c|}{228}      \\ \hline

 \multicolumn{1}{|c|}{Trials} & \multicolumn{1}{c|}{18600599} & \multicolumn{1}{c|}{18598891} & \multicolumn{1}{c|}{18599346} & \multicolumn{1}{c|}{18599346} & \multicolumn{1}{c|}{18601164} & \multicolumn{1}{c|}{18601164} \\ \hline

                              %
 \end{tabular}
 \end{table}

It is also possible to carry out a hypothesis test to see how well a local-realistic theory could reproduce the correlations we observed in our experiment. Such a test does not require assumptions about memory effects and the statistical distributions of each trial's outcomes. We devised the following analysis protocol before running the experiment. The first 5\% of the data was used for training purposes to estimate the rate of relevant events, given by terms in equation \ref{E:BellArgument}. From this, an estimate of the total number of remaining events was made. A stopping criteria, $n_{cut}$, defined as 90\% of the estimated remaining relevant events was defined. Once $n_{cut}$ relevant events were processed the analysis was stopped, and the p-value was computed using a binomial analysis of equation \ref{E:CH} (see \BBTcite{PhysRevLett.115.250402} for more details).

Our hypothesis test protocol was flawed, and failed to observe a violation. When we initially ran the protocol on our data we obtained a p-value $3.3 x 10^{-70} $, with $n_{cut} = 30,916$ events. This is a much smaller p-value than expected. Looking more closely at the data, we realized that there is an approximately $5\%$ bias of excess zeros compared to ones submitted by the Bellsters. Given that over 100,000 individuals from around the world independently submitted zeros and ones, such a large bias was surprising. We do not have a definitive explanation for this, but we suspect it might be related to the majority of humans being right-hand dominant. Most keyboards use the right hand to input a zero. On most specialized number pads, the zero key is also substantially larger. For our setup the $ab$ setting, which contributes to events that violate equation \ref{E:CH}, occurs more frequently than the other settings due to the bias. Since our hypothesis test protocol was devised under the assumption that all the settings are equiprobable, we must correct our p-value to account for this large excess bias. After performing this correction, we obtain a p-value of $\approx 1$, and we are unable to reject the null hypothesis that local realism governs our experiment whose random inputs were generated by humans. The data from the computer-based random number generator, however, had the unbiased settings distribution required by our hypothesis test protocol. From this data set we obtained a p-value of $2.6x10^{-27}$ with $n_{cut} = 54,720$, corresponding to a strong rejection of local realism.

The null result from the hypothesis test with human-based randomness does not mean that local realism governed the experiment. The hypothesis test protocol we followed assumed that there was no bias in both Alice's and Bob's settings choices. An improved protocol, for future experiments, would include a step that first estimates the bias from the training data in order to determine the optimal Bell inequality to test. As long as there are only small deviations from this estimated bias (on the order of $< 0.2\%$) over the remainder of the experiment, it would be possible to compute a rigorous p-value significantly smaller than 1. Additionally, our measurement of the Bell parameter value is good evidence that local realism is not consistent with the experimental correlations we observed. The Bell parameter is normalized by the total trials for each settings choice, and is therefore insensitive to bias in the settings distribution. Under the assumptions that each trial was independent and identically distributed and that the Bell parameter was normally distributed, the measured Bell parameter exceeds the maximum bound predicted by local realism by a statistically significant amount. Even if these assumption are not true, it would take a local realistic theory with a contrived distribution to appreciably lower this significance. For some applications, like using a loophole-free Bell test to extract randomness, it is important to use a rigorous hypothesis testing framework to guard against a malicious adversary who might be able to produce such contrived distributions to attack the system. However, one hopes that nature is not so malicious when testing local realism.

We wanted to see if there was a difference between the results of a Bell test where humans chose the settings at Alice and Bob versus a Bell test that used non-human random number generators. After accounting for the bias in the human inputs, data from both the human-based input and the computer-based input were able to violate a Bell inequality.


%
%


\begin{thebibliography}{100}

\makeatletter
\addtocounter{\@listctr}{114}
\makeatother

\bibitem{knoll2014remote}
L.~T. Knoll, C.~T. Schmiegelow, M.~A. Larotonda, Remote state preparation of a
  photonic quantum state via quantum teleportation, {\it Applied Physics B\/}
  {\bf 115}, 541 (2014).

\bibitem{kwiat1999ultrabright}
P.~G. Kwiat, E.~Waks, A.~G. White, I.~Appelbaum, P.~H. Eberhard, Ultrabright
  source of polarization-entangled photons, {\it Physical Review A\/} {\bf 60},
  R773 (1999).

\end{thebibliography}


\begin{thebibliography}{0}%
\makeatletter
\providecommand \@ifxundefined [1]{%
 \@ifx{#1\undefined}
}%
\providecommand \@ifnum [1]{%
 \ifnum #1\expandafter \@firstoftwo
 \else \expandafter \@secondoftwo
 \fi
}%
\providecommand \@ifx [1]{%
 \ifx #1\expandafter \@firstoftwo
 \else \expandafter \@secondoftwo
 \fi
}%
\providecommand \natexlab [1]{#1}%
\providecommand \enquote  [1]{``#1''}%
\providecommand \bibnamefont  [1]{#1}%
\providecommand \bibfnamefont [1]{#1}%
\providecommand \citenamefont [1]{#1}%
\providecommand \href@noop [0]{\@secondoftwo}%
\providecommand \href [0]{\begingroup \@sanitize@url \@href}%
\providecommand \@href[1]{\@@startlink{#1}\@@href}%
\providecommand \@@href[1]{\endgroup#1\@@endlink}%
\providecommand \@sanitize@url [0]{\catcode `\\12\catcode `\$12\catcode
  `\&12\catcode `\#12\catcode `\^12\catcode `\_12\catcode `\%12\relax}%
\providecommand \@@startlink[1]{}%
\providecommand \@@endlink[0]{}%
\providecommand \url  [0]{\begingroup\@sanitize@url \@url }%
\providecommand \@url [1]{\endgroup\@href {#1}{\urlprefix }}%
\providecommand \urlprefix  [0]{URL }%
\providecommand \Eprint [0]{\href }%
\providecommand \doibase [0]{http://dx.doi.org/}%
\providecommand \selectlanguage [0]{\@gobble}%
\providecommand \bibinfo  [0]{\@secondoftwo}%
\providecommand \bibfield  [0]{\@secondoftwo}%
\providecommand \translation [1]{[#1]}%
\providecommand \BibitemOpen [0]{}%
\providecommand \bibitemStop [0]{}%
\providecommand \bibitemNoStop [0]{.\EOS\space}%
\providecommand \EOS [0]{\spacefactor3000\relax}%
\providecommand \BibitemShut  [1]{\csname bibitem#1\endcsname}%
\let\auto@bib@innerbib\@empty
\end{thebibliography}%


%


\begin{thebibliography}{10}

\makeatletter
\addtocounter{\@listctr}{53}
\makeatother

\bibitem{Bell1964}
J.~S. Bell, {On the Einstein Podolsky Rosen Paradox}, {\it Physics\/} {\bf 1},
  195 (1964).

\bibitem{Brukner2004}
C.~Brukner, S.~Taylor, S.~Cheung, V.~Vedral, {Quantum Entanglement in Time},
  {\it arXiv:quant-ph/0402127\/}  (2004).

\bibitem{Wood2015CausalDiscovery}
C.~J. Wood, R.~W. Spekkens, {The lesson of causal discovery algorithms for
  quantum correlations: causal explanations of Bell-inequality violations
  require fine-tuning}, {\it New J.\ Phys.\/} {\bf 17}, 033002 (2015).

\bibitem{TemporalFramework}
F.~Costa, M.~Ringbauer, M.~E. Goggin, A.~G. White, A.~Fedrizzi, {A Unifying
  Framework for spatial and temporal quantum correlations}, {\it
  arXiv:1710.01776\/}  (2017).

\bibitem{Ringbauer2017TemporalCM}
M.~Ringbauer, R.~Chaves, {Probing the Non-Classicality of Temporal
  Correlations}, {\it Quantum\/} {\bf 1}, 35 (2017).

\bibitem{Coffman2000}
V.~Coffman, J.~Kundu, W.~K. Wootters, {Distributed entanglement}, {\it Phys.\
  Rev.\ A\/} {\bf 61}, 052306 (2000).

\bibitem{Markiewicz2014}
M.~Markiewicz, A.~Przysi{\c{e}}{\.{z}}na, S.~Brierley, T.~Paterek, {Genuinely
  multipoint temporal quantum correlations and universal measurement-based
  quantum computing}, {\it Phys.\ Rev.\ A\/} {\bf 89}, 062319 (2014).

\end{thebibliography}


\begin{thebibliography}{10}

\makeatletter
\addtocounter{\@listctr}{85}
\makeatother


\bibitem{Bouchiat1998}
V.~Bouchiat, D.~Vion, P.~Joyez, D.~Esteve, M.~H. Devoret, Quantum coherence
  with a single {Cooper} pair, {\it Phys. Scr.\/} {\bf T76}, 165 (1998).

\bibitem{Koch2007}
J.~Koch, {\it et~al.\/}, Charge-insensitive qubit design derived from the
  {Cooper} pair box, {\it Phys. Rev. A\/} {\bf 76}, 042319 (2007).

\bibitem{Houck2008}
A.~A. Houck, {\it et~al.\/}, Controlling the spontaneous emission of a
  superconducting transmon qubit, {\it Phys. Rev. Lett.\/} {\bf 101}, 080502
  (2008).

\bibitem{Motzoi2009}
F.~Motzoi, J.~M. Gambetta, P.~Rebentrost, F.~K. Wilhelm, Simple pulses for
  elimination of leakage in weakly nonlinear qubits, {\it Phys. Rev. Lett.\/}
  {\bf 103}, 110501 (2009).

\bibitem{Gambetta2011a}
J.~M. Gambetta, F.~Motzoi, S.~T. Merkel, F.~K. Wilhelm, Analytic control
  methods for high-fidelity unitary operations in a weakly nonlinear
  oscillator, {\it Phys. Rev. A\/} {\bf 83}, 012308 (2011).

\bibitem{Strauch2003}
F.~W. Strauch, {\it et~al.\/}, Quantum logic gates for coupled superconducting
  phase qubits, {\it Phys. Rev. Lett.\/} {\bf 91}, 167005 (2003).

\bibitem{Macklin2015}
C.~Macklin, {\it et~al.\/}, A near-quantum-limited {Josephson} traveling-wave
  parametric amplifier, {\it Science\/} {\bf 350}, 307 (2015).

\bibitem{CHSH1969}
J.~F. Clauser, M.~A. Horne, A.~Shimony, R.~A. Holt, Proposed experiment to test
  local hidden-variable theories, {\it Phys. Rev. Lett.\/} {\bf 23}, 880
  (1969).

\bibitem{Larsson2014}
J.-A. Larsson, Loopholes in {Bell} inequality tests of local realism, {\it J.
  Phys. A: Math. Theor.\/} {\bf 47}, 424003 (2014).

\bibitem{bentkus2004}
V.~Bentkus, On {Hoeffding's} inequalities, {\it Ann. Probab.\/} {\bf 32}, 1650
  (2004).

\bibitem{Elkouss2016}
D.~Elkouss, S.~Wehner, ({Nearly}) optimal {P} values for all {Bell}
  inequalities, {\it npj Quantum Information\/} {\bf 2}, 16026 (2016).

\end{thebibliography}


\begin{thebibliography}{10}
\makeatletter
\addtocounter{\@listctr}{50}
\makeatother

\bibitem{schrodinger}
E.~Schr\"{o}dinger, Discussion of probability relations between separated
  systems, {\it Mathematical Proceedings of the Cambridge Philosophical
  Society\/} {\bf 31}, 555 (1935).

\bibitem{dylan}
D.~J. Saunders, S.~J. Jones, H.~M. Wiseman, G.~J. Pryde, Experimental
  {EPR}-steering using {Bell-local} states, {\it Nature Physics\/} {\bf 6}, 845
  (2010).

\bibitem{adamprx}
A.~J. Bennet, {\it et~al.\/}, Arbitrarily loss-tolerant
  {E}instein-{P}odolsky-{R}osen steering allowing a demonstration over 1 km of
  optical fiber with no detection loophole, {\it Phys. Rev. X\/} {\bf 2},
  031003 (2012).

\end{thebibliography}


\begin{thebibliography}{100}

\makeatletter
\addtocounter{\@listctr}{103}
\makeatother

\bibitem{Matsukevich2005}
D.~N. Matsukevich, {\it et~al.\/}, Entanglement of a photon and a collective
  atomic excitation, {\it Phys. Rev. Lett.\/} {\bf 95}, 040405 (2005).

\bibitem{FarreraPRL2018}
P.~Farrera, G.~Heinze, H.~de~Riedmatten, Entanglement between a photonic
  time-bin qubit and a collective atomic spin excitation, {\it Phys. Rev.
  Lett.\/} {\bf 120}, 100501 (2018).

\bibitem{Duan2001}
L.~M. Duan, M.~D. Lukin, J.~I. Cirac, P.~Zoller, {Long-distance quantum
  communication with atomic ensembles and linear optics.}, {\it Nature\/} {\bf
  414}, 413 (2001).

\bibitem{Albrecht2015}
B.~Albrecht, P.~Farrera, G.~Heinze, M.~Cristiani, H.~de~Riedmatten, Controlled
  rephasing of single collective spin excitations in a cold atomic quantum
  memory, {\it Phys. Rev. Lett.\/} {\bf 115}, 160501 (2015).

\bibitem{Marcikic2004}
I.~Marcikic, {\it et~al.\/}, Distribution of time-bin entangled qubits over 50
  km of optical fiber, {\it Phys. Rev. Lett.\/} {\bf 93}, 180502 (2004).

\bibitem{Clauser1969}
J.~F. Clauser, M.~A. Horne, A.~Shimony, R.~A. Holt, Proposed experiment to test
  local hidden-variable theories, {\it Physical Review Letters\/} {\bf 23}, 880
  (1969).

\bibitem{Peres1996}
A.~Peres, Separability criterion for density matrices, {\it Phys. Rev. Lett.\/}
  {\bf 77}, 1413 (1996).

\end{thebibliography}


\begin{thebibliography}{100}

\makeatletter
\addtocounter{\@listctr}{110}
\makeatother

\bibitem{Olislager2012}
L.~Olislager, {\it et~al.\/}, Implementing two-photon interference in the
  frequency domain with electro-optic phase modulators, {\it New Journal of
  Physics\/} {\bf 14}, 043015 (2012).

\bibitem{Rielander2016}
D.~Riel\"ander, A.~Lenhard, M.~Mazzera, H.~de~Riedmatten, Cavity enhanced
  telecom heralded single photons for spin-wave solid state quantum memories,
  {\it New Journal of Physics\/} {\bf 18}, 123013 (2016).

\bibitem{Clauser1974}
J.~F. Clauser, M.~A. Horne, Experimental consequences of objective local
  theories, {\it Phys. Rev. D\/} {\bf 10}, 526 (1974).

\bibitem{RielanderQST2018}
D.~Riel{\"a}nder, {\it et~al.\/}, Frequency-bin entanglement of ultra-narrow
  band non-degenerate photon pairs, {\it Quantum Science and Technology\/} {\bf
  3}, 014007 (2018).

\end{thebibliography}


\begin{thebibliography}{}

\end{thebibliography}


\begin{thebibliography}{10}

\makeatletter
\addtocounter{\@listctr}{81}
\makeatother

\bibitem{Volz2006}
J.~Volz, {\it et~al.\/}, {Observation of Entanglement of a Single Photon with a
  Trapped Atom}, {\it Phys. Rev. Lett.\/} {\bf 96}, 030404 (2006).

\bibitem{Fuerst2010}
M.~F\"urst, {\it et~al.\/}, {High speed optical quantum random number
  generation}, {\it Opt. Express\/} {\bf 18}, 13029 (2010).

\bibitem{Rosenfeld2017}
W.~Rosenfeld, {\it et~al.\/}, {Event-Ready Bell Test Using Entangled Atoms
  Simultaneously Closing Detection and Locality Loopholes}, {\it Phys. Rev.
  Lett.\/} {\bf 119}, 010402 (2017).

\bibitem{vermessungsamt}
{Bayerisches Landesamt fuer Digitalisierung, Breitband und Vermessung}.

\end{thebibliography}


\begin{thebibliography}{100}

\makeatletter
\addtocounter{\@listctr}{96}
\makeatother

\bibitem{Bell_EPR_1964}
J.~S. Bell, On the {E}instein {P}odolsky {R}osen paradox, {\it Physics\/} {\bf
  1}, 195 (1964).

\bibitem{CHSH_1969}
J.~F. Clauser, M.~A. Horne, A.~Shimony, R.~A. Holt, Proposed experiment to test
  local hidden-variable theories, {\it Phys. Rev. Lett.\/} {\bf 23}, 880
  (1969).

\bibitem{The_Quantum_Challenge_Book}
G.~Greenstein, A.~Zajonc, {\it The Quantum Challenge: Modern Research on the
  Foundations of Quantum Mechanics\/}, Jones and Bartlett series in physics and
  astronomy (Jones and Bartlett Publishers, 2006).

\bibitem{Nielsen_book_2011}
M.~A. Nielsen, I.~L. Chuang, {\it Quantum Computation and Quantum Information:
  10th Anniversary Edition\/} (Cambridge University Press, New York, NY, USA,
  2011), 10th edn.

\bibitem{VergyrisQST2017}
P.~Vergyris, {\it et~al.\/}, Fully guided-wave photon pair source for quantum
  applications, {\it Quantum Science and Technology\/} {\bf 2}, 024007 (2017).

\bibitem{Lim_Sagnac_2010}
H.~C. Lim, A.~Yoshizawa, H.~Tsuchida, K.~Kikuchi, Wavelength-multiplexed
  entanglement distribution, {\it Optical Fiber Technology\/} {\bf 16}, 225
  (2010).

\bibitem{TBBT_website}
ICFO-Website  (2016). \texttt{www.thebigbelltest.org}.

\end{thebibliography}


\begin{thebibliography}{100}

\makeatletter
\addtocounter{\@listctr}{123}
\makeatother

\bibitem{marsili2013detecting}
F.~Marsili, {\it et~al.\/}, Detecting single infrared photons with 93\% system
  efficiency, {\it Nature Photonics\/} {\bf 7}, 210 (2013).

\bibitem{PhysRevLett.115.250402}
L.~K. Shalm, {\it et~al.\/}, Strong loophole-free test of local realism, {\it
  Phys. Rev. Lett.\/} {\bf 115}, 250402 (2015).

\bibitem{PhysRevD.10.526}
J.~F. Clauser, M.~A. Horne, Experimental consequences of objective local
  theories, {\it Phys. Rev. D\/} {\bf 10}, 526 (1974).

\end{thebibliography}


\begin{thebibliography}{10}

\makeatletter
\addtocounter{\@listctr}{64}
\makeatother

\bibitem{Branciard2010}
C.~Branciard, N.~Gisin, S.~Pironio, Characterizing the nonlocal correlations
  created via entanglement swapping, {\it Phys. Rev. Lett.\/} {\bf 104}, 170401
  (2010).

\bibitem{Branciard2012}
C.~Branciard, D.~Rosset, N.~Gisin, S.~Pironio, Bilocal versus nonbilocal
  correlations in entanglement-swapping experiments, {\it Phys. Rev. A\/} {\bf
  85}, 032119 (2012).

\bibitem{polinomial}
R.~Chaves, Polynomial bell inequalities, {\it Phys. Rev. Lett.\/} {\bf 116},
  010402 (2016).

\bibitem{fritz2012}
T.~Fritz, Beyond bell's theorem: correlation scenarios, {\it New Journal of
  Physics\/} {\bf 14}, 103001 (2012).

\bibitem{takavoli2014}
A.~Tavakoli, P.~Skrzypczyk, D.~Cavalcanti, A.~Ac\'{\i}n, Nonlocal correlations
  in the star-network configuration, {\it Phys. Rev. A\/} {\bf 90}, 062109
  (2014).

\bibitem{rosset}
D.~Rosset, {\it et~al.\/}, Nonlinear bell inequalities tailored for quantum
  networks, {\it Phys. Rev. Lett.\/} {\bf 116}, 010403 (2016).

\bibitem{takavoli2016}
A.~Tavakoli, Bell-type inequalities for arbitrary noncyclic networks, {\it
  Phys. Rev. A\/} {\bf 93}, 030101 (2016).

\bibitem{Gisin2017}
N.~Gisin, Q.~Mei, A.~Tavakoli, M.~O. Renou, N.~Brunner, All entangled pure
  quantum states violate the bilocality inequality, {\it Phys. Rev. A\/} {\bf
  96}, 020304 (2017).

\bibitem{AndreoliNJP2017}
F.~Andreoli, G.~Carvacho, L.~Santodonato, R.~Chaves, F.~Sciarrino, Maximal
  qubit violation of n-locality inequalities in a star-shaped quantum network,
  {\it New Journal of Physics\/} {\bf 19}, 113020 (2017).

\bibitem{distribution}
S.~Perseguers, G.~J.~L. Jr, D.~Cavalcanti, M.~Lewenstein, A.~Ac{\'\i}n,
  Distribution of entanglement in large-scale quantum networks, {\it Reports on
  Progress in Physics\/} {\bf 76}, 096001 (2013).

\bibitem{control}
S.~M. Hein, F.~Schulze, A.~Carmele, A.~Knorr, Entanglement control in quantum
  networks by quantum-coherent time-delayed feedback, {\it Phys. Rev. A\/} {\bf
  91}, 052321 (2015).

\bibitem{Cavalcanti2011}
D.~Cavalcanti, M.~L. Almeida, V.~Scarani, A.~Ac{\'\i}n, Quantum networks reveal
  quantum nonlocality, {\it Nature comms.\/} {\bf 2}, 184 (2011).

\bibitem{markovian}
T.~Ramos, B.~Vermersch, P.~Hauke, H.~Pichler, P.~Zoller, Non-markovian dynamics
  in chiral quantum networks with spins and photons, {\it Phys. Rev. A\/} {\bf
  93}, 062104 (2016).

\bibitem{clocks}
P.~Komar, {\it et~al.\/}, A quantum network of clocks, {\it Nat Phys\/} {\bf
  10}, 582 (2014).

\bibitem{bilocality_nostro}
G.~Carvacho, {\it et~al.\/}, Experimental violation of local causality in a
  quantum network, {\it Nature comms.\/} {\bf 8}, 14775 (2017).

\bibitem{saunders}
D.~J. Saunders, A.~J. Bennet, C.~Branciard, G.~J. Pryde, Experimental
  demonstration of nonbilocal quantum correlations, {\it Science Advances\/}
  {\bf 3} (2017).

\bibitem{expandreoli}
F.~Andreoli, {\it et~al.\/}, Experimental bilocality violation without shared
  reference frames, {\it Phys. Rev. A\/} {\bf 95}, 062315 (2017).

\end{thebibliography}


\begin{thebibliography}{10}

\makeatletter
\addtocounter{\@listctr}{60}
\makeatother

\bibitem{RukhinNIST2010}
A.~Rukhin, {\it et~al.\/}, A statistical test suite for random and pseudorandom
  number generators for cryptographic applications, {\it {\normalfont National
  Institutes of Standards and Technology, Special Publication {(NIST SP)} -
  800-22 Rev 1a}\/}  (2010).

\bibitem{CHSH}
J.~F. Clauser, M.~A. Horne, A.~Shimony, R.~A. Holt, Proposed experiment to test
  local hidden-variable theories, {\it Phys. Rev. Lett.\/} {\bf 23}, 880
  (1969).

\bibitem{putz14}
G.~P\"utz, D.~Rosset, T.~J. Barnea, Y.-C. Liang, N.~Gisin, Arbitrarily small
  amount of measurement independence is sufficient to manifest quantum
  nonlocality, {\it Phys. Rev. Lett.\/} {\bf 113}, 190402 (2014).

\bibitem{Yuan15b}
X.~Yuan, Q.~Zhao, X.~Ma, Clauser-horne bell test with imperfect random inputs,
  {\it Phys. Rev. A\/} {\bf 92}, 022107 (2015).

\end{thebibliography}


\begin{thebibliography}{100}

\makeatletter
\addtocounter{\@listctr}{116}
\makeatother

\bibitem{Franson89}
J.~D. Franson, Bell inequality for position and time, {\it Phys. Rev. Lett.\/}
  {\bf 62}, 2205 (1989).

\bibitem{Aerts99}
S.~Aerts, P.~Kwiat, J.-{\AA}. Larsson, M.~\.{Z}ukowski, Two-photon
  {F}ranson-type experiments and local realism, {\it Phys. Rev. Lett.\/} {\bf
  83}, 2872 (1999).

\bibitem{Adan09}
A.~Cabello, A.~Rossi, G.~Vallone, F.~De~Martini, P.~Mataloni, Proposed {B}ell
  experiment with genuine energy-time entanglement, {\it Phys. Rev. Lett.\/}
  {\bf 102}, 040401 (2009).

\bibitem{Cuevas13}
A.~Cuevas, {\it et~al.\/}, Long-distance distribution of genuine energy-time
  entanglement, {\it Nature Communications\/} {\bf 4}, 2871 (2013).

\bibitem{Cuevas15}
G.~Carvacho, {\it et~al.\/}, Postselection-loophole-free {B}ell test over an
  installed optical fiber network, {\it Phys. Rev. Lett.\/} {\bf 115}, 030503
  (2015).

\bibitem{Xavier11}
G.~B. Xavier, J.~P. von~der Weid, Stable single-photon interference in a 1 km
  fiber-optic {M}ach--{Z}ehnder interferometer with continuous phase
  adjustment, {\it Optics Letters\/} {\bf 36}, 1764 (2011).

\bibitem{CHSH69}
J.~F. Clauser, M.~A. Horne, A.~Shimony, R.~A. Holt, Proposed experiment to test
  local hidden-variable theories, {\it Phys. Rev. Lett.\/} {\bf 23}, 880
  (1969).

\end{thebibliography}


\begin{thebibliography}{10}

\bibitem{BellP1964}
J.~S. Bell, On the {E}instein-{P}odolsky-{R}osen paradox, {\it Physics\/} {\bf
  1}, 195 (1964).

\bibitem{LarssonJPA2014Official}
J.-{\AA}. Larsson, Loopholes in {B}ell inequality tests of local realism, {\it
  Journal of Physics A: Mathematical and Theoretical\/} {\bf 47}, 424003
  (2014).

\bibitem{KoflerPRA2016}
J.~Kofler, M.~Giustina, J.-{\AA}. Larsson, M.~W. Mitchell, Requirements for a
  loophole-free photonic {Bell} test using imperfect setting generators, {\it
  Phys. Rev. A\/} {\bf 93}, 032115 (2016).

\bibitem{HensenN2015}
B.~Hensen, {\it et~al.\/}, Loophole-free {Bell} inequality violation using
  electron spins separated by 1.3 kilometres, {\it Nature\/} {\bf 526}, 682
  (2015).

\bibitem{GiustinaPRL2015}
M.~Giustina, {\it et~al.\/}, Significant-loophole-free test of {B}ell's theorem
  with entangled photons, {\it Phys. Rev. Lett.\/} {\bf 115}, 250401 (2015).

\bibitem{ShalmPRL2015}
L.~K. Shalm, {\it et~al.\/}, Strong loophole-free test of local realism, {\it
  Phys. Rev. Lett.\/} {\bf 115}, 250402 (2015).

\bibitem{RosenfeldPRL2017}
W.~Rosenfeld, {\it et~al.\/}, Event-ready {Bell} test using entangled atoms
  simultaneously closing detection and locality loopholes, {\it Phys. Rev.
  Lett.\/} {\bf 119}, 010402 (2017).

\bibitem{BellBook2004Ch7}
J.~Bell, {\it Free variables and local causality\/}, Speakable and Unspeakable
  in Quantum Mechanics: Collected Papers on Quantum Philosophy (Cambridge
  University Press, 2004), chap.~7.

\bibitem{BBTWebsite}
{http://thebigbelltest.org}.

\bibitem{FarreraNComms2016}
P.~Farrera, {\it et~al.\/}, Generation of single photons with highly tunable
  wave shape from a cold atomic ensemble, {\it Nature Communications\/} {\bf 7}
  (2016).

\bibitem{WallraffN2004}
A.~Wallraff, {\it et~al.\/}, Strong coupling of a single photon to a
  superconducting qubit using circuit quantum electrodynamics, {\it Nature\/}
  {\bf 431}, 162 (2004).

\bibitem{CarvachoNC2017}
G.~Carvacho, {\it et~al.\/}, Experimental violation of local causality in a
  quantum network, {\it Nature Communications\/} {\bf 8}, 14775 (2017).

\bibitem{ScheidlPNAS2010}
T.~Scheidl, {\it et~al.\/}, Violation of local realism with freedom of choice,
  {\it Proceedings of the National Academy of Sciences of the United States of
  America\/} {\bf 107}, 19708 (2010).

\bibitem{SorensenN2016}
J.~S\o{}rensen, {\it et~al.\/}, Exploring the quantum speed limit with computer
  games, {\it Nature\/} {\bf 532}, 210 (2016).

\bibitem{ShimonySEP2005}
A.~Shimony, {\it Bell's Theorem\/}, The Stanford Encyclopedia of Philosophy
  (Metaphysics Research Lab, Stanford University, 2005), winter 2016 edn.

\bibitem{ColbeckThesis2007}
R.~Colbeck, Quantum and relativistic protocols for secure multi-party
  computation, {\it Ph.D. Thesis, Cambridge University\/}  (2007).

\bibitem{HoeferSEP2005}
C.~Hoefer, {\it Causal Determinism\/}, The Stanford Encyclopedia of Philosophy
  (Metaphysics Research Lab, Stanford University, 2005), spring 2016 edn.

\bibitem{AcinN2016}
A.~Ac\'{i}n, L.~Masanes, Certified randomness in quantum physics, {\it
  Nature\/} {\bf 540}, 213 (2016).

\bibitem{AaronsonAS2014}
S.~Aaronson, Quantum randomness, {\it American Scientist\/} {\bf 102}, 266
  (2014).

\bibitem{AbellanPRL2015}
C.~Abell\'an, W.~Amaya, D.~Mitrani, V.~Pruneri, M.~W. Mitchell, Generation of
  fresh and pure random numbers for loophole-free {Bell} tests, {\it Phys. Rev.
  Lett.\/} {\bf 115}, 250403 (2015).

\bibitem{FurstMOE2010}
M.~F\"{u}rst, {\it et~al.\/}, High speed optical quantum random number
  generation, {\it Optics Express\/} {\bf 18}, 13029 (2010).

\bibitem{GallicchioPRL2014}
J.~Gallicchio, A.~S. Friedman, D.~I. Kaiser, Testing bell's inequality with
  cosmic photons: Closing the setting-independence loophole, {\it Phys. Rev.
  Lett.\/} {\bf 112}, 110405 (2014).

\bibitem{HandsteinerPRL2017}
J.~Handsteiner, {\it et~al.\/}, Cosmic {B}ell test: Measurement settings from
  milky way stars, {\it Phys. Rev. Lett.\/} {\bf 118}, 060401 (2017).

\bibitem{WuPRL2017}
C.~Wu, {\it et~al.\/}, Random number generation with cosmic photons, {\it Phys.
  Rev. Lett.\/} {\bf 118}, 140402 (2017).

\bibitem{BeraRPP2017}
M.~N. Bera, A.~Ac{\'\i}n, M.~K. M.~W. Mitchell, M.~Lewenstein, Randomness in
  quantum mechanics: philosophy, physics and technology, {\it Reports on
  Progress in Physics\/} {\bf 80}, 124001 (2017).

\bibitem{PironioN2010}
S.~Pironio, {\it et~al.\/}, Random numbers certified by {B}ell's theorem, {\it
  Nature\/} {\bf 464}, 1021 (2010).

\bibitem{BarHillelAAM1991}
M.~Bar-Hillel, W.~A. Wagenaar, The perception of randomness, {\it Advances in
  Applied Mathematics\/} {\bf 12}, 428 (1991).

\bibitem{BierhorstJPA2015}
P.~{Bierhorst}, {A robust mathematical model for a loophole-free Clauser-Horne
  experiment}, {\it Journal of Physics A: Mathematical and Theoretical\/} {\bf
  48}, 195302 (2015).

\bibitem{ElkoussNPJQI2016}
D.~Elkouss, S.~Wehner, ({N}early) optimal {P} values for all {B}ell
  inequalities, {\it npj Quantum Information\/} {\bf 2}, 16026 EP  (2016).

\bibitem{PutzPRL2014}
G.~P\"utz, D.~Rosset, T.~J. Barnea, Y.-C. Liang, N.~Gisin, Arbitrarily small
  amount of measurement independence is sufficient to manifest quantum
  nonlocality, {\it Phys. Rev. Lett.\/} {\bf 113}, 190402 (2014).

\end{thebibliography}

\begin{thebibliography}{10}
\makeatletter
\addtocounter{\@listctr}{30}
\makeatother

\bibitem{EinsteinPR1935}
A.~Einstein, B.~Podolsky, N.~Rosen, Can quantum-mechanical description of
  physical reality be considered complete?, {\it Phys. Rev.\/} {\bf 47}, 777
  (1935).

\bibitem{HallPRA2011}
M.~J.~W. Hall, Relaxed Bell inequalities and Kochen-Specker theorems, {\it
  Phys. Rev. A\/} {\bf 84}, 022102 (2011).

\bibitem{BarrettPRL2011}
J.~Barrett, N.~Gisin, How much measurement independence is needed to
  demonstrate nonlocality?, {\it Phys. Rev. Lett.\/} {\bf 106}, 100406 (2011).


\bibitem{BellD1985}
J.~S.~Bell, Bell, J., Clauser, M. Horne, and A. Shimony, An exchange on local beables, {\it Dialectica\/} {\bf 39}, 85--110 (1985).

\bibitem{ConwayFP2006}
J.~Conway, S.~Kochen, The free will theorem, {\it Foundations of Physics\/}
  {\bf 36}, 1441 (2006).

\bibitem{BellBook2004}
J.~Bell, {\it Speakable and Unspeakable in Quantum Mechanics: Collected Papers
  on Quantum Philosophy\/}, Collected papers on quantum philosophy (Cambridge
  University Press, 2004).

\bibitem{ErvenNPhoton2014}
C.~C. Erven, {\it et~al.\/}, Experimental three-photon quantum nonlocality
  under strict locality conditions, {\it Nat Photon\/} {\bf 8}, 292 (2014).

\bibitem{WeihsPRL1998}
G.~Weihs, T.~Jennewein, C.~Simon, H.~Weinfurter, A.~Zeilinger, Violation of
  {B}ell's inequality under strict {E}instein locality conditions, {\it Phys.
  Rev. Lett.\/} {\bf 81}, 5039 (1998).

\bibitem{Wagenaar72generationof}
W.~A. Wagenaar, Generation of random sequences by human subjects: A critical
  survey of the literature, {\it Psychological Bulletin\/} pp. 65--72 (1972).

\bibitem{rapoport1992generation}
A.~Rapoport, D.~V. Budescu, Generation of random series in two-person strictly
  competitive games., {\it Journal of Experimental Psychology: General\/} {\bf
  121}, 352 (1992).

\bibitem{gibbons1992game}
R.~Gibbons, {\it Game theory for applied economists\/} (Princeton University
  Press, 1992).

\bibitem{MOOKHERJEE199462}
D.~Mookherjee, B.~Sopher, Learning behavior in an experimental matching pennies
  game, {\it Games and Economic Behavior\/} {\bf 7}, 62  (1994).

\bibitem{serfozo2009basics}
R.~Serfozo, {\it Basics of applied stochastic processes\/} (Springer Science \&
  Business Media, 2009).

\bibitem{HeckARX2017}
R.~{Heck}, {\it et~al.\/}, {Do physicists stop searches too early? A
  remote-science, optimization landscape investigation} {\bf { }} (2017).

\end{thebibliography}
\end{document}